\documentclass[twocolumn,english]{revtex4-1}
\usepackage[T1]{fontenc}
\usepackage[latin9]{inputenc}
\usepackage{babel}
\usepackage{textcomp}
\usepackage{bm}
\usepackage{amsmath}
\usepackage{amssymb}
\usepackage{graphicx}
\usepackage[unicode=true,pdfusetitle,
 bookmarks=true,bookmarksnumbered=false,bookmarksopen=false,
 breaklinks=false,pdfborder={0 0 1},backref=false,colorlinks=false]
 {hyperref}
 \usepackage{fixltx2e}
\begin{document}

\title{Antiferromagnetic cavity optomagnonics}

\author{T. S. Parvini} 
\email{tahereh.parvini@mpl.mpg.de}
\affiliation{Max Planck Institute for the Science of Light, Staudtstr. 2, PLZ
91058, Erlangen, Germany}

\author{V. A. S. V. Bittencourt}
\email{victor.bittencourt@mpl.mpg.de}
\affiliation{Max Planck Institute for the Science of Light, Staudtstr. 2, PLZ
91058, Erlangen, Germany}

\author{Silvia {Viola Kusminskiy}}
\email{silvia.viola-kusminskiy@mpl.mpg.de}
\affiliation{Max Planck Institute for the Science of Light, Staudtstr. 2, PLZ
91058, Erlangen, Germany}
\affiliation{Institute for Theoretical Physics, University Erlangen-N\"urnberg, Staudtstr. 7, 91058 Erlangen, Germany}

\begin{abstract}
Currently there is a growing interest in studying the coherent interaction between magnetic systems and electromagnetic radiation in a cavity, prompted partly by possible applications in hybrid quantum systems. We propose a multimode cavity optomagnonic system based on antiferromagnetic insulators, where optical photons couple coherently to the two homogeneous magnon modes of the antiferromagnet. These have frequencies typically in the THz range, a regime so far mostly unexplored in the realm of coherent interactions, and which makes antiferromagnets attractive for quantum transduction from THz to optical frequencies. We derive the theoretical model for the coupled system, and show that it presents unique characteristics. In particular, if the antiferromagnet presents hard-axis magnetic anisotropy, the optomagnonic coupling can be tuned by a magnetic field applied along the easy axis. This allows to bring a selected magnon mode into and out of a dark mode, providing an alternative for a quantum memory protocol. The dynamical features of the driven system present unusual behavior due to optically induced magnon-magnon interactions, including regions of magnon heating for a red detuned driving laser. The multimode character of the system is evident in a substructure of the optomagnonically induced transparency window.  
\end{abstract}

\maketitle

\emph{Introduction.--} The interaction between light and magnetism at the quantum level holds promise for future information technologies. In seminal recent experiments, the coherent coupling of magnons (the spin-wave quanta) to optical photons has been demonstrated in solid state optomagnonic cavities \citep{zhang_optomagnonic_2016,osada_cavity_2016,haigh_triple-resonant_2016}. These are dielectric magnetic structures capable of simultaneously confining light and magnons, providing an enhancement of the magnon-photon coupling and enabling the study of cavity effects in a new platform. Magnons in these structures exhibit good coherence properties and tunable frequencies, and can couple strongly to microwave (MW) cavity fields \citep{soykal_strong_2010,huebl_high_2013,zhang_strongly_2014,tabuchi_hybridizing_2014,goryachev_high_2014,lambert_identification_2015,bourhill_ultrahigh_2016}. Quantum memories \citep{afzelius_demonstration_2010,sangouard_quantum_2011,timoney_single_2013,lambert_all-optical_2014,jobez_cavity_2014,jobez_coherent_2015,gundogan_solid_2015,zhangmagnondark2015} or transducers allowing for coherent information transfer between MW and optics are envisioned applications \citep{lachance-quirion_hybrid_2019,hisatomi_bidirectional_2016}. 

\begin{figure}[t]
\begin{centering}
\includegraphics[width=0.8\columnwidth]{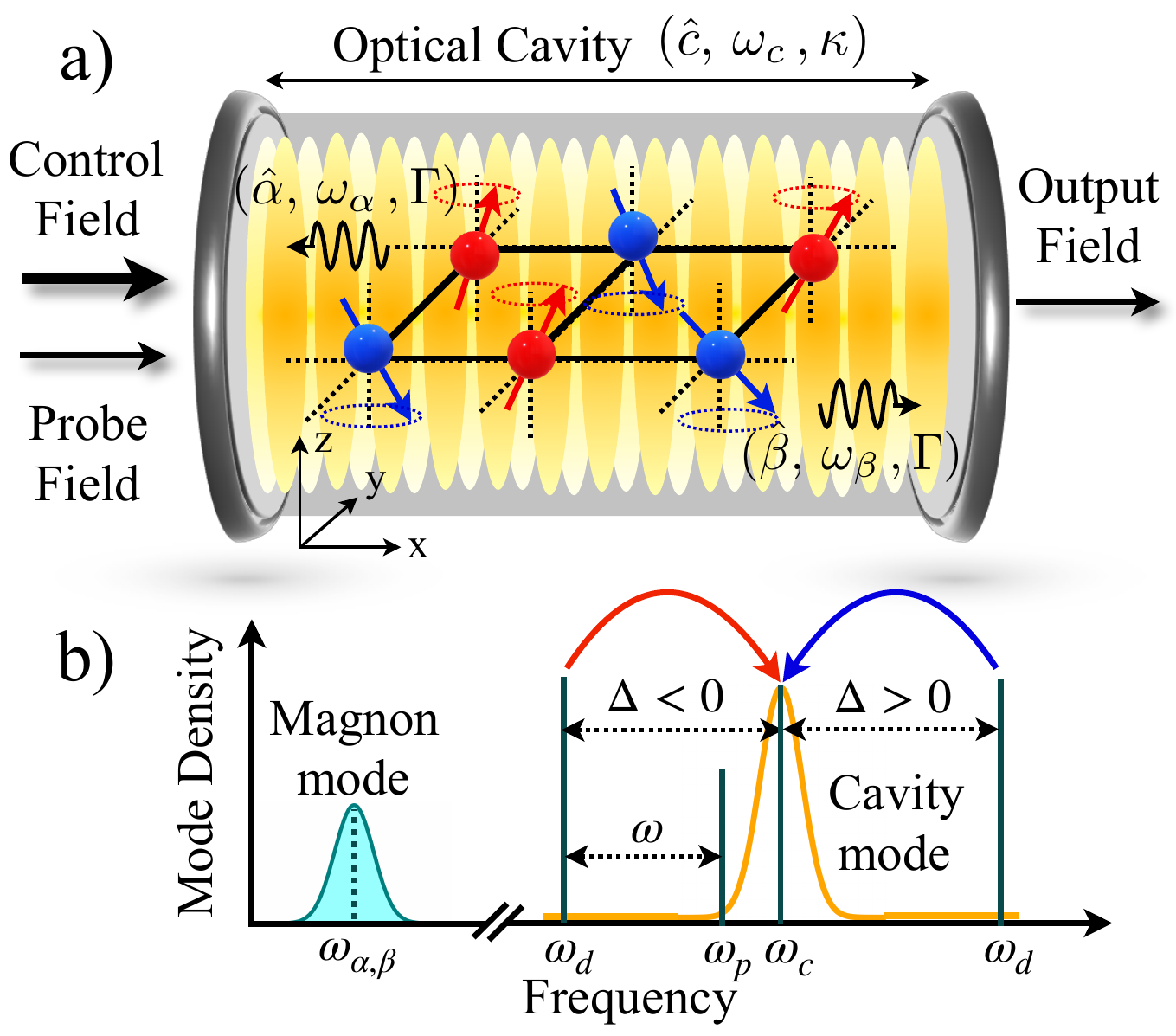}
\par\end{centering}
\centering{}\caption{(a) Schematics of the antiferromagnetic optomagnonic cavity. The homogeneous magnon modes $\hat{\alpha}$ and $\hat{\beta}$ with frequencies $\omega_{\alpha,\beta}$ and decay rates $\Gamma_{\alpha,\beta}$ couple to a cavity mode $\hat{c}$ with frequency $\omega_{c}$ and decay rate $\kappa$. (b) Pump-probe scheme: $\Omega_{m}$, $\omega_{c}$, $\omega_{d}$, $\omega_{p}$, magnon, optical cavity resonance, drive, and probe frequencies respectively. The detuning of the drive is $\Delta=\omega_{d}-\omega_{c}$, and $\omega=\omega_{p}-\omega_{d}$.}
\label{Figure01}
\end{figure}

Exciting developments have also emerged very recently in the THz regime, a frequency range which has been historically challenging due to the absence of efficient sources and detectors (the ``THz gap'') \citep{hesleradvancesin2019}. Strong light-matter coupling in the THz regime has been achieved employing cavities \cite{scalariultrastrongcoupling2012, livacuumbloch2018, paravicinimagnetotransport2019}, including strong coupling to magnons in antiferromagnets (AFM) \citep{sivarajahthz2019, bialekspinwave2020}, opening the door for quantum applications in the THz domain. Antiferromagnetic materials support magnons that can be described as excitations of a spin anti-aligned ground state (the N\`eel state) \citep{keffer_theory_1952}. Their frequencies are typically in the THz range, making AFMs ideal candidates to incorporate in THz platforms \citep{gomonay_antiferromagnetic_2018,jungwirth_the_2018,jungfleisch_perspectives_2018,baltz_antiferromagnetic_2018}. The physics of AFMs in combination with electromagnetic cavities is just starting to be explored. Besides the mentioned experiments in the THz regime \citep{sivarajahthz2019, bialekspinwave2020}, strong coupling between MW photons and AFM magnons has been reported \citep{mergenthaler_strong_2017}, while magnon dark modes \citep{xiao_magnon_2019} and coupling to ferromagnets \citep{johansen_nonlocal_2018} via a MW cavity have been proposed theoretically. In turn, methods involving light to probe and control AFMs are being developed \citep{higuchicontrolofantiferromagnetic2016,nemec_antiferromagnetic_2018}. These developments are a great incentive to study the coupling of AFM magnons to {\it optical} cavities, which could lead to quantum transducers from the THz to the optical regime.

In this letter, we propose a cavity optomagnonic system based on an antiferromagnetic insulator, see Fig.~\ref{Figure01}. Optomagnonic cavities have been investigated so far exclusively within the scope of ferromagnetic (FM) magnons with GHz frequencies, both experimentally \citep{haigh_magneto-optical_2015,zhang_optomagnonic_2016,haigh_triple-resonant_2016,osada_cavity_2016,osada_brillouin_2018,osada_orbital_2018} (although the material of choice Yttrium Iron Garnet (YIG) is a \emph{ferri}magnet, one sublattice has a much larger spin and is dominant) and theoretically  \citep{liu_optomagnonics_2016, almpanisdielectric2018, pantazopoulosphotomagnonic2017,pantazopouloshighefficiency2019,viola_kusminskiy_coupled_2016,graf_cavity_2018,sharma_optical_2018,bittencourt_magnon_2019}. We show that for an AFM new phenomenology emerges. We derive the Hamiltonian governing the system and show that, in the presence of hard-axis anisotropy, the optomagnonic coupling to both supported homogeneous magnon modes can be tuned by an external magnetic field. In particular, we show that the magnon modes can be selectively decoupled from the cavity, rendering them dark. This is unique to AFM cavity optomagnonics. Based on this tunability, we sketch a quantum memory protocol. We further characterize the dynamical response of the system and show that the cavity-mediated interaction between the AFM magnon modes leads to unusual optically induced magnon cooling and heating. 

\emph{Model.--} We consider an AFM insulator with two magnetic sublattices $A$ and $B$ of opposite spin. The AFM hosts two homogeneous magnon modes $\alpha$ and $\beta$, see Fig.~\ref{Figure01}, and we assume it acts also as an optical cavity by total internal reflection, analogous to dielectric optomechanical \citep{aspelmeyer_cavity_2014} or optomagnonic \citep{haigh_triple-resonant_2016,osada_cavity_2016,zhang_optomagnonic_2016} cavities. AFMs with a high index of refraction and low absorption in the optical range, such as ${\rm NiO}$ ($n\approx2.4$) \citep{powell_optical_1970} would serve the purpose, or heterostructures containing ${\rm MnF_{2}}$ ($n\thickapprox1.4$) \citep{dodge_refractive_1984} or ${\rm FeF_{2}}$ ($n\thickapprox1.5$) \citep{jahn_linear_1973}. The Hamiltonian of the coupled system is 
\begin{equation}
\hat{\mathcal{H}}=\hat{\mathcal{H}}_{{\rm ph}}+\hat{\mathcal{H}}_{{\rm AFM}}+\hat{\mathcal{H}}_{{\rm OM}}\,,\label{eq:Htot}
\end{equation}
with $\hat{\mathcal{H}}_{{\rm ph}}$ and $\hat{\mathcal{H}}_{{\rm AFM}}$ the free photonic and AFM Hamiltonians, respectively. $\hat{\mathcal{H}}_{{\rm OM}}$ contains the coupling between the AFM magnons and the cavity photons, and is our main result in this section. 

The quantized optical field in the cavity is $\hat{\mathbf{E}}(\mathbf{r},t)=1/2\sum_{\xi}\left(\mathbf{E}_{\xi}(\mathbf{r})\hat{c}_{\xi}(t)+\mathbf{E}_{\xi}^{*}(\mathbf{r})\hat{c}_{\xi}^{\dagger}(t)\right)$ with $\hat{c}_{\xi}^{(\dagger)}$ annihilation (creation) operator of mode $\xi$ with resonance frequency $\omega_{\xi}$, hence
$\mathbf{\hat{\mathcal{H}}_{{\rm ph}}=\hbar}\sum_{\xi}\omega_{\xi}\hat{c}_{\xi}^{\dagger}\hat{c}_{\xi}
$. $\hat{\mathcal{H}}_{{\rm AFM}}$ consists of (i) the exchange interaction between nearest-neighbor spins $J\sum_{\langle i\neq j\rangle}\hat{\mathbf{S}}_{i}\cdot\hat{\mathbf{S}}_{j}$ ($J>0$), (ii) the Zeeman interaction between spins and an external DC magnetic field $\bm{B}_{0}$ along $\bm{e}_{z}$, $\left|\gamma\right|B_{0}\sum_{i}\hat{\mathbf{S}}_{i}^{z}$ ($\gamma$ gyromagnetic ratio), and (iii) easy $\text{\textminus}\frac{K_{\parallel}}{2}\sum_{i}\left(S_{i}^{z}\right)^{2}$ ($K_{\parallel}>0$) and hard $\frac{K_{\perp}}{2}\sum_{i}\left(S_{i}^{x}\right)^{2}$ ($K_{\bot}\geq0$) axis anisotropy in the $\bm{e}_{z}$ and $\bm{e}_{x}$ directions respectively. For small magnetization fluctuations around the N\`eel ordered state, the Holstein-Primakoff (HP) transformations \citep{kittelQuantumTheorySolids1963,holstein_field_1940} can be used to express $\hat{\mathcal{H}}_{{\rm AFM}}$ in terms of bosonic operators $\hat{a}_{\bm{k}}$ and $\hat{b}_{\bm{k}}$ associated with the sublattices $A$ and $B$. To first order, the HP transformations are given in terms of spin ladder operators as $\hat{S}_{A(B)}^{+}=\sqrt{2S/N}\sum_{\bm{k}}e^{-i\bm{k}.\bm{x}_{i}}\hat{a}_{\bm{k}}(\hat{b}_{\bm{k}}^{\dagger})$, where $N$ is the total number of sites per sublattice and $S$ the spin on each site. $\hat{\mathcal{H}}_{{\rm AFM}}$ is diagonalized via a 4D Bogoliubov transformation to the bosonic operators $\hat{\alpha}_{\bm{k}}=u_{\alpha,a}\hat{a}_{\bm{k}}+v_{\alpha,b}\hat{b}_{-\bm{k}}^{\dagger}+v_{\alpha,a}\hat{a}_{-\bm{k}}^{\dagger}+u_{\alpha,b}\hat{b}_{\bm{k}}$ and $\hat{\beta}_{\bm{k}}=u_{\beta,a}\hat{a}_{\bm{k}}+v_{\beta,b}\hat{b}_{-\bm{k}}^{\dagger}+v_{\beta,a}\hat{a}_{-\bm{k}}^{\dagger}+u_{\beta,b}\hat{b}_{\bm{k}}$ (see Sup. Mat. \citep{parviniSupplementalMaterial}): 
$
\mathcal{\hat{H}}_{{\rm AFM}}=\hbar\sum_{\bm{k}}\left[\omega_{\alpha\bm{k}}\hat{\alpha}_{\bm{k}}^{\dagger}\hat{\alpha}_{\bm{k}}+\omega_{\beta\bm{k}}\hat{\beta}_{\bm{k}}^{\dagger}\hat{\beta}_{\bm{k}}\right],
$
with $\omega_{\alpha,\beta\bm{k}}$ the respective eigenfrequencies. We restrict our analysis to the two homogeneous ($\bm{k}=0$)
AFM magnon modes, hence from now onwards we drop the index $\bm{k}$. The corresponding magnon frequencies $\omega_{\alpha,\beta}$ are functions of the characteristic frequencies $\omega_{E}=\hbar JSN$, $\omega_{\parallel,\bot}=\hbar SNK_{\parallel,\bot}$, and $\omega_{H}=\vert\gamma\vert B_{0}$: $\omega_{\alpha(\beta)}^{2}=\omega_{H}^{2}+\omega_{E}\omega_{\perp}+2\omega_{E}\omega_{\parallel}+\omega_{\perp}\omega_{\parallel}+\omega_{\parallel}^{2}\pm\sqrt{\omega_{E}^{2}\omega_{\perp}^{2}+4\omega_{E}\omega_{H}^{2}(\omega_{\perp}+2\omega_{\parallel})+\omega_{H}^{2}(\omega_{\perp}+2\omega_{\parallel})^{2}}$ \citep{kamra_noninteger-spin_2017,kamra_spin_2017,johansen_nonlocal_2018}. Note that $\omega_{\alpha}\ge\omega_{\beta}$ and hence $\alpha$ ($\beta$) labels the upper (lower) mode. Whereas $\omega_{\alpha}$ increases with the magnetic field, $\omega_{\beta}$ decreases and goes to zero at the onset of the spin-flop phase at $\omega_{H} = \omega_{SF}\approx\sqrt{2\omega_{E}\omega_{\parallel}}$ \citep{machado_spin-flop_2017}. 

The interaction between light and magnetization is described by the magneto-optical coupling $\mathcal{H}_{{\rm OM}}=\sum_{\mu,\nu}\int{\rm d}\bm{r}E_{\mu}^{*}(\bm{r})\varepsilon_{\mu\nu}(S_{\bm{r}})E_{\nu}(\bm{r})/4$, where $\varepsilon_{\mu\nu}\:(\mu,\nu=x,y,z)$ is the spin-dependent part of the permittivity tensor. In this letter we consider simple cubic and rutile-structure AFMs, other more complex structures will be considered elsewhere. For these materials, within linear response in the deviations from the magnetic equilibrium, $\mathcal{H}_{{\rm OM}}$ reduces to \citep{cottam_on_1975,cottamLightScatteringMagnetic1986,parviniSupplementalMaterial}
\begin{align} 
\mathcal{H}_{{\rm OM}}=  \frac{K_{+} V}{4 N}\sum_{i\in A,B} \left(P_{i}^{+} S_{-}^{i}-P_{i}^{-}S_{+}^{i} \right)&  \label{eq:interaction}\\
+  \frac{K_{-} V}{4 N} \Big[ \displaystyle \sum_{i \in A}\left(P_{i}^{+}S_{+}^{i}-P_{i}^{-}S_{-}^{i}\right) 
- \displaystyle \sum_{j \in B}&\left(P_{j}^{+}S_{+}^{j}-P_{j}^{-}S_{-}^{j}\right) \Big], \nonumber
\end{align}
where we have discretized the interaction and $P_{i}^{\pm}=E_{z}^{*}(\bm{r}_{i})E_{\pm}(\bm{r}_{i})-E_{\mp}^{*}(\bm{r}_{i})E_{z}(\bm{r}_{i})$, with $E_{\pm}=E_{x}\pm iE_{y}$ and $S_{\pm}=S_{x}\pm iS_{y}$. The linear magneto-optic coefficients $K_{\pm}$ correspond to processes in which the two sublattices scatter the light in-phase ($+$) or out-of-phase ($-$). For our purposes, one-magnon processes coming from quadratic terms in the spin (e.g. $\propto\hat{S}^{z}\hat{S}^{\pm}$) can be absorbed in the definition of $K_{\pm}$. This model applies e.g. to the uniaxial AFMs ${\rm MnF}_{2}$ or ${\rm FeF}_{2}$ \citep{lockwood_magnetooptic_2012,ariai_effects_1982}, and for the simple cubic AFM NiO for which $K_{-}=0$. 

We obtain the optomagnonic coupling Hamiltonian $\hat{\mathcal{H}}_{{\rm OM}}$ by quantizing Eq.~(\ref{eq:interaction}) assuming that the electric field varies smoothly, such that $P_{i}^{\pm}\approx P_{j}^{\pm}$ for nearest neighbors. We focus on the interaction between the homogeneous AFM magnon modes $\hat{\alpha}$ and $\hat{\beta}$ with a \emph{single} optical mode $\hat{c}$ with frequency $\omega_{c}$. For an optical mode with circular polarization in the $yz$ plane, from Eq.~(\ref{eq:interaction}) we obtain \citep{parviniSupplementalMaterial}
\begin{align}
\hat{\mathcal{H}}_{{\rm OM}} & =-\hbar G\hat{c}^{\dagger}\hat{c}(g_{\alpha}\hat{\alpha}^{\dagger}+g_{\beta}\hat{\beta}^{\dagger}+{\rm h.c.}),\label{eq:OMCouplingBogoliubov-1}
\end{align}
with ($\varepsilon$ is the AFM dielectric constant)
\begin{equation}
G=\frac{\omega_{c}K_{+}}{8\varepsilon}\sqrt{\frac{2S}{N}}\,.\label{eq:couplingG}
\end{equation}
The AFM optomagnonic coupling depends on the Bogoliubov coefficients through
\begin{align}
g_{\alpha(\beta)} & =\left(u_{\alpha(\beta)}^{+}+v_{\alpha(\beta)}^{+}\right)+K\left(u_{\alpha(\beta)}^{-}+v_{\alpha(\beta)}^{-}\right)\label{eq:CouplingStrength-1}
\end{align}
where $K=K_{-}/K_{+}$ quantifies the intrinsic magneto-optical asymmetry between the sublattices, and we have defined $u_{\alpha(\beta)}^{\pm}=u_{a,\alpha(\beta)}\pm u_{b,\alpha(\beta)}$ and $v_{\alpha(\beta)}^{\pm}=v_{a,\alpha(\beta)}\pm v_{b,\alpha(\beta)}$. Hence, the two AFM magnon modes $\hat{\alpha}$ and $\hat{\beta}$ couple, in general, with different strength to the cavity mode.

\emph{Optomagnonic Coupling.--} The constant $G$ describes the coupling to the magnetization's fluctuations sector and is consistent with the one derived in Ref. \citep{viola_kusminskiy_coupled_2016} for the optomagnonic coupling in a \emph{ferro}magnetically ordered system. Assuming equivalent sublattices with Faraday rotation per unit length $\theta_{{\rm F}}$, then $K_+ =  c \sqrt{\varepsilon} \theta_{\rm{F}}/(\omega_c S)$ (with $c$ the speed of light) and thus $G=\left(1/\sqrt{2NS}\right)\left(c\theta_{{\rm F}}/4\sqrt{\varepsilon}\right)$. The $1/\sqrt{N}$ dependence indicates that the \emph{density} of excitations is relevant for the coupling, favoring small magnetic volumes. Due to the lack of data on absolute values for $K_{+}$ (or $\theta_{{\rm F}}$) for simple AFMs, we take as an estimate for $G$ the value for $\left(1\mu{\rm m}\right)^{3}$ YIG (diffraction limit volume), $G_{{\rm YIG}}=0.1{\rm MHz}$ \citep{viola_kusminskiy_coupled_2016}. Some measurements indicate that the Faraday rotation coefficient in AFMs can be quite large, e.g. similar values as for YIG have been reported for ${\rm BiFeO}_{3}$ \citep{bi_structural_2008}.  Note that $G$ given in Eq.~(\ref{eq:couplingG}) assumes perfect mode matching. Imperfect mode overlap can be accounted for by a mode-volume ratio factor \citep{viola_kusminskiy_coupled_2016} and it is responsible for a suppression of the coupling in current experiments with YIG \citep{sharma_2019,lachance-quirion_hybrid_2019}. The second term in Eq.~(\ref{eq:CouplingStrength-1}) gives a contribution proportional to $KG$ and describes the coupling to fluctuations of the N\`eel vector. Typical values of $K$ are $K\approx0.01$ (\emph{e.g. }for ${\rm MnF}_{2}$ or ${\rm FeF}_{2}$ \citep{lockwood_magnetooptic_2012}).

The reduced couplings $g_{\alpha,\beta}$ can be found analytically, but the general solution is lengthy. Simple expressions can be given in certain cases. Since the exchange energy is usually the largest energy scale in the AFM, the condition $\omega_{\perp,\parallel}\ll\omega_{E}$ holds. For an easy-axis AFM ($\omega_{\bot}=0$), we obtain \citep{parviniSupplementalMaterial}
\begin{equation}
g_{\alpha,\beta}^{\omega_{\bot}=0}\approx\left(\frac{\omega_{\parallel}}{2\omega_{E}}\right)^{1/4}\pm K\left(\frac{2\omega_{E}}{\omega_{\parallel}}\right)^{1/4}\,.\label{eq:coupling_g_a_b}
\end{equation}
Eq.~(\ref{eq:coupling_g_a_b}) is independent of the magnetic field $B_{0}$, a consequence of the axial symmetry of the system in this case \citep{parviniSupplementalMaterial}. In the absence of magneto-optical asymmetry ($K=0$) both modes couple equally to the light field, while for finite $K$, $g_{\alpha}^{\omega_{\bot}=0}$ ($g_{\beta}^{\omega_{\bot}=0}$) increases (decreases) linearly. If the condition $K=\sqrt{\omega_{\parallel}/2\omega_{E}}$ is met, $g_{\beta}^{\omega_{\bot}=0}=0$ and $\beta$ is a dark mode, completely decoupled from the cavity. Whereas this requires
fine tuning, it could be achievable in cold atoms realizations where the relevant parameters can be tuned \citep{brahms_spin_2010,kohler_cavity-assisted_2017,landini_formation_2018,kroeze_spinor_2018,mivehvar_cavity-quantum-electrodynamical_2019}. The situation nevertheless changes in the presence of hard-axis anisotropy, where the coupling to the modes can be tuned externally by the magnetic field as we show below. From Eq.~(\ref{eq:coupling_g_a_b}) we obtain $g_{\alpha,\beta}^{{\rm MnF}_{2}}\approx0.5,0.4$ ($\omega_{E}=9.3\,{\rm THz}$, $\omega_{\parallel}=0.15\,{\rm THz}$, $K=0.007$ \citep{barak_magnetic_nodate,lockwood_magnetooptic_2012}) and $g_{\alpha,\beta}^{{\rm FeF}_{2}}\approx0.6,0.7$ ($\omega_{E}=9.5\,{\rm THz}$, $\omega_{\parallel}=3.5\,{\rm THz}$, $K=0.01$ \citep{ohlmann_antiferromagnetic_1961,cottamLightScatteringMagnetic1986}).

\begin{figure}[t]
\begin{centering}
\includegraphics[width=1\columnwidth]{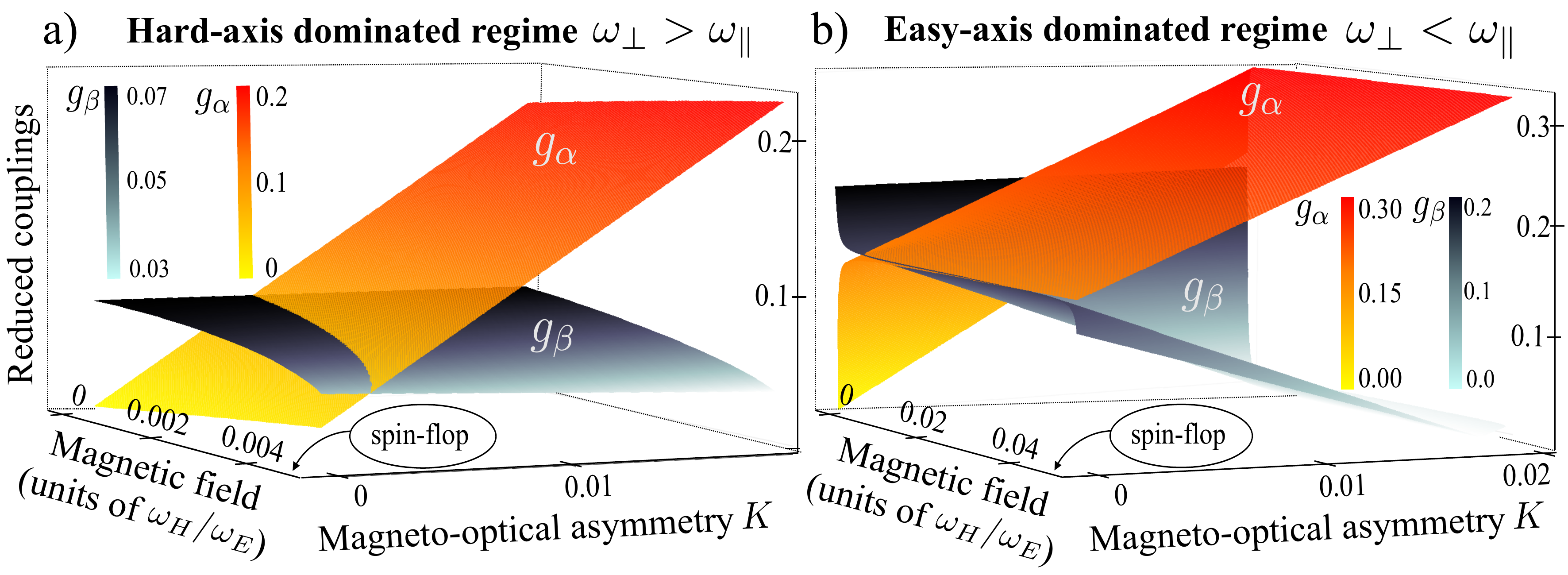}
\par\end{centering}
\caption{Reduced optomagnonic coupling coefficients $g_{\alpha}$ and $g_{\beta}$, as a function of external bias magnetic field (as $\omega_{H}/\omega_{E}$)
and of the magneto-optical asymmetry $K$. Parameters: (a) $\omega_{\parallel}/\omega_{E}=1.3\times10^{-5}$, $\omega_{\bot}/\omega_{E}=7.6\times10^{-4}$ (${\rm NiO}$ \citep{satoh_spin_2010}); (b) $\omega_{\perp}/\omega_{E}=1.3\times10^{-5}$, $\omega_{\parallel}/\omega_{E}=7.6\times10^{-4}$.}
\label{Figure02}
\end{figure}

In the absence of a magnetic field, $\hat{\mathcal{H}}_{{\rm AFM}}$ is invariant under  $\hat{a}_{\bm{k}}\longleftrightarrow\hat{b}_{\bm{-k}}$. For finite hard-axis anisotropy ($\omega_{\perp}\neq0$), imposing this symmetry we obtain \citep{parviniSupplementalMaterial} 
\begin{align*}
g_{\alpha}^{\omega_{H}=0,\omega_{\perp}\neq0} & =2K\left(u_{\alpha,a}-v_{\alpha,a}\right)\,,\\
g_{\beta}^{\omega_{H}=0,\omega_{\perp}\neq0} & =2(u_{\beta,b}-v_{\beta,b})\,,
\end{align*}
and hence for $K=0$ $\hat{\alpha}$ is a dark mode ($g_{\alpha}=0$) while $\hat{\beta}$ is independent of $K$. The case $\omega_{\bot}=B_{0}=0$ is however pathological, since $\hat{\alpha}$ and $\hat{\beta}$ are degenerate. Then Eq.~(\ref{eq:coupling_g_a_b}) holds, with $g_{\alpha}=g_{\beta}\neq0$ (the Bogoliubov coefficients present a discontinuity at $\omega_{\bot}=0$).

Fig.~\ref{Figure02} shows $|g_{\alpha}|$ and $|g_{\beta}|$ as a function of $B_0$ and $K$ for representative finite anisotropy values $\omega_{\bot}$ and $\omega_{\parallel}$. In Fig.~\ref{Figure02}~(a) we took these as for NiO  \citep{satoh_spin_2010}, and in Fig.~\ref{Figure02}~(b) we exchanged them such that $\omega_{\parallel}>\omega_{\perp}$. In both cases the coupling strengths $g_{\alpha,\beta}$ can be tuned by $B_{0}$, although with some qualitative differences. For $K=0$ the $\alpha$-mode can be tuned from dark to bright by increasing $B_{0}$, with a slow linear increase for $\omega_{\parallel}<\omega_{\perp}$ and rapidly but saturating for $\omega_{\parallel}>\omega_{\perp}$. For both cases there is a threshold $K_{{\rm th}}$ such that for $K>K_{{\rm th}}$, there exists a finite $B_{0}$ for which the $\beta$-mode is rendered dark ($g_{\beta}=0$). In the regime considered for Fig.~\ref{Figure02}, $g_{\alpha,\beta}<1$ for all fields, since the maximum $B_{0}$ is limited by the spin-flop transition. This suppresses the corresponding optomagnonic coupling $(Gg_{\alpha,\beta})$. $g_{\alpha}$
increases nevertheless rapidly with $K$, so materials with a larger magneto-optical asymmetry would be favorable for larger coupling values. Our calculations indicate that $K\gtrsim0.1$ would be sufficient for $g_{\alpha}>1$ \citep{parviniSupplementalMaterial}. 

\begin{figure}[t]
\begin{centering}
\includegraphics[width=1\columnwidth]{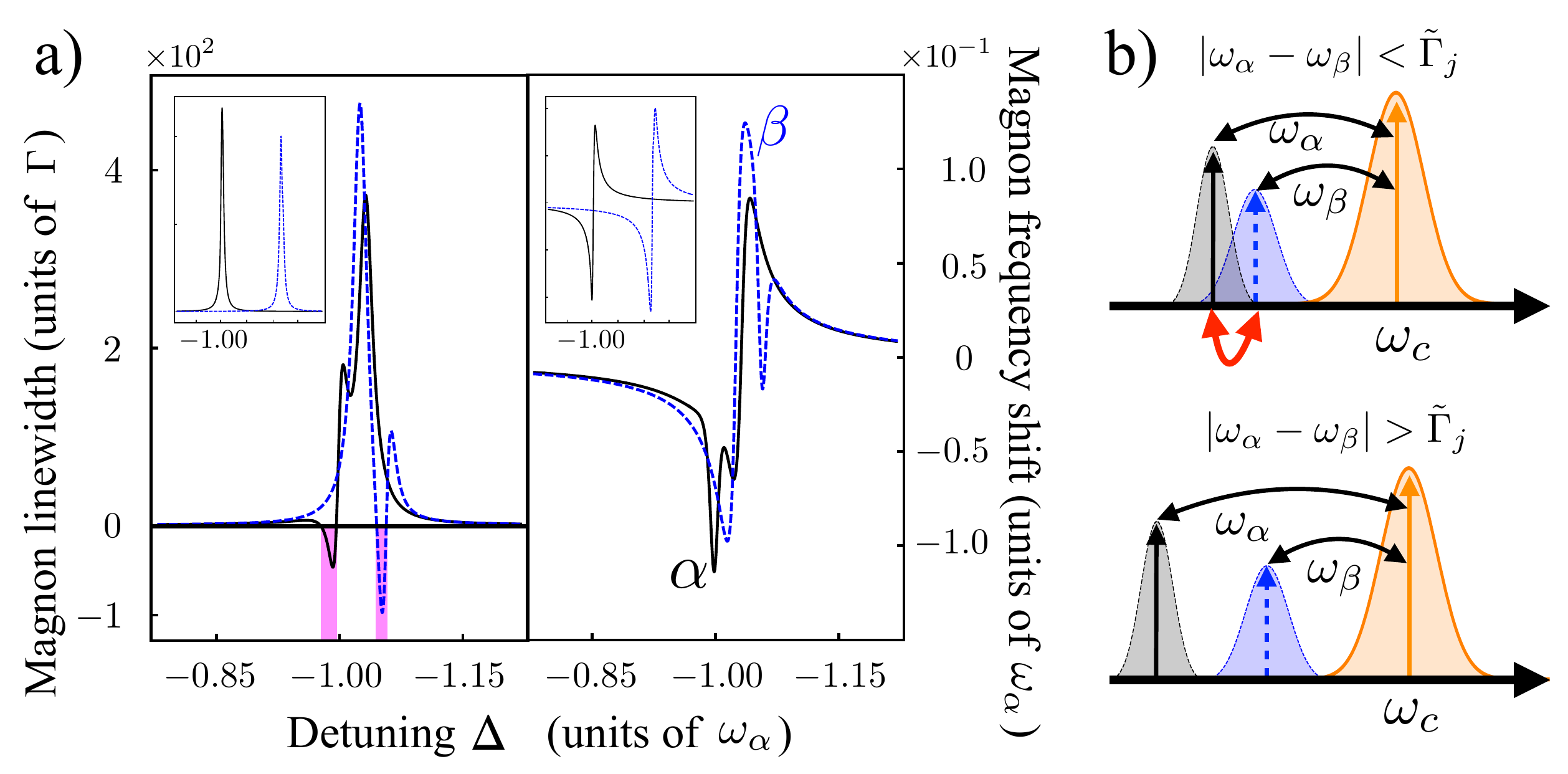}
\par\end{centering}
\caption{(a) Effective linewidth (left) and frequency shift (right) of the magnon modes vs. detuning $\Delta/\omega_{\alpha}$ for near degenerate modes and (inset) well separated modes. The highlighted area indicates an unusual amplification region in the red detuned regime. (b) Frequency scheme for near degenerate and well separated frequencies. The red arrow indicates indirect magnon-magnon interactions, more relevant in the near-degenerate case. Parameters for MnF\textsubscript{2} and $\Gamma/\omega_{E}=1.6\times10^{-4}$, $\kappa/\omega_{E}= 3.7 \times10^{-3}$, $\omega_H/\omega_E = 5.4 \times 10^{-3}$ for the main figures and  $\omega_H/\omega_E =3.2 \times 10^{-2}$ for the inset.}
\label{Figure03}
\end{figure}

A figure of merit for determining the strength of the coupling is the cooperativity \citep{aspelmeyer_cavity_2014}. Taking $G=0.1{\rm MHz}$ as noted above, and typical values for the magnon ($\Gamma\approx1{\rm GHz}$ \citep{kampfrath_coherent_2011,zhou_giant_2018}) and optical cavity decay rates ($\kappa\approx100{\rm MHz}$ \citep{zhang_optomagnonic_2016}), for $g_{\alpha,\beta}=1$ we obtain a single-photon cooperativity $\mathcal{C}_{\alpha,\beta}^{0}=4G^{2}g_{\alpha,\beta}^{2}/\Gamma\kappa\approx4\times10^{-6}$. For an estimated maximum photon density of $10^{5}$/$\mu{\rm m}^{3}$ allowed in the cavity \citep{viola_kusminskiy_coupled_2016}, the cooperativity $\mathcal{C}_{\alpha,\beta}=n_{c}\mathcal{C}_{\alpha,\beta}^{0}$ (with $n_{c}=\langle\hat{c}^{\dagger}\hat{c}\rangle$ steady state number of photons circulating in the cavity) could be therefore tuned into the strong coupling regime ($\mathcal{C}_{\alpha,\beta}>1$) by reaching $g_{\alpha,\beta}>1$. Improved cavity and magnon decay rates would boost this value further. In this regime, magnons and photons hybridize and coherent exchange of information is possible.

\emph{Dynamical Response.--} We now consider a cavity driven by a strong control laser with amplitude $s_{d}$ and frequency $\omega_{d}$, and a weak probe laser with amplitude $s_{p}$ and frequency $\omega_{p}$, see Fig.~\ref{Figure01}. Correspondingly, we add a driving term $\hat{\mathcal{H}}_{{\rm D}}=i\hbar\sqrt{\eta\kappa}(\hat{c}_{\xi}^{\dagger}s_{in}+H.c.)$ to the Hamiltonian in Eq.~(\ref{eq:Htot}), where $s_{in}=s_{d}e^{-i\omega_{l}t}+s_{p}e^{-i\omega_{p}t}$. The total loss rate of the optical cavity is $\kappa=\kappa_{ex}+\kappa_{0}$, where $\kappa_{ex}$ and $\kappa_{0}$ correspond to the loss rates due to external coupling and intrinsic dissipation, respectively. The coupling efficiency $\eta=\kappa_{ex}/\kappa_{0}$ is adjustable in experiments \citep{weis2010optomechanically,xiong_fundamentals_2018}.

The cavity leads to the modification of both the magnon resonance frequency and the magnon damping. Both effects are quantified through the magnon self energy, which also includes a cavity-mediated coupling between the two magnon modes. This term becomes relevant in the strong coupling regime ($g_j\sqrt{n_c}>\Gamma_j,\,\kappa$, with $j=\alpha ,\,\beta$ and $\Gamma_{j}$ the magnon linewidth of mode $j$) for near degenerate magnon modes $\vert \omega_\alpha - \omega_\beta \vert < \Gamma_j$ , see Sup. Mat. \citep{parviniSupplementalMaterial}. Together with hybridization effects \cite{genessimultaneous2008, ockeloensidebandcooling2019, sommerpartialoptomechanical2019} and counter-rotating terms that cannot be neglected in this regime, the optically induced magnon-magnon interaction is responsible for unusual behavior, for example amplification in the red detuned regime, see Fig.~\ref{Figure03}. The  AFM cavity provides a unique platform to probe such regimes for materials that exhibit degenerate modes at zero magnetic field (such as MnF\textsubscript{2}), since $\vert \omega_{\alpha} - \omega_{\beta} \vert$ can be tuned via an external magnetic field. 

We turn now our attention to the transmission and reflection properties of the AFM optomagnonic cavity. Following the standard procedure (see Sup. Mat. \citep{parviniSupplementalMaterial}) we obtain the cavity mode spectra $\delta c[\omega]$ in the frame rotating at the control light frequency
\begin{align}
\delta c[\omega]=\frac{(1+F(\omega))\sqrt{\eta\kappa}\delta s_{{\rm in}}[\omega]}{-i(\tilde{\Delta}+\omega)+\frac{\kappa}{2}-2i\tilde{\Delta}F(\omega)},
\label{eq:cavityspectra}
\end{align}
where $\omega=\omega_{p}-\omega_{d}$ is the pump-probe detuning, $\tilde{\Delta}=\Delta+2G(g_{\alpha}{\rm Re}[\langle\hat{\alpha}\rangle]+g_{\beta}{\rm Re}[\langle\hat{\beta}\rangle])$ is the renormalized detuning due to the magnon induced cavity frequency shift with $\langle\hat{j}\rangle=iGg_{j}n_{c}/\left(i\omega_{j}+\Gamma_{j}/2\right)$ (for $j=\alpha,\beta$), and $F(\omega) = \Sigma(\omega) /(i (\tilde{\Delta} - \omega) +\kappa/2)$ with the photon self-energy term $\Sigma(\omega)=\Sigma_{\alpha}(\omega)+\Sigma_{\beta}(\omega)$ given in terms of
\begin{align}
\Sigma_{j}\left(\omega\right) & =\left[\frac{G^{2}\vert g_{j}\vert^{2}n_{c}}{-i\left(\omega_{j}+\omega\right)+\frac{\Gamma_{j}}{2}}-\frac{G^{2}\vert g_{j}\vert^{2}n_{c}}{i\left(\omega_{j}-\omega\right)+\frac{\Gamma_{j}}{2}}\right].
\end{align}

\begin{figure}[b]
\begin{centering}
\includegraphics[width=1\columnwidth]{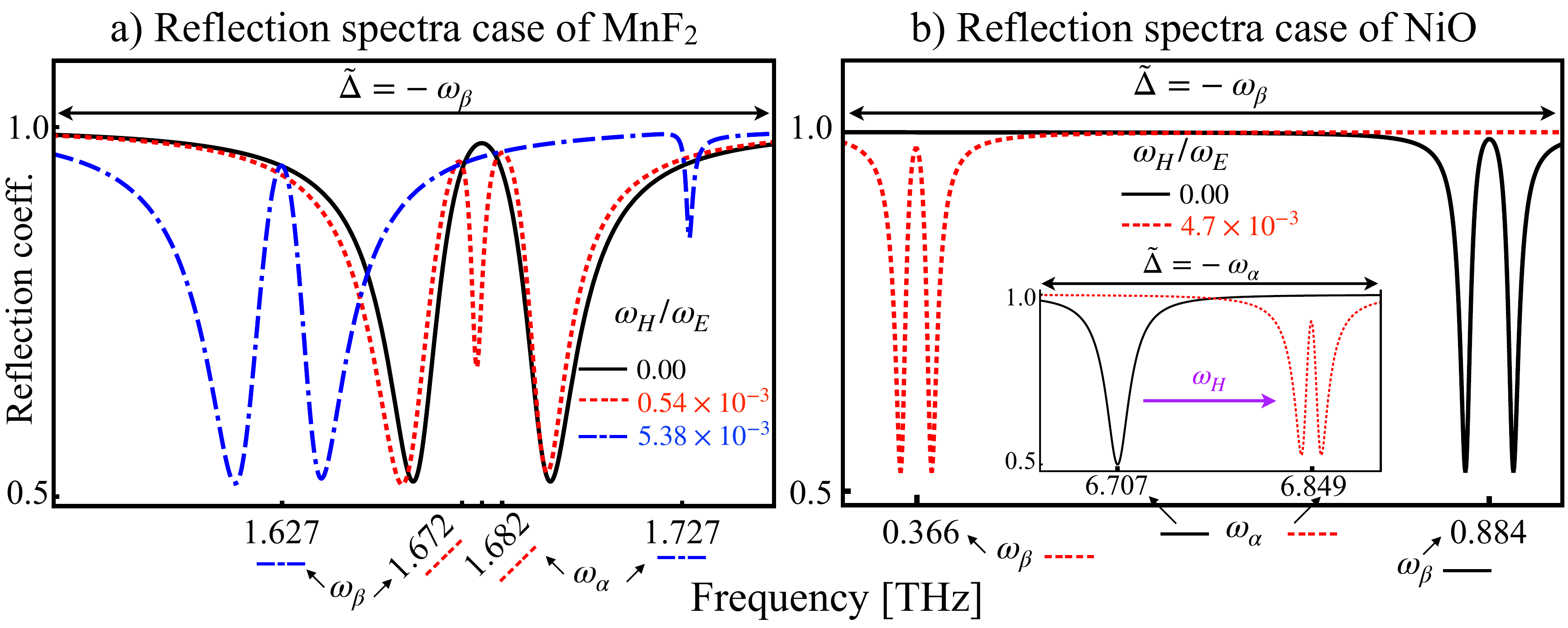}
\par\end{centering}
\caption{Reflection spectra for a red detuned control laser as a function of the probe-pump detuning $\omega$ for a material with (a) degenerate magnon modes at zero magnetic field  and (b) non-degenerate modes. Parameters: $\eta = 0.25$, $\kappa = 3.5 \times 10^{-2}$ THz, $\Gamma = 1.5 \times 10^{-3}$ THz, (a) parameters for MnF\textsubscript{2} and  $G \sqrt{n_{c}}/\omega_{E}=4.0\times10^{-3}$; (b) parameters for NiO and $G \sqrt{n_{c}}/\omega_{E} = 1.8\times10^{-3}$.}
\label{Figure04}
\end{figure}

The transmission and reflection spectra are obtained  from Eq.~(\ref{eq:cavityspectra}) by using the input-output boundary conditions $\delta c_{out}(\omega)=\delta c_{in}(\omega)+\left(\kappa_{ex}/2\right)^{1/2}\delta c(\omega)$. In Fig.~\ref{Figure04} we plot the reflection spectra for the fast-cavity regime ($\Gamma < g_{\alpha}G\sqrt{n_{c}} < \kappa$ and for simplicity we assume $\Gamma_{\alpha}=\Gamma_{\beta}\equiv\Gamma$). Due to destructive interference between the up-converted control field and the probe field, an optomagnonically induced transparency (OMIT) window opens in the transmission spectrum around the corresponding magnon resonance. In the near degenerate regime, depicted Fig.~\ref{Figure04}~(a) for representative parameters of MnF\textsubscript{2} (easy-axis AFM), the OMIT window has an additional structure due to the closeness of the $\alpha$ and $\beta$ modes' sidebands. Increasing $B_0$ increases $|\omega_\beta-\omega_\alpha|$ (but has no effect on the optomagnonic coupling, see Eq.~\ref{eq:coupling_g_a_b}) and the usual OMIT behavior is recovered. In Fig.~\ref{Figure04}~(b) we show results for representative parameters of NiO (finite hard-axis anisotropy), for which the magnon frequencies are well separated even at zero magnetic field. In this case the OMIT behavior can be tuned by $B_{0}$ through the optomagnonic coupling, see Fig.~\ref{Figure02}. 

Finally, the dark-to-bright tunability of the magnon modes can be used for a quantum memory protocol. Driving the system with a strong control red detuned laser, the cavity-magnon coupling can be controlled by $B_0$ such that the (linearized) Hamiltonian is $\sim g(t) (\delta \hat{c}^\dagger \hat{\alpha} + \delta \hat{c} \hat{\alpha}^\dagger)$ \citep{aspelmeyer_cavity_2014}. An arbitrary initial cavity state can then be stored in the magnon mode by bringing $g(t)$ from its initial value $g_0$ to $g(T) = 0$ such that $\int_0^T dt g(t) = \pi g_0$ (analogous to a $\pi$-pulse protocol \cite{mcgeemechanicalresonators2013}). This swaps the state of the cavity with the magnon mode, which is then rendered dark for $t> T$. The state can be transferred with high fidelity for strong coupling and $T \gg 1/\kappa$, and stored up to the magnon lifetime. Alternatively, the OMIT could be employed, in a similar fashion to memories implemented in cold atoms \cite{lukinentanglementof2000, fleischhauerdarkstate2000, fleischhauerquantummemory2002, gorshkovphotonstorage2007, maopticalquantum2017}. The AFM permits to tune the OMIT window via $B_0$, allowing a broad bandwidth storage. 

\emph{Conclusions.--} We proposed a solid state optomagnonic cavity system in which optical photons are coupled to long wavelength AFM magnons and derived its governing Hamiltonian and dynamical features. We showed that the AFM system presents unique characteristics, such as tunability of the coupling with a magnetic field, and unusual dynamical effects due to cavity-induced interactions between the two homogeneous magnon modes. We estimated the values for the coupling and showed that, although challenging, the strong coupling regime could be reached in micron sized single-domain AFMs cavities \citep{kondoh1964observation,baruchel_antiferromagnetic_1981}.  AFMs optical cavities could therefore provide a new platform to study light-matter interaction, and possibly a new tool to probe AFMs due to the enhanced light-magnon coupling. The tunability with a magnetic field, in particular for tuning a magnon mode from dark to bright, shows promise for quantum protocols for quantum information storage and retrieval. The coherent coupling of THz magnons to optical photons could allow the implementation of a quantum transducer \cite{bialekspinwavecoupling2020}.  In this first work we focused on one-magnon processes, two-magnon processes will be treated elsewhere. 

\emph{Acknowledgments.--}The authors acknowledge the financial support from the Max Planck Gesellschaft through an independent Max Planck Research group. We thank  A. Aiello, T. Nova, L. Petruzziello, L. Smaldone and C. Tzschaschel for enlightening discussions.

\section*{Supplemental Material}

\subsection{\label{sec:Antiferromagnetic-Hamiltonian}Antiferromagnetic Hamiltonian}

For completeness we present here how to diagonalize the antiferromagnetic
Hamiltonian 

\begin{align}
\hat{H}_{{\rm AFM}} & =\hbar\sum_{\langle i\neq j\rangle}J\hat{\mathbf{S}}^{i}\cdot\hat{\mathbf{S}}^{j}\text{+}\hbar\left|\gamma\right|\mathbf{B}_{0} \cdot \sum_{i}\hat{\mathbf{S}}^{i}\label{eq:HAFMApp}\\
 & \text{\textminus}\frac{\hbar K_{\parallel}}{2}\sum_{i}\left(\hat{S}_{z}^{i}\right)^{2}+\frac{\hbar K_{\perp}}{2}\sum_{i}\left(\hat{S}_{x}^{i}\right)^{2},\nonumber 
\end{align}
following Ref. \citep{johansen_nonlocal_2018}.
Considering small fluctuations around equilibrium, which we set to
be in the $\bm{e}_{z}$ direction, the spin ladder operators $\hat{S}_{\pm}^{i}=\hat{S}_{x}^{i}\pm i\hat{S}_{y}^{i}$,
and $\hat{S}_{z}^{i}$ are given by

\begin{align}
\hat{S}_{+}^{i\in A} & =\left(2S/N\right)^{1/2}\sum_{\bm{k}}e^{-i\bm{k}\cdot\bm{r}_{i}}\hat{a}_{\bm{k}},\nonumber \\
\hat{S}_{+}^{j\in B} & =\left(2S/N\right)^{1/2}\sum_{\bm{k}}e^{-i\bm{k}\cdot\bm{r}_{i}}\hat{b}_{\bm{k}}^{\dagger},\nonumber \\
S_{z}^{i\in A} & =S-N^{-1}\sum_{\bm{k}\bm{k}^{\prime}}e^{i(\bm{k}-\bm{k}^{\prime})\cdot\bm{r}_{i}}\hat{a}_{\bm{k}}^{\dagger}\hat{a}_{\bm{k}^{\prime}},\nonumber \\
S_{z}^{j\in B} & =-S+N^{-1}\sum_{\bm{k}\bm{k}^{\prime}}e^{-i(\bm{k}-\bm{k}^{\prime})\cdot\bm{r}_{i}}b_{\bm{k}}^{\dagger}b_{\bm{k}^{\prime}},\label{eq:HP_Transformation}
\end{align}
where $\hat{a}_{\bm{k}}$ and $\hat{b}_{\bm{k}}$ are the collective
bosonic operators associated to the sublattices A and B respectively,
satisfying the commutation relations $\left[\hat{a}_{\bm{k}},\hat{a}_{\bm{k}^{\prime}}^{\dagger}\right]=\left[\hat{b}_{\bm{k}},\hat{b}_{\bm{k}^{\prime}}^{\dagger}\right]=\delta_{\bm{k}\bm{k}^{\prime}}$,
$\left[\hat{a}_{\bm{k}},\hat{a}_{\bm{k}^{\prime}}\right]=\left[\hat{a}_{\bm{k}}^{\dagger},\hat{a}_{\bm{k}^{\prime}}^{\dagger}\right]=\left[\hat{b}_{\bm{k}},\hat{b}_{\bm{k}^{\prime}}\right]=\left[\hat{b}_{\bm{k}}^{\dagger},\hat{b}_{\bm{k}^{\prime}}^{\dagger}\right]=0$.
$S$ is the total spin per lattice site and $N$ is the number of
sites in each sublattice. Using Eq. \ref{eq:HP_Transformation} and
keeping only terms up to two bosonic operators, $\hat{H}_{{\rm AFM}}$
is written in $k$-space as

\begin{align}
\hat{H}_{{\rm AFM}} & =\hbar\sum_{k}\Big[\frac{\mathcal{A}}{2}\hat{a}_{\bm{k}}^{\dagger}\hat{a}_{\bm{k}}+\frac{\mathcal{B}}{2}\hat{b}_{\bm{k}}^{\dagger}\hat{b}_{\bm{k}}+\mathcal{C}_{\bm{k}}\hat{a}_{\bm{k}}\hat{b}_{\text{\textminus}\bm{k}}\nonumber \\
 & +\mathcal{D}\left(\hat{a}_{\bm{k}}\hat{a}_{-\bm{k}}+\hat{b}_{\bm{k}}\hat{b}_{-\bm{k}}\right)+h.c\Big]\,.\label{eq:H_AFM}
\end{align}
The coefficients $\mathcal{A}$, $\mathcal{B}$, $\mathcal{C}_{k}$
and $\mathcal{D}$ are given in terms of the characteristic angular
frequencies $\omega_{E}=SJZ$ (with Z the number of nearest neighbors),
$\omega_{\parallel}=SK_{\parallel}$, $\omega_{\perp}=SK_{\perp}$,
and $\omega_{H}=\left|\gamma\right|B_{0}$ as
\begin{align}
\mathcal{A} & =(\omega_{E}+\omega_{\parallel}+\frac{\omega_{\perp}}{2}-\omega_{H}),\label{eq:coeffsAFM}\\
\mathcal{B} & =(\omega_{E}+\omega_{\parallel}+\frac{\omega_{\perp}}{2}+\omega_{H}),\nonumber \\
\mathcal{C}_{\bm{k}} & =\frac{\omega_{E}}{Z}f(\bm{k},\bm{\delta}),\,\,\,\,\mathcal{\,\,D}=\frac{\omega_{\perp}}{4},\nonumber 
\end{align}
where $f(\bm{k},\bm{\delta})=\sum_{j}e^{-i\bm{k}\cdot\bm{\delta}_{j}}$,
with the sum carried over all nearest-neighbor vectors $\bm{\delta}_{j}$.
For long wavelength magnons $\mathcal{C}_{k}\sim\omega_{E}$.

In order to diagonalize the Hamiltonian, we use the four-dimensional
Bogoliubov transformation \citep{kamra_noninteger-spin_2017,johansen_nonlocal_2018} 

\begin{equation}
\left[\begin{array}{c}
\hat{\alpha}_{\bm{k}}\\
\hat{\beta}_{-\bm{k}}^{\dagger}\\
\hat{\alpha}_{-\bm{k}}^{\dagger}\\
\hat{\beta}_{\bm{k}}
\end{array}\right]=\left[\begin{array}{cccc}
u_{\alpha,a} & v_{\alpha,b} & v_{\alpha,a} & u_{\alpha,b}\\
v_{\beta,a}^{*} & u_{\beta,b}^{*} & u_{\beta,a}^{*} & v_{\beta,b}^{*}\\
v_{\alpha,a}^{*} & u_{\alpha,b}^{*} & u_{\alpha,a}^{*} & v_{\alpha,b}^{*}\\
u_{\beta,a} & v_{\beta,b} & v_{\beta,a} & u_{\beta,b}
\end{array}\right]\left[\begin{array}{c}
\hat{a}_{\bm{k}}\\
\hat{b}_{-\bm{k}}^{\dagger}\\
\hat{a}_{-\bm{k}}^{\dagger}\\
\hat{b}_{\bm{k}}
\end{array}\right],\label{eq:Bogoliubov}
\end{equation}
where $\hat{\alpha}_{\bm{k}}$ and $\hat{\beta}_{\bm{k}}$ are the
destruction operators of the bosonic magnon modes which satisfy $[\hat{\alpha}_{\bm{k}},\hat{H}_{{\rm AFM}}]=\hbar\omega_{\alpha k}\hat{\alpha}_{\bm{k}}$
and $[\hat{\beta}_{\bm{k}},\hat{H}_{{\rm AFM}}]=\hbar\omega_{\beta k}\hat{\beta}_{\bm{k}}$.
These commutation relations lead to an eigenvalue problem that, together
with the bosonic commutation rules for $\hat{\alpha}_{\bm{k}}$ and
$\hat{\beta}_{\bm{k}}$, determine the coefficients of (\ref{eq:Bogoliubov})
and the eigenfrequencies $\omega_{\alpha,\beta}$\citep{kamra_noninteger-spin_2017}.
The diagonalized form of (\ref{eq:H_AFM}) is given by

\begin{equation}
\hat{H}_{{\rm AFM}}=\hbar\sum_{k}\left[\omega_{\alpha k}\hat{\alpha}_{\bm{k}}^{\dagger}\hat{\alpha}_{\bm{k}}+\omega_{\beta k}\hat{\beta}_{\bm{k}}^{\dagger}\hat{\beta}_{\bm{k}}\right].
\end{equation}

\subsection{\label{sec:Magneto-optical-Hamiltonian-for}Magneto-optical Hamiltonian
for Antiferromagnets}

Here we present the details on the derivation of Eq. (3) in the main. The starting point is the time-average energy of the electromagnetic field on a magnetized material \cite{landauElectrodynamicsContinuousMedia1984}
\begin{equation}
\mathcal{H}_{{\rm OM}} = \frac{1}{4} \int d^3 \bm{r} \sum_{\mu,\nu}E_{\mu}^{*}(\bm{r})\varepsilon_{\mu\nu}(\mathbf{r})E_{\nu}(\bm{r}), \nonumber
\end{equation}
where the integration is performed over all the cavity volume. We discretize the integral as a sum over all lattices as
\begin{equation}
\mathcal{H}_{{\rm OM}}=\frac{V}{4 N}{\displaystyle \sum_{i}} \sum_{\mu,\nu}E_{\mu}^{*}(\bm{r}_{i})\varepsilon_{\mu\nu}(\mathbf{r}_{i})E_{\nu}(\bm{r}_{i}),\label{eq:H_MO}
\end{equation}
To first order in the spins, the permittivity tensor is given by
\begin{equation}
\varepsilon_{\mu\nu}(\mathbf{r}_{i})=\sum_{\zeta}K_{\mu\nu\zeta}S_{\zeta}^{i},\label{eq:Epsilon}
\end{equation}
with $K_{\mu\nu\zeta}$ the magneto-optical coefficients which are in general restricted by symmetry conditions. Further terms can be included to describe second order magneto-optical effects \citep{cottam_on_1975,cottamLightScatteringMagnetic1986}, which we do not consider in this work. 

Following Ref. \citep{cottam_on_1975}, we assume system with a rutile crystal structure, where the two magnetic sublattices are arranged in a body-centered cubic geometry. One of the sublattices occupies the central sites, while the other occupies the corner sites. The components of the permittivity tensor are then given in terms of three imaginary constants $K_{1}$, $K_{2}$ and $K_{3}$ \citep{cottam_on_1975}, such that for sublattice A 
\begin{align}
K_{yzx}^{(A)} & =-K_{zyx}^{(A)}=K_{1},\nonumber \\
K_{zxy}^{(A)} & =-K_{xzy}^{(A)}=K_{2},\nonumber \\
K_{xyz}^{(A)} & =-K_{yxz}^{(A)}=K_{3}.\label{eq:KA}
\end{align}
Given the considered geometry, $K_{\mu\nu\zeta}^{(B)}$ can be obtained
from $K_{\mu\nu\zeta}^{(A)}$ by $K_{\mu\nu\zeta}^{(B)}=\sum_{\mu^{\prime},\nu^{\prime},\zeta^{\prime}}R_{\mu\mu^{\prime}}R_{\nu\nu^{\prime}}R_{\zeta\zeta^{\prime}}K_{\mu^{\prime}\nu^{\prime}\zeta^{\prime}}^{(A)}$
where $R$ is the $\pi/2$ rotation matrix relating the symmetry of
the sublattice A to the symmetry of the sublattice B. Therefore, for
sublattice B
\begin{align}
K_{yzx}^{(B)} & =-K_{zyx}^{(B)}= -K_{xzy}^{(A)} = K_{2},\nonumber \\
K_{zxy}^{(B)} & =-K_{xzy}^{(B)}= - K_{zyx}^{(A)}=K_{1},\nonumber \\
K_{xyz}^{(B)} & =-K_{yxz}^{(B)}= -K_{yxz}^{(A)}=K_{3}.\label{eq:KB}
\end{align}
If the sublattices are equivalent, $K_{1}=K_{2}$.

We are interested in the optomagnonic coupling Hamiltonian, which
represents the coupling of an optical field to the magnon excitations
on top of the static ground state spin configuration. In correspondence
with the setup of Section \ref{sec:Antiferromagnetic-Hamiltonian},
we assume that the N\`eel equilibrium of the AFM takes place along the
$\bm{e}_{z}$ direction. Therefore, to first order in the magnon excitations,
the coupling between light and the deviations from the magnetic equilibrium
is encoded in the terms $\propto S_{x,y}$, which correspond to scattering
processes involving one magnon. We thus do not consider the terms
$\propto S_{z}$, which would correspond to higher order processes.
Substituting (\ref{eq:Epsilon}), (\ref{eq:KA}) and (\ref{eq:KB})
in (\ref{eq:H_MO}), the optomagnonic Hamiltonian can be written as
\begin{align*}
\mathcal{H}_{{\rm OM}}= &\frac{V}{4N} {\displaystyle \sum_{i\in A}}   \Big[K_{2}\left(E_{z}^{*}E_{x}-E_{x}^{*}E_{z}\right)S_{y}^{i}\\
 & {\displaystyle -K_{1}\left(E_{z}^{*}E_{y}-E_{y}^{*}E_{z}\right)S_{x}^{i}}\Big]\\
 & +\frac{V}{4 N}{\displaystyle \sum_{j\in B}} \Big[K_{1}\left(E_{z}^{*}E_{x}-E_{x}^{*}E_{z}\right)S_{y}^{j}\\
 & -K_{2}\left(E_{z}^{*}E_{y}-E_{y}^{*}E_{z}\right)S_{x}^{j}\Big],
\end{align*}
which includes contributions from each sublattice. Defining $K_{\pm}=i(K_{1}\pm K_{2})/4$,
we obtain Eq. (2) in the main text. Quantizing it we obtain

\begin{align*}
\hat{\mathcal{H}}_{{\rm OM}}= & \frac{K_{+} V}{4 N}\sum_{\beta,\gamma}\sum_{s,s^{\prime}}\hat{c}_{\beta,s}^{\dagger}\hat{c}_{\gamma,s^{\prime}}\Big[\\
 & +\displaystyle \sum_{i\in A, B}  \left(P_{\beta\gamma,ss^{\prime}}^{i+} \hat{S}_{-}^{i} -P_{\beta\gamma,ss^{\prime}}^{i-}\hat{S}_{+}^{i} \right) \nonumber \\
 &+ K  \displaystyle \sum_{i\in A}  \left( P_{\beta\gamma,ss^{\prime}}^{i+} \hat{S}_{+}^{i} - P_{\beta\gamma,ss^{\prime}}^{i-} \hat{S}_{-}^{i} \right) \nonumber \\
 & - K \displaystyle \sum_{j \in B}  \left( P_{\beta\gamma,ss^{\prime}}^{j+} \hat{S}_{+}^{j} - P_{\beta\gamma,ss^{\prime}}^{j-} \hat{S}_{-}^{j} \right) \Big],
\end{align*}
where $\mathbf{\hat{E}}(\mathbf{r},t)=\sum_{\gamma,s}\mathbf{E}_{\gamma s}(\mathbf{r})\hat{c}_{\gamma,s}(t)$
($\gamma$ denotes the mode indices and $s$ the polarizations), $P_{\beta\gamma,ss^{\prime}}^{i\pm}=E_{\beta s,z}^{*}(\bm{r}_{i})E_{\gamma s^{\prime},\pm}(\bm{r}_{i})-E_{\beta s,\mp}^{*}(\bm{r}_{i})E_{\gamma s^{\prime},z}(\bm{r}_{i})$, with $E_{\beta s, \pm} = E_x \pm i E_y$, 
and $K=K_{-}/K_{+}$. For modes polarized in the $yz$ plane (as in Fig. 1 in the main text) $P_{\beta\gamma,ss^{\prime}}^{i+}=-P_{\beta\gamma,ss^{\prime}}^{i-}=\mathcal{G}_{\beta\gamma,ss^{\prime}}^{i}$, where the last equality defines the coefficients $\mathcal{G}_{\beta\gamma,ss^{\prime}}^{i}\equiv \mathcal{G}_{\beta\gamma,ss^{\prime}}(\bm{r}_{i})$. Thus
\begin{align}
\hat{\mathcal{H}}_{{\rm OM}}= & \frac{K_{+} V}{4 N}\sum_{\beta s,\gamma s^\prime} \hat{c}_{\beta,s}^{\dagger}\hat{c}_{\gamma,s^{\prime}}\Big[\label{eq:HMOGi}\\
 & \displaystyle \sum_{i \in A, B} \mathcal{G}_{\beta\gamma,ss^{\prime}}^{i} \left( \hat{S}_{-}^{i}+ \hat{S}_{+}^{i} \right)\nonumber \\
 & +K \displaystyle \sum_{i \in A} \mathcal{G}_{\beta\gamma,ss^{\prime}}^{i} \left(\hat{S}_{+}^{i} + \hat{S}_{-}^{i} \right)    \nonumber \\
 &- K \displaystyle \sum_{j \in B} \mathcal{G}_{\beta\gamma,ss^{\prime}}^{j} \left(\hat{S}_{+}^{j} + \hat{S}_{-}^{j} \right) \Big].\nonumber 
\end{align}
In general, the total coupling coefficients will be determined by taking the continuum limit of Eq. (\ref{eq:HMOGi}), such that the sum over the lattice $\sum_{i,j} \frac{V}{N}$ is replaced by the corresponding volume integrals. The total coupling will therefore also depend on the degree of overlap of the corresponding optical and magnon modes \citep{graf_cavity_2018}. 

We further specify the model by considering that only one relevant cavity mode interacts with AFM magnons with $\bm{k}=0$. Moreover, we assume for simplicity that the cavity mode with polarization $\chi$ has a plane wave profile $\mathbf{E}_\chi(\mathbf{r})=i \sqrt{\frac{\hbar\omega_{c}}{2\varepsilon V}}e^{i\bm{k}_{c}\cdot\mathbf{r}}\mathbf{e}_{\chi}$, where $\varepsilon$ is the electric permittivity of the material, $V$ is the cavity volume, $\bm{k}_{c}$ is the wave vector of the mode, and $\bm{e}_{R(L)}=(\bm{e}_{y}\mp i\bm{e}_{z})/\sqrt{2}$ denotes right and left circularly polarized modes. For this case, using the Holstein-Primakoff approximation for the spin operators (see Eq. \ref{eq:HP_Transformation}), the Hamiltonian for the right circular polarization component reads (from now on $\hat{a}_{\bm{k}=0}\equiv\hat{a}$ and $\hat{b}_{\bm{k}=0}\equiv\hat{b}$) 

\begin{align}
\mathcal{\hat{H}}_{{\rm MO}}^{R}= & -\hbar G\hat{c}_{R}^{\dagger}\hat{c}_{R}\Big[\hat{a}^{\dagger}+\hat{b}+\hat{a}+\hat{b}^{\dagger}+\nonumber \\
 & +K\big(\hat{a}-\hat{b}^{\dagger}+\hat{a}^{\dagger}-\hat{b}\big)\Big]\,,\label{eq:H_MO_3}
\end{align}
with 
\begin{equation}
G=\frac{\omega_{c}K_{+}}{8\varepsilon}\sqrt{\frac{2 S}{ N}}\,.\label{eq:couplingG}
\end{equation}
We notice that the integration over the plane wave factors gives a volume factor that cancels the volume dependence on the denominator of the electric field modes.
For the left circular polarization component $\mathcal{\hat{H}}_{{\rm MO}}^{L}=-\mathcal{\hat{H}}_{{\rm MO}}^{R}$.
Since the system is diagonal in the R/L basis, we further consider
only one circular polarization component of the optomagnonic Hamiltonian
(R for definiteness) and drop the index $R$ of (\ref{eq:H_MO_3}). We also notice that if the two sublattices are equivalent and since we defined $K_{1,2}$ via the expansion of the permittivity tensor in terms of the spin in each lattice, see eq. \eqref{eq:Epsilon}, the relation of $K_+$ to the Faraday rotation angle per length $\theta_{\rm{F}}$ is given by $K = c \sqrt{\varepsilon} \theta_{\rm{F}}/(\omega_c S)$ such that
\begin{equation}
G_{K_- = 0} = \frac{c \theta_{\rm{F}}}{4 \sqrt{\varepsilon}} \frac{1}{\sqrt{2 S N}},
\end{equation}
which is equivalent to the coupling in a ferromagnetically ordered system \cite{viola_kusminskiy_coupled_2016}.

In order to express the optomagnonic Hamiltonian in terms of the magnon
modes $\hat{\alpha}\equiv\hat{\alpha}_{\bm{k}=0}$ and $\hat{\beta}\equiv\hat{\beta}_{\bm{k}=0}$,
we use the inverse of the transformation (\ref{eq:Bogoliubov}): $\hat{a}=u_{a,\alpha}\hat{\alpha}+u_{a,\beta}\hat{\beta}+v_{a,\alpha}\hat{\alpha}^{\dagger}+v_{a,\beta}\hat{\beta}^{\dagger}$
and $\hat{b}=u_{b,\alpha}\hat{\alpha}+u_{b,\beta}\hat{\beta}+v_{b,\alpha}\hat{\alpha}^{\dagger}+v_{b,\beta}\hat{\beta}^{\dagger}$.
Substituting these expressions in Eq. (\ref{eq:H_MO_3}) we obtain
Eqs. (4) and (5) of the main text.

In the absence of hard axis anisotropy ($\omega_{\bot}=0$) the non-vanishing
Bogoliubov coefficients are independent of the external magnetic field
and given by \citep{kamra_noninteger-spin_2017}
\begin{align*}
u_{a,\alpha}=u_{b,\beta} & =\sqrt{\frac{\omega_{E}+\omega_{\parallel}}{\sqrt{\omega_{\parallel}(2\omega_{E}+\omega_{\parallel})}}+1},\\
v_{a,\beta}=v_{b,\alpha} & =-\sqrt{\frac{\omega_{E}+\omega_{\parallel}}{\sqrt{\omega_{\parallel}(2\omega_{E}+\omega_{\parallel})}}-1}.
\end{align*}
In the limit $\omega_{\parallel}\ll\omega_{E}$ the couplings $g_{\alpha,\beta}$
take the simple form given in Eq. (8) of the main text.

\begin{figure}[tph]
\centering{}\includegraphics[width=1\columnwidth]{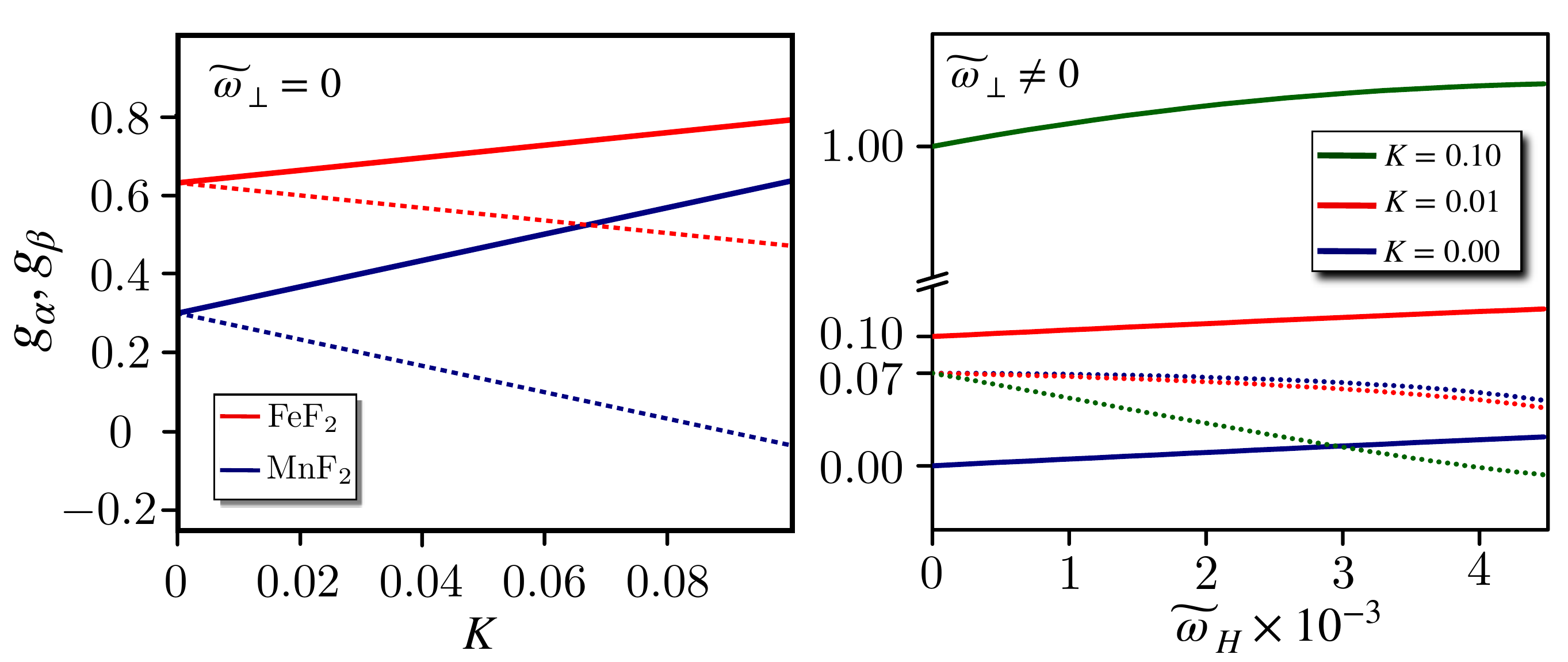}\caption{Couplings $g_{\alpha}$ and $g_{\beta}$ as a function of $K$ for
$\tilde{\omega}_{\perp}=\omega_{\perp}/\omega_{{\rm E}}=0$ (left
plot) and as a function of $\tilde{\omega}_{H}=\omega_{H}/\omega_{E}$
for $\tilde{\omega}_{\perp}=7.6\times10^{-4}$ (right plot). Continuous lines correspond
to $g_{\alpha}$ and dotted lines to $g_{\beta}$.}
\label{FigSupp:01}
\end{figure}

In figure \ref{FigSupp:01} we show the exact $g_{\alpha,\beta}$
as a function of $K$ for $\omega_{\perp}=0$ and representative values
of $\omega_{\parallel}$ (left plot) and as a function of $\omega_{H}$
for different values of $K$ (right plot). In the left plot, $\omega_{\parallel}\ll\omega_{E}$
and we observe a linear behavior with $K$, in agreement with Eq.
(8) of the main text. 

\subsection{Symmetry Considerations}

\subsubsection{Zero external magnetic field case ($\omega_{H}=0$ and $\omega_{\perp}\protect\neq0$
)}

For zero external magnetic field, the antiferromagnetic Hamiltonian
is invariant under the transformation $\hat{a}_{\bm{k}}\longleftrightarrow\hat{b}_{\bm{-k}}$,
see Eqs. \eqref{eq:H_AFM} and \eqref{eq:coeffsAFM}. For $\bm{k}=0$,
this corresponds simply to swapping the sublattices $A$ and $B$.
Under this transformation $\hat{\mathcal{S}}:\hat{a}\rightarrow\hat{b}$
, the Bogoliubov modes read (remembering that for our case the Bogoliubov
coefficients are all real)
\begin{equation}
\left[\begin{array}{c}
\hat{\alpha}^{\prime}\\
\hat{\beta}^{\prime\dagger}\\
\hat{\alpha}^{\prime\dagger}\\
\hat{\beta}^{\prime}
\end{array}\right]=\left[\begin{array}{cccc}
u_{\alpha,b} & v_{\alpha,a} & v_{\alpha,b} & u_{\alpha,a}\\
v_{\beta,b} & u_{\beta,a} & u_{\beta,b} & v_{\beta,a}\\
v_{\alpha,b} & u_{\alpha,a} & u_{\alpha,b} & v_{\alpha,a}\\
u_{\beta,b} & v_{\beta,a} & v_{\beta,b} & u_{\beta,a}
\end{array}\right]\left[\begin{array}{c}
\hat{a}\\
\hat{b}^{\dagger}\\
\hat{a}^{\dagger}\\
\hat{b}
\end{array}\right]\label{eq:BogTransTrans}
\end{equation}
where the prime denotes the transformed modes. Since \eqref{eq:H_AFM}
is invariant under $\hat{\mathcal{S}}$, the transformed Bogoliubov
modes fulfill 
\begin{equation}
\hbar\left[\omega_{\alpha}\hat{\alpha}^{\prime\dagger}\hat{\alpha}^{\prime}+\omega_{\beta}\hat{\beta}^{\prime\dagger}\hat{\beta}^{\prime}\right]=\hbar\left[\omega_{\alpha}\hat{\alpha}^{\dagger}\hat{\alpha}+\omega_{\beta}\hat{\beta}^{\dagger}\hat{\beta}\right]\,.\label{eq:hamprime}
\end{equation}
Considering Eqs. (\ref{eq:BogTransTrans}) and (\ref{eq:hamprime}),
for non-degenerate modes $\omega_{\alpha}\neq\omega_{\beta}$ (this
requires $\omega_{\perp}\neq0$) we obtain the following conditions
on the Bogoliubov coefficients: $u_{j,b}=\pm u_{j,a}$ and $v_{j,b}=\pm v_{j,a}$
(for $j=\alpha,\beta$). In our case, $\hat{\alpha}$ ($\hat{\beta}$)
corresponds to the antisymmetric (symmetric) mode under the transformation:

\begin{align*}
u_{\alpha,b} & =-u_{\alpha,a}=-U_{\alpha},\\
v_{\alpha,b} & =-v_{\alpha,a}=-V_{\alpha},\\
u_{\beta,b} & =u_{\beta,a}=U_{\beta},\\
v_{\beta,b} & =v_{\beta,a}=V_{\beta}.
\end{align*}
Therefore
\[
\left[\begin{array}{c}
\hat{\alpha}\\
\hat{\beta}^{\dagger}\\
\hat{\alpha}^{\dagger}\\
\hat{\beta}
\end{array}\right]=\left[\begin{array}{cccc}
U_{\alpha} & -V_{\alpha} & V_{\alpha} & -U_{\alpha}\\
V_{\beta} & U_{\beta} & U_{\beta} & V_{\beta}\\
V_{\alpha} & -U_{\alpha} & U_{\alpha} & -V_{\alpha}\\
U_{\beta} & V_{\beta} & V_{\beta} & U_{\beta}
\end{array}\right]\left[\begin{array}{c}
\hat{a}\\
\hat{b}^{\dagger}\\
\hat{a}^{\dagger}\\
\hat{b}
\end{array}\right],
\]
 and the inverse transformation reads
\begin{equation}
\left[\begin{array}{c}
\hat{a}\\
\hat{b}^{\dagger}\\
\hat{a}^{\dagger}\\
\hat{b}
\end{array}\right]=\left[\begin{array}{cccc}
U_{\alpha} & -V_{\beta} & -V_{\alpha} & U_{\beta}\\
V_{\alpha} & U_{\beta} & -U_{\alpha} & -V_{\beta}\\
-V_{\alpha} & U_{\beta} & U_{\alpha} & -V_{\beta}\\
-U_{\alpha} & -V_{\beta} & V_{\alpha} & U_{\beta}
\end{array}\right]\left[\begin{array}{c}
\hat{\alpha}\\
\hat{\beta}^{\dagger}\\
\hat{\alpha}^{\dagger}\\
\hat{\beta}
\end{array}\right],\label{eq:InverseOMCoupling}
\end{equation}
where we have used that $2\left(\vert U_{j}\vert^{2}-\vert V_{j}\vert^{2}\right)=1$.
From $g_{j}=\left(u_{j}^{+}+v_{j}^{+}\right)+K\left(u_{j}^{-}+v_{j}^{-}\right)$
we obtain 
\begin{align*}
g_{\alpha} & =2K\left(U_{\alpha}-V_{\alpha}\right),\\
g_{\beta} & =2(U_{\beta}-V_{\beta})
\end{align*}
and hence for $K=0$ the $\alpha$-mode is decoupled from the light,
while the $\beta$-mode coupling is independent of $K$.

\subsubsection{Easy axis AFM case ($\omega_{\perp}=0$)}

In the absence of hard axis anisotropy, the Hamiltonian \ref{eq:HAFMApp}
is invariant under rotations around the $\bm{e}_{z}$ axis and \ref{eq:H_AFM}
reads (for $\bm{k}=0$, and $\hbar=1$)
\begin{align}
\hat{H}_{{\rm AFM}} & =\mathcal{A}\hat{a}^{\dagger}\hat{a}+\mathcal{B}\hat{b}^{\dagger}\hat{b}+\mathcal{C}\left(\hat{a}\hat{b}+\hat{a}^{\dagger}\hat{b}^{\dagger}\right),\label{eq:HAFM_SYM}
\end{align}
A rotation by $\theta$ around the $\bm{e}_{z}$ axis is given by
$\hat{\mathcal{R}}:\hat{S}_{+}\rightarrow e^{i\theta}\hat{S}_{+}$,
thus at the level of the bosonic operators $\hat{a}\rightarrow e^{i\theta}\hat{a}$
and $\hat{b}\rightarrow e^{-i\theta}\hat{b}$ The Bogoliubov modes
transform as
\begin{align*}
\hat{\alpha} & =e^{i\theta}\left(u_{\alpha,a}\hat{a}+v_{\alpha,b}\hat{b}^{\dagger}\right)+e^{-i\theta}\left(u_{\alpha,b}\hat{b}+v_{\alpha,a}\hat{a}^{\dagger}\right),\\
\hat{\beta} & =e^{i\theta}\left(u_{\beta,a}\hat{a}+v_{\beta,b}\hat{b}^{\dagger}\right)+e^{-i\theta}\left(u_{\beta,b}\hat{b}+v_{\beta,a}\hat{a}^{\dagger}\right).
\end{align*}
We conclude that, in order for $\hat{H}_{{\rm AFM}}=\omega_{\alpha}\hat{\alpha}^{\dagger}\hat{\alpha}+\omega_{\beta}\hat{\beta}^{\dagger}\hat{\beta}$
to be invariant we have (for $j=\alpha,\beta$) $u_{j,a}=v_{j,b}=0$
or $u_{j,b}=v_{j,a}=0.$ In our case, $\hat{\mathcal{R}}$: $\hat{\alpha}\rightarrow e^{i\theta}\hat{\alpha}$
and $\hat{\beta}\rightarrow e^{-i\theta}\hat{\beta}$, thus fixing
$u_{\alpha,a}=U_{\alpha}$, $v_{\alpha,b}=V_{\alpha}$, $u_{\beta,b}=U_{\beta}$,
$v_{\beta,a}=V_{\beta}$ and $u_{\alpha,b}=u_{\beta,a}=v_{\alpha,a}=v_{\beta,b}=0$.
We can then write $\hat{H}_{{\rm AFM}}$ in terms of the Bogoliubov
coefficients (besides a constant term) as
\begin{align}
\hat{H}_{AFM} & =(\omega_{\alpha}U_{\alpha}^{2}+\omega_{\beta}V_{\beta}^{2})\hat{a}^{\dagger}\hat{a}+(\omega_{\beta}U_{\beta}^{2}+\omega_{\alpha}V_{\alpha}^{2})\hat{b}^{\dagger}\hat{b}\nonumber \\
 & +(\omega_{\alpha}U_{\alpha}V_{\alpha}+\omega_{\beta}U_{\beta}V_{\beta})(\hat{a}\hat{b}+\hat{a}^{\dagger}\hat{b}^{\dagger}).\label{HAFM_SYM_NEW}
\end{align}
Comparing the above expression with (\ref{eq:HAFM_SYM}) we have
\begin{align*}
\omega_{\alpha}U_{\alpha}^{2}+\omega_{\beta}V_{\beta}^{2} & =\mathcal{A},\\
\omega_{\beta}U_{\beta}^{2}+\omega_{\alpha}V_{\alpha}^{2} & =\mathcal{B},
\end{align*}
and thus, using $U_{j}^{2}-V_{j}^{2}=1$
\begin{equation}
\omega_{\alpha}-\omega_{\beta}=\mathcal{A}-\mathcal{B}=-2\omega_{H}.\label{eq:RelationAB}
\end{equation}

To obtain further information on the form of the Bogoliubov coefficients
we use the eigenvalue equations to obtain

\begin{equation}
\begin{cases}
(\mathcal{A}-\omega_{\alpha})U_{\alpha}-\mathcal{C}V_{\alpha} & =0,\\
\mathcal{C}U_{\alpha}-(\mathcal{B}+\omega_{\alpha})V_{\alpha} & =0,
\end{cases}\label{eq:SoE}
\end{equation}
\begin{equation}
\begin{cases}
-(\mathcal{A}+\omega_{\beta})V_{\beta}+\mathcal{C}U_{\beta} & =0,\\
-\mathcal{C}V_{\beta}+(\mathcal{B}-\omega_{\beta})U_{\beta} & =0.
\end{cases}\label{eq:SoE2}
\end{equation}
Using (\ref{eq:RelationAB}) we can rearrange Eq. (\ref{eq:SoE2})
into
\[
\begin{cases}
(\mathcal{A}-\omega_{\alpha})U_{\beta}-\mathcal{C}V_{\beta} & =0,\\
\mathcal{C}U_{\beta}-(\mathcal{B}+\omega_{\alpha})V_{\beta} & =0.
\end{cases}
\]
Comparing with (\ref{eq:SoE}) we conclude that $U_{\beta}=U_{\alpha}=U$
and $V_{\beta}=V_{\alpha}=V$. Finally we notice that since $\omega_{\alpha}=-\omega_{H}+\frac{1}{2}\sqrt{(\mathcal{A}+\mathcal{B})^{2}-4\mathcal{C}^{2}}$,
$\mathcal{A}=-\omega_{H}+\omega_{E}+\omega_{\parallel}$ and $\mathcal{B}=\omega_{H}+\omega_{E}+\omega_{\parallel}$,
then $\mathcal{A}-\omega_{\alpha}$ and $\mathcal{B}+\omega_{\alpha}$
are independent of $\omega_{H}$, and hence the Bogoliubov coefficients
$U$ and $V$ are independent of $\omega_{H}$. Therefore the couplings 

\begin{align}
g_{\alpha} & =U-V+K\left(U+V\right),\label{eq:gagb}\\
g_{\beta} & =U-V-K(U+V)\nonumber 
\end{align}
are independent of the magnetic field for $\omega_{\perp}=0$ and
at $K=0$ they have equal strength $g_{\alpha}=g_{\beta}$. In this
derivation the important fact was $u_{\alpha,b}=u_{\beta,a}=v_{\alpha,a}=v_{\beta,b}=0$,
which is consequence of the invariance under rotations around the
$\bm{e}_{z}$ axis.

\subsubsection{Degenerate case $\omega_{H}=\omega_{\perp}=0$}

For\textbf{ $\omega_{H}=\omega_{\perp}=0$}, under $\hat{\mathcal{S}}:\hat{a}\rightarrow\hat{b}$
we have $\hat{\mathcal{S}}\hat{\alpha}\hat{S}^{-1}=\hat{\beta}$,
and the diagonalized Hamiltonian Eq. (\ref{eq:hamprime}) is invariant
since $\omega_{\alpha}=\omega_{\beta}$. This falls into the previous
case and the couplings $g_{\alpha,\beta}$ are given by Eqs. (\ref{eq:gagb}). 

Note that the Bogoliubov coefficients present a discontinuity at $\omega_{\perp}=0$
and therefore also the $g_{\alpha,\beta}$. In particular, $g_{\alpha}(\omega_{H}=0,K=0)=0$
for $\omega_{\perp}\neq0$ as we showed in Subsection A, but it is
finite for $\omega_{\perp}=0$, see Subsection B and Eq. (8) in the
main text.

\subsection{Equations of motion for the control-probe pump scheme}

In this section we derive the cavity spectra of the antiferromagnetic
optomagnonic system by solving the linearized quantum Langevin equations.
The total Hamiltonian of the driven optomagnonic cavity, under the
simplifications of the previous sections, reads
\begin{align*}
\hat{H}= & \hbar\omega_{c}\hat{c}^{\dagger}\hat{c}+\hbar\omega_{\alpha}\hat{\alpha}^{\dagger}\hat{\alpha}+\hbar\omega_{\beta}\hat{\beta}^{\dagger}\hat{\beta}\\
 & -\hbar G\hat{c}^{\dagger}\hat{c}\left(g_{\alpha}\hat{\alpha}^{\dagger}+g_{\beta}\beta^{\dagger}+{\rm h.c.}\right)+\hat{H}_{{\rm drive}}.
\end{align*}
The intracavity field is driven by input lasers which can be described
by the Hamiltonian $\hat{H}_{{\rm drive}}=i\hbar\sqrt{\eta\kappa}\left(s_{{\rm in}}(t)\hat{c}^{\dagger}-s_{{\rm in}}^{*}(t)\hat{c}\right)$,
where $s_{in}(t)=\left(s_{d}e^{-i\omega_{d}t}+s_{p}e^{-i\omega_{p}t}\right)$
is the amplitude of the drive normalized to the input photon flux
and $\omega_{d}\left(\omega_{p}\right)$ is the control (probe) laser
frequency, with $s_{d,p}=\sqrt{\frac{2\mathcal{P}_{d,p}}{\hbar\omega_{d,p}}}$
given in terms of the drive powers $\mathcal{P}_{d,p}$. The total
cavity loss rate is denoted by $\kappa=\kappa_{0}+\kappa_{ex}$. where
$\kappa_{0}$ denotes the intrinsic loss rate and $\kappa_{ex}$ the
external loss rate. The dimensionless parameter $\eta\equiv\kappa_{ex}/(\kappa_{0}+\kappa_{ex})$,
can be continuously adjusted in experiments \citep{caiObservationCriticalCoupling2000,spillaneIdealityFiberTaperCoupledMicroresonator2003}.
In this control-probe scheme, the control laser has a stronger intensity
than the probe laser $s_{d}\gg s_{p}$ \citep{safavi-naeiniElectromagneticallyInducedTransparency2011a,weis2010optomechanically,aspelmeyer_cavity_2014}.
In a frame rotating with the control laser frequency, the total Hamiltonian
reads
\begin{align*}
\hat{H}= & -\hbar\Delta\hat{c}^{\dagger}\hat{c}+\hbar\omega_{\alpha}\hat{\alpha}^{\dagger}\hat{\alpha}+\hbar\omega_{\beta}\hat{\beta}^{\dagger}\hat{\beta}\\
 & -\hbar G\hat{c}^{\dagger}\hat{c}\left(g_{\alpha}\hat{\alpha}^{\dagger}+g_{\beta}\beta^{\dagger}+{\rm h.c.}\right)\\
 & +i\hbar\sqrt{\eta\kappa}\left[\left(s_{d}+s_{p}e^{-i(\omega_{p}-\omega_{d})t}\right)\hat{c}^{\dagger}-{\rm h.c.}\right],
\end{align*}
where $\Delta=\omega_{d}-\omega_{c}$ is the detuning between the
cavity and the control laser frequency. The Langevin equations of
motion for the operators $\hat{\alpha}$, $\hat{\beta}$ and $\hat{c}$
are thus

\begin{align*}
\dot{\hat{\alpha}} & =-\left(i\omega_{\alpha}+\frac{\Gamma_\alpha}{2}\right)\hat{\alpha}+iGg_{\alpha}\hat{c}^{\dagger}\hat{c}+\sqrt{\Gamma_\alpha}\hat{\alpha}_{{\rm noise}}(t),\\
\dot{\hat{\beta}} & =-\left(i\omega_{\beta}+\frac{\Gamma_\beta}{2}\right)\hat{\beta}+iGg_{\beta}\hat{c}^{\dagger}\hat{c}+\sqrt{\Gamma_\beta}\hat{\beta}_{{\rm noise}}(t),\\
\dot{\hat{c}} & =\left(i\Delta-\frac{\kappa}{2}\right)\hat{c}+iG\hat{c}\left[g_{\alpha}\left(\hat{\alpha}+\hat{\alpha}^{\dagger}\right)+g_{\beta}\left(\hat{\beta}+\hat{\beta}^{\dagger}\right)\right]\\
 & +\sqrt{\eta\kappa}s_{d}+\sqrt{(1-\eta)\kappa}\hat{c}_{{\rm noise}}(t),
\end{align*}
with $\Gamma_{\alpha, \beta}$ is the intrinsic magnon damping rates which we assume to be equal $\Gamma_\alpha = \Gamma_\beta = \Gamma$. We have disregarded,
for now, the effects of the weak probe pump since $s_{d}\gg s_{p}$.
Here and in what follows, the noise operators describe random fluctuations
of the system and have vanishing expectation values \citep{gardinerQuantumNoiseHandbook2000}.
We consider the equations for the expectations values $\langle\hat{\alpha}\rangle$,
$\langle\hat{\beta}\rangle$ and $\langle\hat{c}\rangle$, disregarding
quantum fluctuations, such that for example $\langle\hat{c}^{\dagger}\hat{c}\rangle=\vert\langle\hat{c}\rangle\vert^{2}$.
The steady state is obtained by setting $\langle\dot{\hat{c}}\rangle\equiv\dot{\bar{c}}=0$,
$\langle\dot{\hat{\alpha}}\rangle=0$ and $\langle\dot{\hat{\beta}}\rangle=0$,
such that
\begin{align*}
\langle\hat{\alpha}\rangle & =\frac{iGg_{\alpha}\vert\bar{c}\vert^{2}}{i\omega_{\alpha}+\frac{\Gamma}{2}},\\
\langle\hat{\beta}\rangle & =\frac{iGg_{\beta}\vert\bar{c}\vert^{2}}{i\omega_{\beta}+\frac{\Gamma}{2}},\\
\bar{c} & =\frac{-\sqrt{\eta\kappa}s_{d}}{(i\Delta-\frac{\kappa}{2})+2iG\left(g_{\alpha}{\rm Re}[\langle\hat{\alpha}\rangle]+g_{\beta}{\rm Re}[\langle\hat{\beta}\rangle]\right)}.
\end{align*}

We linearize the dynamics of the system by considering the fluctuations
over the steady state values $\hat{\alpha}(t)=\langle\hat{\alpha}\rangle+\delta\hat{\alpha}(t)$,
$\hat{\beta}(t)=\langle\hat{\beta}\rangle+\delta\hat{\beta}(t)$ and
$\hat{c}(t)=\bar{c}+\delta\hat{c}(t)$. Correspondingly for the input
field $s_{{\rm in}}(t)=\bar{s}+\delta s_{{\rm in}}(t)$, where we
identify $\bar{s}=s_{d}$ as the amplitude of the control field due
to $s_{d}\gg s_{p}$. The linearized Langevin equations for the fields'
fluctuations read
\begin{align*}
\delta\dot{\hat{\alpha}}= & -\left(i\omega_{\alpha}+\frac{\Gamma}{2}\right)\delta\hat{\alpha}+iGg_{\alpha}(\bar{c}\delta\hat{c}^{\dagger}+\bar{c}^{*}\delta\hat{c})\\
 & +\sqrt{\Gamma}\delta\hat{\alpha}_{{\rm noise}}(t),\\
\delta\dot{\hat{\beta}}= & -\left(i\omega_{\beta}+\frac{\Gamma}{2}\right)\delta\hat{\beta}+iGg_{\beta}(\bar{c}\delta\hat{c}^{\dagger}+\bar{c}^{*}\delta\hat{c})\\
 & +\sqrt{\Gamma}\delta\hat{\beta}_{{\rm noise}}(t),\\
\delta\dot{\hat{c}}= & \left(i\tilde{\Delta}-\frac{\kappa}{2}\right)\delta\hat{c}+iG\bar{c}\left(g_{\alpha}\delta\hat{\alpha}^{\dagger}+g_{\beta}\delta\hat{\beta}^{\dagger}+{\rm h.c.}\right)\\
 & +\sqrt{\eta\kappa}\delta s_{{\rm in}}(t)+\sqrt{\kappa}\delta\hat{c}_{{\rm noise}}(t).
\end{align*}
In these equations $\tilde{\Delta}=\Delta+2G(g_{\alpha}{\rm Re}[\langle\hat{\alpha}\rangle]+g_{\beta}{\rm Re}[\langle\hat{\beta}\rangle])$
takes into account the cavity frequency shift due to the coupling
to the magnon modes.

Finally, we obtain the mode spectra via the Langevin equations for
the average values of the fluctuations in frequency space. Defining
the Fourier transform as $\delta X[\omega]=\int_{-\infty}^{\infty}dte^{-i\omega t}\langle\delta\hat{X}(t)\rangle$
for $X=\delta\alpha^{\left(\dagger\right)},\delta\beta^{\left(\dagger\right)},\delta c^{\left(\dagger\right)}$
\citep{purdyStrongOptomechanicalSqueezing2013} the Langevin equations
read
\begin{align*}
\left[i(\omega_{\alpha}-\omega)+\frac{\Gamma}{2}\right]\delta\alpha[\omega] & =iGg_{\alpha}\left(\bar{c}\delta c^{*}[-\omega]+\bar{c}^{*}\delta c[\omega]\right),\\
\left[i(\omega_{\beta}-\omega)+\frac{\Gamma}{2}\right]\delta\beta[\omega] & =iGg_{\beta}\left(\bar{c}\delta c^{*}[-\omega]+\bar{c}^{*}\delta c[\omega]\right),\\
\left[-i(\tilde{\Delta}+\omega)+\frac{\kappa}{2}\right]\delta c[\omega] & =iG\bar{c}\Big(g_{\alpha}\left(\delta\alpha^{*}[-\omega]+\delta\alpha[\omega]\right)\\
 & +g_{\beta}\left(\delta\beta^{*}[-\omega]+\delta\beta[\omega]\right)\Big)\\
 & \,\,\,+\sqrt{\eta\kappa}\delta s_{{\rm in}}[\omega].
\end{align*}
The cavity spectrum is thus 
\[
\delta c[\omega]=\frac{(1+F(\omega))\sqrt{\eta\kappa}\delta s_{{\rm in}}[\omega]}{-i(\tilde{\Delta}+\omega)+\frac{\kappa}{2}-2i\tilde{\Delta}F(\omega)},
\]
where 
\[
F(\omega)=\frac{\Sigma(\omega)}{i(\tilde{\Delta}-\omega)+\frac{\kappa}{2}},
\]
with $\Sigma(\omega)$ the self-energy term given by

\begin{align*}
\Sigma \left(\omega\right) & =G^{2}\vert\bar{c}\vert^{2}\\
 & \left[-\frac{g_{\alpha}^{2}}{i\left(\omega_{\alpha}-\omega\right)+\frac{\Gamma}{2}}+\frac{g_{\alpha}^{2}}{-i\left(\omega_{\alpha}+\omega\right)+\frac{\Gamma}{2}}\right.\\
 & \left.-\frac{g_{\beta}^{2}}{i\left(\omega_{\beta}-\omega\right)+\frac{\Gamma}{2}}+\frac{g_{\beta}^{2}}{-i\left(\omega_{\beta}+\omega\right)+\frac{\Gamma}{2}}\right].
\end{align*}

Since we are working in a rotating frame with the control laser, the
frequency $\omega$ is the sideband shift of the probe $\omega_{p}$
from the control light frequency, $\omega=\omega_{p}-\omega_{d}$.
In the following and in the main text we restrict our results to the
resolved sideband regime case, $\omega_{\alpha,\beta}\gg\kappa$).
For a red detuned control drive ($\tilde{\Delta}<0$), Stoke's processes
are far off-resonance and the relevant resonance is $\tilde{\Delta}\approx-\omega\approx-\omega_{\alpha,\beta}$.
In this case we can approximate $F(\omega)\sim\Sigma(\omega)/2i\tilde{\Delta}$
and, if $\tilde{\Delta}\gg\Sigma(\omega)$, then the cavity spectrum
is given by
\begin{equation}
\delta c_{{\rm \tilde{\Delta}<0}}[\omega]=\frac{\sqrt{\eta\kappa}\delta s_{{\rm in}}[\omega]}{-i(\tilde{\Delta}+\omega)+\frac{\kappa}{2}-\Sigma_{{\rm AS}}(\omega)},\label{eq:CavitySpectrumRed}
\end{equation}
where $\Sigma_{{\rm AS}}=-\frac{G^{2} \vert\bar{c}\vert^{2} g_{\alpha}^{2}}{i(\omega_{\alpha}-\omega)+\Gamma/2}-\frac{G^{2} \vert\bar{c}\vert^{2} g_{\beta}^{2}}{i(\omega_{\beta}-\omega)+\Gamma/2}$.
The input-output boundary conditions $\delta s_{{\rm out}}[\omega]=\delta s_{{\rm in}}[\omega]-\sqrt{\eta\kappa}\delta c[\omega]$
thus yield to the transmission amplitude $t=\delta s_{{\rm out}}[\omega]/\delta s_{{\rm in}}[\omega]$.

Analogously, by calculating the self energy term of the magnon mode $j$, $\tilde{\Sigma}_{j}[\omega]$ ($j=\alpha,\beta$), we can obtain the corresponding linewidth $\tilde{\Gamma}_{j}=\Gamma+2{\rm Re}[\tilde{\Sigma}_{j}]$
and frequency shift $\delta\omega_{j}={\rm Im}[\tilde{\Sigma}_{j}]$. The magnon self energy is given explicitly by (for $k\neq j$, and omitting the
dependences on $\omega$)
\begin{align}
\tilde{\Sigma}_{j}[\omega] & =-\xi_{j}(1+X_{j})+\xi_{jk}X_{k}\nonumber \\
 & -\frac{\xi_{j}X_{j}+\xi_{jk}(X_{k}+1)}{\chi_{j}^{-1}-\xi_{j}(1+X_{j})+\xi_{jk}X_{k}}\left(\xi_{k}X_{k}+\xi_{jk}(1+X_{j})\right),\label{eq:LongExpression}
\end{align}
with
\begin{align*}
\xi_{j}[\omega] & =G^{2}\vert\bar{c}\vert^{2}g_{j}^{2}\left[\frac{1}{i(\tilde{\Delta}-\omega)+\frac{\kappa}{2}}-\frac{1}{-i(\tilde{\Delta}+\omega)+\frac{\kappa}{2}}\right],\\
\xi_{jk}[\omega] & =\sqrt{\xi_{j}[\omega]\xi_{k}[\omega]},\\
\chi_{j}[\omega] & =\left[i(\omega_{j}-\omega)+\frac{\Gamma}{2}\right]^{-1},\\
X_{j}[\omega] & =-\left[\chi_{j}^{*}[-\omega]+\xi_{j}[\omega]-\frac{\xi_{jk}[\omega]}{\chi_{k}^{*}[-\omega]+\xi_{k}[\omega]}\right]^{-1}\\
 & \times\xi_{j}[\omega]\left[\frac{\xi_{k}}{\chi_{k}^{*}[-\omega]+\xi_{k}}-1\right].
\end{align*}

\subsection{\label{sec:magnon-magnon}Cavity-induced magnon-magnon interactions}

The full self-energy expression \eqref{eq:LongExpression} includes contributions due to bare Stokes and anti-Stokes processes involving the magnon mode and the cavity mode, as one can see directly in the terms $\xi_j$, as well as contributions due to indirect magnon-magnon interactions mediated by the cavity. The effects of these different terms are not immediately clear from \eqref{eq:LongExpression}, thus, we now consider a framework to understand the effects of each contribution of the self-energy by comparing the full expression \eqref{eq:LongExpression} with the magnon self-energy obtained by considering a fixed detuning and performing the rotating wave approximation. Let's consider that the control laser is red-detuned (the same procedure applies to the blue detuning regime with equivalent conclusions). Employing the rotating wave approximation we obtain the following equations of motion
\begin{align*}
\delta\dot{\hat{\alpha}}= & -\left(i\omega_{\alpha}+\frac{\Gamma}{2}\right)\delta\hat{\alpha}+iGg_{\alpha} \bar{c}^{*}\delta\hat{c}\\
 & +\sqrt{\Gamma}\delta\hat{\alpha}_{{\rm noise}}(t),\\
\delta\dot{\hat{\beta}}= & -\left(i\omega_{\beta}+\frac{\Gamma}{2}\right)\delta\hat{\beta}+iGg_{\beta} \bar{c}^{*}\delta\hat{c}\\
 & +\sqrt{\Gamma}\delta\hat{\beta}_{{\rm noise}}(t),\\
\delta\dot{\hat{c}}= & \left(i\tilde{\Delta}-\frac{\kappa}{2}\right)\delta\hat{c}+iG\bar{c}\left(g_{\alpha}\delta\hat{\alpha}+g_{\beta}\delta\hat{\beta} \right)\\
 & +\sqrt{\eta\kappa}\delta s_{{\rm in}}(t)+\sqrt{\kappa}\delta\hat{c}_{{\rm noise}}(t).
\end{align*}
In contrast to the full equations of motion, the above system of equations does not include terms $\propto \delta \alpha^\dagger, \delta \beta^\dagger, \delta c^\dagger$, since for a red-detuned control laser those are counter-rotating terms. From now own we also assume that $\bar{c}^* = \bar{c}$ and $G^* = G$. By following the same procedure leading to \eqref{eq:LongExpression}, i.e. consider the frequency domain system of equations for the average values of the operators, we obtain the rotating wave approximation self-energy of the magnon mode $j=\alpha, \beta$, $\tilde{\Sigma}^{{\rm{RW}}}_j[\omega]$:
\begin{equation}
\label{SigmaRW}
\begin{aligned}
\tilde{\Sigma}^{{\rm{RW}}}_j[\omega] &= \tilde{\Sigma}_j^{{\rm{Bare}}}[\omega] + \tilde{\Sigma}_j^{\rm{MM}}[\omega], \\
\tilde{\Sigma}_j^{{\rm{Bare}}}[\omega] &= \frac{G^2 \bar{c}^2 g_j ^2}{-i \left( \tilde{\Delta} + \omega \right)+\frac{\kappa}{2}}, \\
 \tilde{\Sigma}_j^{\rm{MM}}[\omega] &= -\frac{G^4 \bar{c}^4 g_j ^2 g_k ^2}{\left[-i \left( \tilde{\Delta} + \omega \right)+\frac{\kappa}{2} \right]^2 \left( \chi_{k}[\omega] + \tilde{\Sigma}^{{\rm{Bare}}}_k[\omega] \right)},
\end{aligned}
\end{equation}
with $j \neq k$. In the self-energy $\tilde{\Sigma}^{{\rm{RW}}}_j[\omega] $, the first term $\tilde{\Sigma}_j^{{\rm{Bare}}}[\omega]$ describes the effects of the anti-Stokes processes converting a magnon $j$ into a cavity photon, while the second term $\tilde{\Sigma}_j^{\rm{MM}}[\omega]$ is associated with the magnon-magnon interactions mediated by a cavity photon. 

\begin{figure}[h!]
\centering{}\includegraphics[width=1\columnwidth]{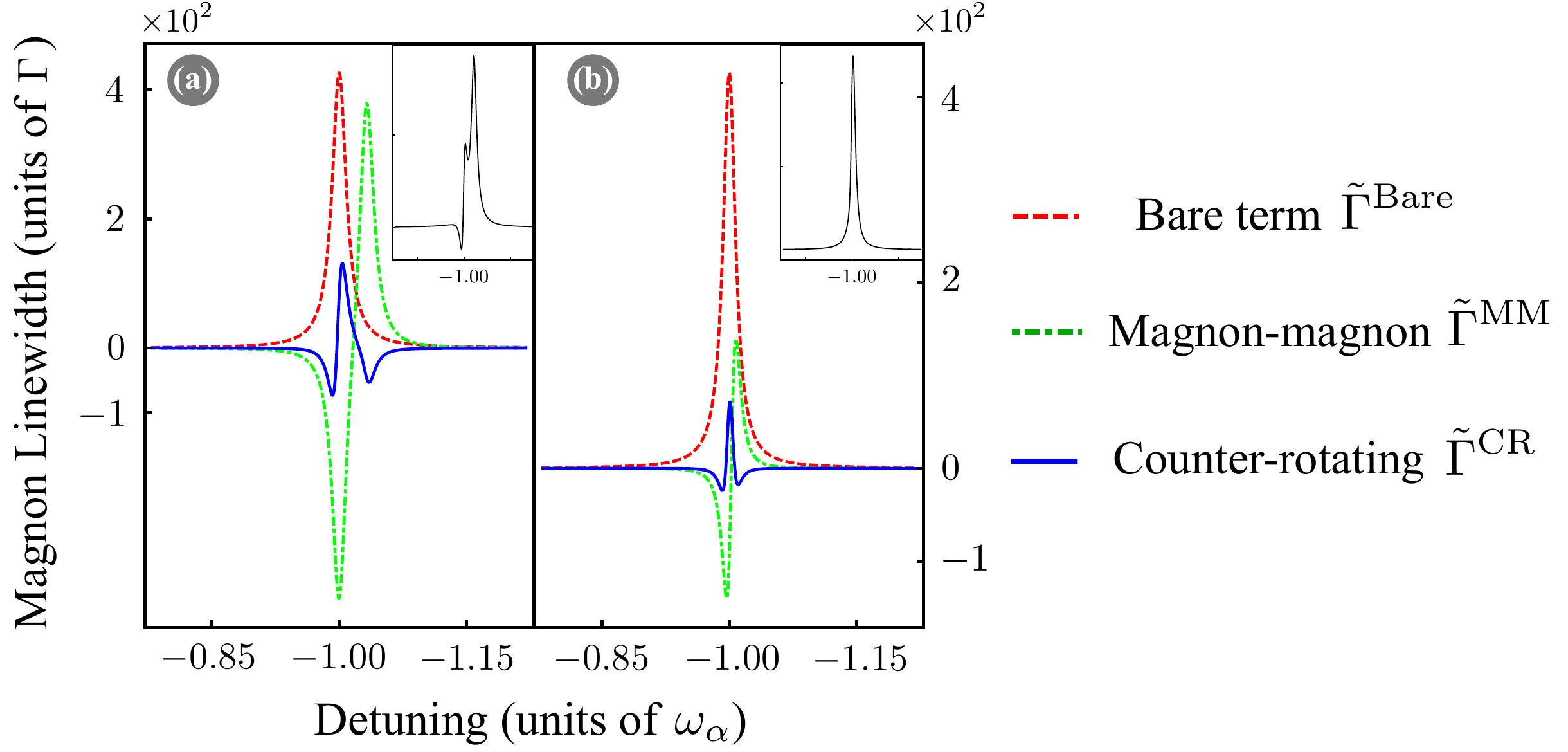}
\caption{Contribution from each term of the magnon self energy \eqref{SigmaRW} and \eqref{SigmaCR} in the red-detuning regime. Parameters as in Figure 3 of the main text for (a) the main figure and (b) the inset. In the inset we show the full magnon decay rate.}
\label{FigSuppNew}
\end{figure}

In this scenario (red detuned control light), the counter-rotating contribution to the self energy $\tilde{\Sigma}^{{\rm{CR}}}_j[\omega]$ can be obtained as the difference between $\tilde{\Sigma}^{{\rm{RW}}}_j[\omega]$ and the full self-energy term \eqref{eq:LongExpression}:
\begin{equation}
\label{SigmaCR}
\tilde{\Sigma}^{{\rm{CR}}}_j[\omega] = \tilde{\Sigma}_{j}[\omega] -\tilde{\Sigma}^{{\rm{RW}}}_j[\omega].
\end{equation}
This term includes all counter-rotating contributions due to Stokes process and the induced counter-rotating magnon-magnon interactions generated by them. We can now define the decay rates and frequency shifts associated with each of those terms (${\rm{A}} = {\rm{Bare}},\, {\rm{MM}},\, {\rm{CR}}$):
\begin{equation}
\begin{aligned}
\tilde{\Gamma}^{{\rm{A}}}[\omega] &= 2 {\rm{Re}} \left[ \tilde{\Sigma}^{{\rm{A}}} [\omega]\right], \\
\delta \omega_j^{{\rm{A}}}[\omega]  &= {\rm{Im}} \left[ \tilde{\Sigma}^{{\rm{A}}}[\omega] \right].
\end{aligned}
\end{equation}
We show in Figure \ref{FigSuppNew} $\tilde{\Gamma}^{{\rm{Bare}}}[\omega]$, $\tilde{\Gamma}^{{\rm{MM}}}[\omega]$ and $\tilde{\Gamma}^{{\rm{CR}}}[\omega]$ for the $\alpha$ mode and same parameters as in Figure 3 of the main text for (a) the main plot and (b) the inset. We see that in the case (a), in which the magnon frequencies are closer than the effective decay rate (the near degenerate case), both magnon-magnon interactions (green curve) and counter-rotating terms (blue curve) contribute substantially to the decay rate. In particular the contribution due to magnon-magnon indirect interactions is as important as the bare term. The importance of those terms depends on the strength of the coupling, the closeness of the magnon modes and the cavity decay rate, and these are also related to the formation of hybrid modes between the cavity and the magnon modes (see \cite{genessimultaneous2008, ockeloensidebandcooling2019, sommerpartialoptomechanical2019}). The combination of those three terms gives the unusual behavior depicted in the Figure 3 of the main text. In case (b) the frequencies are well separated and both magnon-magnon indirect interactions and counter-rotating terms have a small contribution to the overall decay rate.

Finally, we notice that under the same assumptions leading to (\ref{eq:CavitySpectrumRed}),
(\ref{eq:LongExpression}) is given by
\[
\tilde{\Sigma}_{j}[\omega]\approx\frac{G^{2}\vert\bar{c}\vert^{2}g_{j}^{2}}{-i(\tilde{\Delta}+\omega)+\frac{\kappa}{2}},
\]
which corresponds to the bare self energy term in \eqref{SigmaRW}. In this case the magnon linewidth and frequency shift are given by
\begin{align*}
\delta\omega_{j}[\omega] & \approx\frac{G^{2}\vert\bar{c}\vert^{2}g_{j}^{2}(\tilde{\Delta}+\omega)}{(\tilde{\Delta}+\omega)^{2}+\frac{\kappa^{2}}{4}},\\
\tilde{\Gamma}_j & \approx \Gamma+\frac{G^{2}\vert\bar{c}\vert^{2}g_{j}^{2} \kappa}{(\tilde{\Delta}+\omega)^{2}+\frac{\kappa^{2}}{4}}.
\end{align*}

\bibliographystyle{apsrev4-1}
\bibliography{SubminV03}

\begin{thebibliography}{93}%
\makeatletter
\providecommand \@ifxundefined [1]{%
 \@ifx{#1\undefined}
}%
\providecommand \@ifnum [1]{%
 \ifnum #1\expandafter \@firstoftwo
 \else \expandafter \@secondoftwo
 \fi
}%
\providecommand \@ifx [1]{%
 \ifx #1\expandafter \@firstoftwo
 \else \expandafter \@secondoftwo
 \fi
}%
\providecommand \natexlab [1]{#1}%
\providecommand \enquote  [1]{``#1''}%
\providecommand \bibnamefont  [1]{#1}%
\providecommand \bibfnamefont [1]{#1}%
\providecommand \citenamefont [1]{#1}%
\providecommand \href@noop [0]{\@secondoftwo}%
\providecommand \href [0]{\begingroup \@sanitize@url \@href}%
\providecommand \@href[1]{\@@startlink{#1}\@@href}%
\providecommand \@@href[1]{\endgroup#1\@@endlink}%
\providecommand \@sanitize@url [0]{\catcode `\\12\catcode `\$12\catcode
  `\&12\catcode `\#12\catcode `\^12\catcode `\_12\catcode `\%12\relax}%
\providecommand \@@startlink[1]{}%
\providecommand \@@endlink[0]{}%
\providecommand \url  [0]{\begingroup\@sanitize@url \@url }%
\providecommand \@url [1]{\endgroup\@href {#1}{\urlprefix }}%
\providecommand \urlprefix  [0]{URL }%
\providecommand \Eprint [0]{\href }%
\providecommand \doibase [0]{http://dx.doi.org/}%
\providecommand \selectlanguage [0]{\@gobble}%
\providecommand \bibinfo  [0]{\@secondoftwo}%
\providecommand \bibfield  [0]{\@secondoftwo}%
\providecommand \translation [1]{[#1]}%
\providecommand \BibitemOpen [0]{}%
\providecommand \bibitemStop [0]{}%
\providecommand \bibitemNoStop [0]{.\EOS\space}%
\providecommand \EOS [0]{\spacefactor3000\relax}%
\providecommand \BibitemShut  [1]{\csname bibitem#1\endcsname}%
\let\auto@bib@innerbib\@empty
\bibitem [{\citenamefont {Zhang}\ \emph {et~al.}(2016)\citenamefont {Zhang},
  \citenamefont {Zhu}, \citenamefont {Zou},\ and\ \citenamefont
  {Tang}}]{zhang_optomagnonic_2016}%
  \BibitemOpen
  \bibfield  {author} {\bibinfo {author} {\bibfnamefont {X.}~\bibnamefont
  {Zhang}}, \bibinfo {author} {\bibfnamefont {N.}~\bibnamefont {Zhu}}, \bibinfo
  {author} {\bibfnamefont {C.-L.}\ \bibnamefont {Zou}}, \ and\ \bibinfo
  {author} {\bibfnamefont {H.~X.}\ \bibnamefont {Tang}},\ }\href {\doibase
  10.1103/PhysRevLett.117.123605} {\bibfield  {journal} {\bibinfo  {journal}
  {Physical Review Letters}\ }\textbf {\bibinfo {volume} {117}},\ \bibinfo
  {pages} {123605} (\bibinfo {year} {2016})}\BibitemShut {NoStop}%
\bibitem [{\citenamefont {Osada}\ \emph {et~al.}(2016)\citenamefont {Osada},
  \citenamefont {Hisatomi}, \citenamefont {Noguchi}, \citenamefont {Tabuchi},
  \citenamefont {Yamazaki}, \citenamefont {Usami}, \citenamefont {Sadgrove},
  \citenamefont {Yalla}, \citenamefont {Nomura},\ and\ \citenamefont
  {Nakamura}}]{osada_cavity_2016}%
  \BibitemOpen
  \bibfield  {author} {\bibinfo {author} {\bibfnamefont {A.}~\bibnamefont
  {Osada}}, \bibinfo {author} {\bibfnamefont {R.}~\bibnamefont {Hisatomi}},
  \bibinfo {author} {\bibfnamefont {A.}~\bibnamefont {Noguchi}}, \bibinfo
  {author} {\bibfnamefont {Y.}~\bibnamefont {Tabuchi}}, \bibinfo {author}
  {\bibfnamefont {R.}~\bibnamefont {Yamazaki}}, \bibinfo {author}
  {\bibfnamefont {K.}~\bibnamefont {Usami}}, \bibinfo {author} {\bibfnamefont
  {M.}~\bibnamefont {Sadgrove}}, \bibinfo {author} {\bibfnamefont
  {R.}~\bibnamefont {Yalla}}, \bibinfo {author} {\bibfnamefont
  {M.}~\bibnamefont {Nomura}}, \ and\ \bibinfo {author} {\bibfnamefont
  {Y.}~\bibnamefont {Nakamura}},\ }\href {\doibase
  10.1103/PhysRevLett.116.223601} {\bibfield  {journal} {\bibinfo  {journal}
  {Physical Review Letters}\ }\textbf {\bibinfo {volume} {116}},\ \bibinfo
  {pages} {223601} (\bibinfo {year} {2016})}\BibitemShut {NoStop}%
\bibitem [{\citenamefont {Haigh}\ \emph {et~al.}(2016)\citenamefont {Haigh},
  \citenamefont {Nunnenkamp}, \citenamefont {Ramsay},\ and\ \citenamefont
  {Ferguson}}]{haigh_triple-resonant_2016}%
  \BibitemOpen
  \bibfield  {author} {\bibinfo {author} {\bibfnamefont {J.~A.}\ \bibnamefont
  {Haigh}}, \bibinfo {author} {\bibfnamefont {A.}~\bibnamefont {Nunnenkamp}},
  \bibinfo {author} {\bibfnamefont {A.~J.}\ \bibnamefont {Ramsay}}, \ and\
  \bibinfo {author} {\bibfnamefont {A.~J.}\ \bibnamefont {Ferguson}},\ }\href
  {\doibase 10.1103/PhysRevLett.117.133602} {\bibfield  {journal} {\bibinfo
  {journal} {Phys. Rev. Lett.}\ }\textbf {\bibinfo {volume} {117}},\ \bibinfo
  {pages} {133602} (\bibinfo {year} {2016})}\BibitemShut {NoStop}%
\bibitem [{\citenamefont {Soykal}\ and\ \citenamefont
  {Flatt\'e}(2010)}]{soykal_strong_2010}%
  \BibitemOpen
  \bibfield  {author} {\bibinfo {author} {\bibfnamefont {{\"O}.~O.}\
  \bibnamefont {Soykal}}\ and\ \bibinfo {author} {\bibfnamefont {M.~E.}\
  \bibnamefont {Flatt\'e}},\ }\href {\doibase 10.1103/PhysRevLett.104.077202}
  {\bibfield  {journal} {\bibinfo  {journal} {Physical Review Letters}\
  }\textbf {\bibinfo {volume} {104}},\ \bibinfo {pages} {077202} (\bibinfo
  {year} {2010})}\BibitemShut {NoStop}%
\bibitem [{\citenamefont {Huebl}\ \emph {et~al.}(2013)\citenamefont {Huebl},
  \citenamefont {Zollitsch}, \citenamefont {Lotze}, \citenamefont {Hocke},
  \citenamefont {Greifenstein}, \citenamefont {Marx}, \citenamefont {Gross},\
  and\ \citenamefont {Goennenwein}}]{huebl_high_2013}%
  \BibitemOpen
  \bibfield  {author} {\bibinfo {author} {\bibfnamefont {H.}~\bibnamefont
  {Huebl}}, \bibinfo {author} {\bibfnamefont {C.~W.}\ \bibnamefont
  {Zollitsch}}, \bibinfo {author} {\bibfnamefont {J.}~\bibnamefont {Lotze}},
  \bibinfo {author} {\bibfnamefont {F.}~\bibnamefont {Hocke}}, \bibinfo
  {author} {\bibfnamefont {M.}~\bibnamefont {Greifenstein}}, \bibinfo {author}
  {\bibfnamefont {A.}~\bibnamefont {Marx}}, \bibinfo {author} {\bibfnamefont
  {R.}~\bibnamefont {Gross}}, \ and\ \bibinfo {author} {\bibfnamefont
  {S.~T.~B.}\ \bibnamefont {Goennenwein}},\ }\href {\doibase
  10.1103/PhysRevLett.111.127003} {\bibfield  {journal} {\bibinfo  {journal}
  {Phys. Rev. Lett.}\ }\textbf {\bibinfo {volume} {111}},\ \bibinfo {pages}
  {127003} (\bibinfo {year} {2013})}\BibitemShut {NoStop}%
\bibitem [{\citenamefont {Zhang}\ \emph {et~al.}(2014)\citenamefont {Zhang},
  \citenamefont {Zou}, \citenamefont {Jiang},\ and\ \citenamefont
  {Tang}}]{zhang_strongly_2014}%
  \BibitemOpen
  \bibfield  {author} {\bibinfo {author} {\bibfnamefont {X.}~\bibnamefont
  {Zhang}}, \bibinfo {author} {\bibfnamefont {C.-L.}\ \bibnamefont {Zou}},
  \bibinfo {author} {\bibfnamefont {L.}~\bibnamefont {Jiang}}, \ and\ \bibinfo
  {author} {\bibfnamefont {H.~X.}\ \bibnamefont {Tang}},\ }\href {\doibase
  10.1103/PhysRevLett.113.156401} {\bibfield  {journal} {\bibinfo  {journal}
  {Physical Review Letters}\ }\textbf {\bibinfo {volume} {113}},\ \bibinfo
  {pages} {156401} (\bibinfo {year} {2014})}\BibitemShut {NoStop}%
\bibitem [{\citenamefont {Tabuchi}\ \emph {et~al.}(2014)\citenamefont
  {Tabuchi}, \citenamefont {Ishino}, \citenamefont {Ishikawa}, \citenamefont
  {Yamazaki}, \citenamefont {Usami},\ and\ \citenamefont
  {Nakamura}}]{tabuchi_hybridizing_2014}%
  \BibitemOpen
  \bibfield  {author} {\bibinfo {author} {\bibfnamefont {Y.}~\bibnamefont
  {Tabuchi}}, \bibinfo {author} {\bibfnamefont {S.}~\bibnamefont {Ishino}},
  \bibinfo {author} {\bibfnamefont {T.}~\bibnamefont {Ishikawa}}, \bibinfo
  {author} {\bibfnamefont {R.}~\bibnamefont {Yamazaki}}, \bibinfo {author}
  {\bibfnamefont {K.}~\bibnamefont {Usami}}, \ and\ \bibinfo {author}
  {\bibfnamefont {Y.}~\bibnamefont {Nakamura}},\ }\href {\doibase
  10.1103/PhysRevLett.113.083603} {\bibfield  {journal} {\bibinfo  {journal}
  {Phys. Rev. Lett.}\ }\textbf {\bibinfo {volume} {113}},\ \bibinfo {pages}
  {083603} (\bibinfo {year} {2014})}\BibitemShut {NoStop}%
\bibitem [{\citenamefont {Goryachev}\ \emph {et~al.}(2014)\citenamefont
  {Goryachev}, \citenamefont {Farr}, \citenamefont {Creedon}, \citenamefont
  {Fan}, \citenamefont {Kostylev},\ and\ \citenamefont
  {Tobar}}]{goryachev_high_2014}%
  \BibitemOpen
  \bibfield  {author} {\bibinfo {author} {\bibfnamefont {M.}~\bibnamefont
  {Goryachev}}, \bibinfo {author} {\bibfnamefont {W.~G.}\ \bibnamefont {Farr}},
  \bibinfo {author} {\bibfnamefont {D.~L.}\ \bibnamefont {Creedon}}, \bibinfo
  {author} {\bibfnamefont {Y.}~\bibnamefont {Fan}}, \bibinfo {author}
  {\bibfnamefont {M.}~\bibnamefont {Kostylev}}, \ and\ \bibinfo {author}
  {\bibfnamefont {M.~E.}\ \bibnamefont {Tobar}},\ }\href {\doibase
  10.1103/PhysRevApplied.2.054002} {\bibfield  {journal} {\bibinfo  {journal}
  {Phys. Rev. Applied}\ }\textbf {\bibinfo {volume} {2}},\ \bibinfo {pages}
  {054002} (\bibinfo {year} {2014})}\BibitemShut {NoStop}%
\bibitem [{\citenamefont {Lambert}\ \emph {et~al.}(2015)\citenamefont
  {Lambert}, \citenamefont {Haigh},\ and\ \citenamefont
  {Ferguson}}]{lambert_identification_2015}%
  \BibitemOpen
  \bibfield  {author} {\bibinfo {author} {\bibfnamefont {N.~J.}\ \bibnamefont
  {Lambert}}, \bibinfo {author} {\bibfnamefont {J.~A.}\ \bibnamefont {Haigh}},
  \ and\ \bibinfo {author} {\bibfnamefont {A.~J.}\ \bibnamefont {Ferguson}},\
  }\href {\doibase 10.1063/1.4907694} {\bibfield  {journal} {\bibinfo
  {journal} {Journal of Applied Physics}\ }\textbf {\bibinfo {volume} {117}},\
  \bibinfo {pages} {053910} (\bibinfo {year} {2015})},\ \Eprint
  {http://arxiv.org/abs/https://doi.org/10.1063/1.4907694}
  {https://doi.org/10.1063/1.4907694} \BibitemShut {NoStop}%
\bibitem [{\citenamefont {Bourhill}\ \emph {et~al.}(2016)\citenamefont
  {Bourhill}, \citenamefont {Kostylev}, \citenamefont {Goryachev},
  \citenamefont {Creedon},\ and\ \citenamefont
  {Tobar}}]{bourhill_ultrahigh_2016}%
  \BibitemOpen
  \bibfield  {author} {\bibinfo {author} {\bibfnamefont {J.}~\bibnamefont
  {Bourhill}}, \bibinfo {author} {\bibfnamefont {N.}~\bibnamefont {Kostylev}},
  \bibinfo {author} {\bibfnamefont {M.}~\bibnamefont {Goryachev}}, \bibinfo
  {author} {\bibfnamefont {D.~L.}\ \bibnamefont {Creedon}}, \ and\ \bibinfo
  {author} {\bibfnamefont {M.~E.}\ \bibnamefont {Tobar}},\ }\href {\doibase
  10.1103/PhysRevB.93.144420} {\bibfield  {journal} {\bibinfo  {journal}
  {Physical Review B}\ }\textbf {\bibinfo {volume} {93}},\ \bibinfo {pages}
  {144420} (\bibinfo {year} {2016})}\BibitemShut {NoStop}%
\bibitem [{\citenamefont {Afzelius}\ \emph {et~al.}(2010)\citenamefont
  {Afzelius}, \citenamefont {Usmani}, \citenamefont {Amari}, \citenamefont
  {Lauritzen}, \citenamefont {Walther}, \citenamefont {Simon}, \citenamefont
  {Sangouard}, \citenamefont {Min\'a\ifmmode~\check{r}\else \v{r}\fi{}},
  \citenamefont {{de Riedmatten}}, \citenamefont {Gisin},\ and\ \citenamefont
  {Kr\"oll}}]{afzelius_demonstration_2010}%
  \BibitemOpen
  \bibfield  {author} {\bibinfo {author} {\bibfnamefont {M.}~\bibnamefont
  {Afzelius}}, \bibinfo {author} {\bibfnamefont {I.}~\bibnamefont {Usmani}},
  \bibinfo {author} {\bibfnamefont {A.}~\bibnamefont {Amari}}, \bibinfo
  {author} {\bibfnamefont {B.}~\bibnamefont {Lauritzen}}, \bibinfo {author}
  {\bibfnamefont {A.}~\bibnamefont {Walther}}, \bibinfo {author} {\bibfnamefont
  {C.}~\bibnamefont {Simon}}, \bibinfo {author} {\bibfnamefont
  {N.}~\bibnamefont {Sangouard}}, \bibinfo {author} {\bibfnamefont
  {J.}~\bibnamefont {Min\'a\ifmmode~\check{r}\else \v{r}\fi{}}}, \bibinfo
  {author} {\bibfnamefont {H.}~\bibnamefont {{de Riedmatten}}}, \bibinfo
  {author} {\bibfnamefont {N.}~\bibnamefont {Gisin}}, \ and\ \bibinfo {author}
  {\bibfnamefont {S.}~\bibnamefont {Kr\"oll}},\ }\href {\doibase
  10.1103/PhysRevLett.104.040503} {\bibfield  {journal} {\bibinfo  {journal}
  {Phys. Rev. Lett.}\ }\textbf {\bibinfo {volume} {104}},\ \bibinfo {pages}
  {040503} (\bibinfo {year} {2010})}\BibitemShut {NoStop}%
\bibitem [{\citenamefont {Sangouard}\ \emph {et~al.}(2011)\citenamefont
  {Sangouard}, \citenamefont {Simon}, \citenamefont {de~Riedmatten},\ and\
  \citenamefont {Gisin}}]{sangouard_quantum_2011}%
  \BibitemOpen
  \bibfield  {author} {\bibinfo {author} {\bibfnamefont {N.}~\bibnamefont
  {Sangouard}}, \bibinfo {author} {\bibfnamefont {C.}~\bibnamefont {Simon}},
  \bibinfo {author} {\bibfnamefont {H.}~\bibnamefont {de~Riedmatten}}, \ and\
  \bibinfo {author} {\bibfnamefont {N.}~\bibnamefont {Gisin}},\ }\href
  {\doibase 10.1103/RevModPhys.83.33} {\bibfield  {journal} {\bibinfo
  {journal} {Reviews of Modern Physics}\ }\textbf {\bibinfo {volume} {83}},\
  \bibinfo {pages} {33} (\bibinfo {year} {2011})}\BibitemShut {NoStop}%
\bibitem [{\citenamefont {Timoney}\ \emph {et~al.}(2013)\citenamefont
  {Timoney}, \citenamefont {Usmani}, \citenamefont {Jobez}, \citenamefont
  {Afzelius},\ and\ \citenamefont {Gisin}}]{timoney_single_2013}%
  \BibitemOpen
  \bibfield  {author} {\bibinfo {author} {\bibfnamefont {N.}~\bibnamefont
  {Timoney}}, \bibinfo {author} {\bibfnamefont {I.}~\bibnamefont {Usmani}},
  \bibinfo {author} {\bibfnamefont {P.}~\bibnamefont {Jobez}}, \bibinfo
  {author} {\bibfnamefont {M.}~\bibnamefont {Afzelius}}, \ and\ \bibinfo
  {author} {\bibfnamefont {N.}~\bibnamefont {Gisin}},\ }\href {\doibase
  10.1103/PhysRevA.88.022324} {\bibfield  {journal} {\bibinfo  {journal} {Phys.
  Rev. A}\ }\textbf {\bibinfo {volume} {88}},\ \bibinfo {pages} {022324}
  (\bibinfo {year} {2013})}\BibitemShut {NoStop}%
\bibitem [{\citenamefont {Lambert}\ \emph {et~al.}(2014)\citenamefont
  {Lambert}, \citenamefont {Mangin}, \citenamefont {Varaprasad}, \citenamefont
  {Takahashi}, \citenamefont {Hehn}, \citenamefont {Cinchetti}, \citenamefont
  {Malinowski}, \citenamefont {Hono}, \citenamefont {Fainman}, \citenamefont
  {Aeschlimann},\ and\ \citenamefont {Fullerton}}]{lambert_all-optical_2014}%
  \BibitemOpen
  \bibfield  {author} {\bibinfo {author} {\bibfnamefont {C.-H.}\ \bibnamefont
  {Lambert}}, \bibinfo {author} {\bibfnamefont {S.}~\bibnamefont {Mangin}},
  \bibinfo {author} {\bibfnamefont {B.~S. D. C.~S.}\ \bibnamefont
  {Varaprasad}}, \bibinfo {author} {\bibfnamefont {Y.~K.}\ \bibnamefont
  {Takahashi}}, \bibinfo {author} {\bibfnamefont {M.}~\bibnamefont {Hehn}},
  \bibinfo {author} {\bibfnamefont {M.}~\bibnamefont {Cinchetti}}, \bibinfo
  {author} {\bibfnamefont {G.}~\bibnamefont {Malinowski}}, \bibinfo {author}
  {\bibfnamefont {K.}~\bibnamefont {Hono}}, \bibinfo {author} {\bibfnamefont
  {Y.}~\bibnamefont {Fainman}}, \bibinfo {author} {\bibfnamefont
  {M.}~\bibnamefont {Aeschlimann}}, \ and\ \bibinfo {author} {\bibfnamefont
  {E.~E.}\ \bibnamefont {Fullerton}},\ }\href {\doibase
  10.1126/science.1253493} {\bibfield  {journal} {\bibinfo  {journal}
  {Science}\ }\textbf {\bibinfo {volume} {345}},\ \bibinfo {pages} {1337}
  (\bibinfo {year} {2014})}\BibitemShut {NoStop}%
\bibitem [{\citenamefont {Jobez}\ \emph {et~al.}(2014)\citenamefont {Jobez},
  \citenamefont {Usmani}, \citenamefont {Timoney}, \citenamefont {Laplane},
  \citenamefont {Gisin},\ and\ \citenamefont {Afzelius}}]{jobez_cavity_2014}%
  \BibitemOpen
  \bibfield  {author} {\bibinfo {author} {\bibfnamefont {P.}~\bibnamefont
  {Jobez}}, \bibinfo {author} {\bibfnamefont {I.}~\bibnamefont {Usmani}},
  \bibinfo {author} {\bibfnamefont {N.}~\bibnamefont {Timoney}}, \bibinfo
  {author} {\bibfnamefont {C.}~\bibnamefont {Laplane}}, \bibinfo {author}
  {\bibfnamefont {N.}~\bibnamefont {Gisin}}, \ and\ \bibinfo {author}
  {\bibfnamefont {M.}~\bibnamefont {Afzelius}},\ }\href {\doibase
  10.1088/1367-2630/16/8/083005} {\bibfield  {journal} {\bibinfo  {journal}
  {New Journal of Physics}\ }\textbf {\bibinfo {volume} {16}},\ \bibinfo
  {pages} {083005} (\bibinfo {year} {2014})}\BibitemShut {NoStop}%
\bibitem [{\citenamefont {Jobez}\ \emph {et~al.}(2015)\citenamefont {Jobez},
  \citenamefont {Laplane}, \citenamefont {Timoney}, \citenamefont {Gisin},
  \citenamefont {Ferrier}, \citenamefont {Goldner},\ and\ \citenamefont
  {Afzelius}}]{jobez_coherent_2015}%
  \BibitemOpen
  \bibfield  {author} {\bibinfo {author} {\bibfnamefont {P.}~\bibnamefont
  {Jobez}}, \bibinfo {author} {\bibfnamefont {C.}~\bibnamefont {Laplane}},
  \bibinfo {author} {\bibfnamefont {N.}~\bibnamefont {Timoney}}, \bibinfo
  {author} {\bibfnamefont {N.}~\bibnamefont {Gisin}}, \bibinfo {author}
  {\bibfnamefont {A.}~\bibnamefont {Ferrier}}, \bibinfo {author} {\bibfnamefont
  {P.}~\bibnamefont {Goldner}}, \ and\ \bibinfo {author} {\bibfnamefont
  {M.}~\bibnamefont {Afzelius}},\ }\href {\doibase
  10.1103/PhysRevLett.114.230502} {\bibfield  {journal} {\bibinfo  {journal}
  {Phys. Rev. Lett.}\ }\textbf {\bibinfo {volume} {114}},\ \bibinfo {pages}
  {230502} (\bibinfo {year} {2015})}\BibitemShut {NoStop}%
\bibitem [{\citenamefont {G{\"u}ndo{\v g}an}\ \emph {et~al.}(2015)\citenamefont
  {G{\"u}ndo{\v g}an}, \citenamefont {Ledingham}, \citenamefont {Kutluer},
  \citenamefont {Mazzera},\ and\ \citenamefont {{de
  Riedmatten}}}]{gundogan_solid_2015}%
  \BibitemOpen
  \bibfield  {author} {\bibinfo {author} {\bibfnamefont {M.}~\bibnamefont
  {G{\"u}ndo{\v g}an}}, \bibinfo {author} {\bibfnamefont {P.~M.}\ \bibnamefont
  {Ledingham}}, \bibinfo {author} {\bibfnamefont {K.}~\bibnamefont {Kutluer}},
  \bibinfo {author} {\bibfnamefont {M.}~\bibnamefont {Mazzera}}, \ and\
  \bibinfo {author} {\bibfnamefont {H.}~\bibnamefont {{de Riedmatten}}},\
  }\href {\doibase 10.1103/PhysRevLett.114.230501} {\bibfield  {journal}
  {\bibinfo  {journal} {Physical Review Letters}\ }\textbf {\bibinfo {volume}
  {114}},\ \bibinfo {pages} {230501} (\bibinfo {year} {2015})}\BibitemShut
  {NoStop}%
\bibitem [{\citenamefont {Zhang}\ \emph {et~al.}(2015)\citenamefont {Zhang},
  \citenamefont {Zou}, \citenamefont {Zhu}, \citenamefont {Marquardt},
  \citenamefont {Jiang},\ and\ \citenamefont {Tang}}]{zhangmagnondark2015}%
  \BibitemOpen
  \bibfield  {author} {\bibinfo {author} {\bibfnamefont {X.}~\bibnamefont
  {Zhang}}, \bibinfo {author} {\bibfnamefont {C.-L.}\ \bibnamefont {Zou}},
  \bibinfo {author} {\bibfnamefont {N.}~\bibnamefont {Zhu}}, \bibinfo {author}
  {\bibfnamefont {F.}~\bibnamefont {Marquardt}}, \bibinfo {author}
  {\bibfnamefont {L.}~\bibnamefont {Jiang}}, \ and\ \bibinfo {author}
  {\bibfnamefont {H.~X.}\ \bibnamefont {Tang}},\ }\href {\doibase
  10.1038/ncomms9914} {\bibfield  {journal} {\bibinfo  {journal} {Nature
  Communications}\ }\textbf {\bibinfo {volume} {6}},\ \bibinfo {pages} {8914}
  (\bibinfo {year} {2015})}\BibitemShut {NoStop}%
\bibitem [{\citenamefont {Lachance-Quirion}\ \emph {et~al.}(2019)\citenamefont
  {Lachance-Quirion}, \citenamefont {Tabuchi}, \citenamefont {Gloppe},
  \citenamefont {Usami},\ and\ \citenamefont
  {Nakamura}}]{lachance-quirion_hybrid_2019}%
  \BibitemOpen
  \bibfield  {author} {\bibinfo {author} {\bibfnamefont {D.}~\bibnamefont
  {Lachance-Quirion}}, \bibinfo {author} {\bibfnamefont {Y.}~\bibnamefont
  {Tabuchi}}, \bibinfo {author} {\bibfnamefont {A.}~\bibnamefont {Gloppe}},
  \bibinfo {author} {\bibfnamefont {K.}~\bibnamefont {Usami}}, \ and\ \bibinfo
  {author} {\bibfnamefont {Y.}~\bibnamefont {Nakamura}},\ }\href {\doibase
  10.7567/1882-0786/ab248d} {\bibfield  {journal} {\bibinfo  {journal} {Applied
  Physics Express}\ }\textbf {\bibinfo {volume} {12}},\ \bibinfo {pages}
  {070101} (\bibinfo {year} {2019})}\BibitemShut {NoStop}%
\bibitem [{\citenamefont {Hisatomi}\ \emph {et~al.}(2016)\citenamefont
  {Hisatomi}, \citenamefont {Osada}, \citenamefont {Tabuchi}, \citenamefont
  {Ishikawa}, \citenamefont {Noguchi}, \citenamefont {Yamazaki}, \citenamefont
  {Usami},\ and\ \citenamefont {Nakamura}}]{hisatomi_bidirectional_2016}%
  \BibitemOpen
  \bibfield  {author} {\bibinfo {author} {\bibfnamefont {R.}~\bibnamefont
  {Hisatomi}}, \bibinfo {author} {\bibfnamefont {A.}~\bibnamefont {Osada}},
  \bibinfo {author} {\bibfnamefont {Y.}~\bibnamefont {Tabuchi}}, \bibinfo
  {author} {\bibfnamefont {T.}~\bibnamefont {Ishikawa}}, \bibinfo {author}
  {\bibfnamefont {A.}~\bibnamefont {Noguchi}}, \bibinfo {author} {\bibfnamefont
  {R.}~\bibnamefont {Yamazaki}}, \bibinfo {author} {\bibfnamefont
  {K.}~\bibnamefont {Usami}}, \ and\ \bibinfo {author} {\bibfnamefont
  {Y.}~\bibnamefont {Nakamura}},\ }\href {\doibase 10.1103/PhysRevB.93.174427}
  {\bibfield  {journal} {\bibinfo  {journal} {Physical Review B}\ }\textbf
  {\bibinfo {volume} {93}},\ \bibinfo {pages} {174427} (\bibinfo {year}
  {2016})}\BibitemShut {NoStop}%
\bibitem [{\citenamefont {Hesler}\ \emph {et~al.}(2019)\citenamefont {Hesler},
  \citenamefont {Prasankumar},\ and\ \citenamefont
  {Tignon}}]{hesleradvancesin2019}%
  \BibitemOpen
  \bibfield  {author} {\bibinfo {author} {\bibfnamefont {J.}~\bibnamefont
  {Hesler}}, \bibinfo {author} {\bibfnamefont {R.}~\bibnamefont {Prasankumar}},
  \ and\ \bibinfo {author} {\bibfnamefont {J.}~\bibnamefont {Tignon}},\ }\href
  {\doibase 10.1063/1.5122975} {\bibfield  {journal} {\bibinfo  {journal}
  {Journal of Applied Physics}\ }\textbf {\bibinfo {volume} {126}},\ \bibinfo
  {pages} {110401} (\bibinfo {year} {2019})}\BibitemShut {NoStop}%
\bibitem [{\citenamefont {Scalari}\ \emph {et~al.}(2012)\citenamefont
  {Scalari}, \citenamefont {Maissen}, \citenamefont {Tur{\v c}inkov{\'a}},
  \citenamefont {Hagenm{\"u}ller}, \citenamefont {De~Liberato}, \citenamefont
  {Ciuti}, \citenamefont {Reichl}, \citenamefont {Schuh}, \citenamefont
  {Wegscheider}, \citenamefont {Beck},\ and\ \citenamefont
  {Faist}}]{scalariultrastrongcoupling2012}%
  \BibitemOpen
  \bibfield  {author} {\bibinfo {author} {\bibfnamefont {G.}~\bibnamefont
  {Scalari}}, \bibinfo {author} {\bibfnamefont {C.}~\bibnamefont {Maissen}},
  \bibinfo {author} {\bibfnamefont {D.}~\bibnamefont {Tur{\v c}inkov{\'a}}},
  \bibinfo {author} {\bibfnamefont {D.}~\bibnamefont {Hagenm{\"u}ller}},
  \bibinfo {author} {\bibfnamefont {S.}~\bibnamefont {De~Liberato}}, \bibinfo
  {author} {\bibfnamefont {C.}~\bibnamefont {Ciuti}}, \bibinfo {author}
  {\bibfnamefont {C.}~\bibnamefont {Reichl}}, \bibinfo {author} {\bibfnamefont
  {D.}~\bibnamefont {Schuh}}, \bibinfo {author} {\bibfnamefont
  {W.}~\bibnamefont {Wegscheider}}, \bibinfo {author} {\bibfnamefont
  {M.}~\bibnamefont {Beck}}, \ and\ \bibinfo {author} {\bibfnamefont
  {J.}~\bibnamefont {Faist}},\ }\href {\doibase 10.1126/science.1216022}
  {\bibfield  {journal} {\bibinfo  {journal} {Science}\ }\textbf {\bibinfo
  {volume} {335}},\ \bibinfo {pages} {1323} (\bibinfo {year}
  {2012})}\BibitemShut {NoStop}%
\bibitem [{\citenamefont {Li}\ \emph {et~al.}(2018)\citenamefont {Li},
  \citenamefont {Bamba}, \citenamefont {Zhang}, \citenamefont {Fallahi},
  \citenamefont {Gardner}, \citenamefont {Gao}, \citenamefont {Lou},
  \citenamefont {Yoshioka}, \citenamefont {Manfra},\ and\ \citenamefont
  {Kono}}]{livacuumbloch2018}%
  \BibitemOpen
  \bibfield  {author} {\bibinfo {author} {\bibfnamefont {X.}~\bibnamefont
  {Li}}, \bibinfo {author} {\bibfnamefont {M.}~\bibnamefont {Bamba}}, \bibinfo
  {author} {\bibfnamefont {Q.}~\bibnamefont {Zhang}}, \bibinfo {author}
  {\bibfnamefont {S.}~\bibnamefont {Fallahi}}, \bibinfo {author} {\bibfnamefont
  {G.~C.}\ \bibnamefont {Gardner}}, \bibinfo {author} {\bibfnamefont
  {W.}~\bibnamefont {Gao}}, \bibinfo {author} {\bibfnamefont {M.}~\bibnamefont
  {Lou}}, \bibinfo {author} {\bibfnamefont {K.}~\bibnamefont {Yoshioka}},
  \bibinfo {author} {\bibfnamefont {M.~J.}\ \bibnamefont {Manfra}}, \ and\
  \bibinfo {author} {\bibfnamefont {J.}~\bibnamefont {Kono}},\ }\href {\doibase
  10.1038/s41566-018-0153-0} {\bibfield  {journal} {\bibinfo  {journal} {Nature
  Photonics}\ }\textbf {\bibinfo {volume} {12}},\ \bibinfo {pages} {324}
  (\bibinfo {year} {2018})}\BibitemShut {NoStop}%
\bibitem [{\citenamefont {Paravicini-Bagliani}\ \emph
  {et~al.}(2019)\citenamefont {Paravicini-Bagliani}, \citenamefont
  {Appugliese}, \citenamefont {Richter}, \citenamefont {Valmorra},
  \citenamefont {Keller}, \citenamefont {Beck}, \citenamefont {Bartolo},
  \citenamefont {R\"{o}ssler}, \citenamefont {Ihn}, \citenamefont {Ensslin},
  \citenamefont {Ciuti}, \citenamefont {Scalari},\ and\ \citenamefont
  {Faist}}]{paravicinimagnetotransport2019}%
  \BibitemOpen
  \bibfield  {author} {\bibinfo {author} {\bibfnamefont {G.~L.}\ \bibnamefont
  {Paravicini-Bagliani}}, \bibinfo {author} {\bibfnamefont {F.}~\bibnamefont
  {Appugliese}}, \bibinfo {author} {\bibfnamefont {E.}~\bibnamefont {Richter}},
  \bibinfo {author} {\bibfnamefont {F.}~\bibnamefont {Valmorra}}, \bibinfo
  {author} {\bibfnamefont {J.}~\bibnamefont {Keller}}, \bibinfo {author}
  {\bibfnamefont {M.}~\bibnamefont {Beck}}, \bibinfo {author} {\bibfnamefont
  {N.}~\bibnamefont {Bartolo}}, \bibinfo {author} {\bibfnamefont
  {C.}~\bibnamefont {R\"{o}ssler}}, \bibinfo {author} {\bibfnamefont
  {T.}~\bibnamefont {Ihn}}, \bibinfo {author} {\bibfnamefont {K.}~\bibnamefont
  {Ensslin}}, \bibinfo {author} {\bibfnamefont {C.}~\bibnamefont {Ciuti}},
  \bibinfo {author} {\bibfnamefont {G.}~\bibnamefont {Scalari}}, \ and\
  \bibinfo {author} {\bibfnamefont {J.}~\bibnamefont {Faist}},\ }\href
  {\doibase 10.1038/s41567-018-0346-y} {\bibfield  {journal} {\bibinfo
  {journal} {Nature Physics}\ }\textbf {\bibinfo {volume} {15}},\ \bibinfo
  {pages} {186} (\bibinfo {year} {2019})}\BibitemShut {NoStop}%
\bibitem [{\citenamefont {Sivarajah}\ \emph {et~al.}(2019)\citenamefont
  {Sivarajah}, \citenamefont {Steinbacher}, \citenamefont {Dastrup},
  \citenamefont {Lu}, \citenamefont {Xiang}, \citenamefont {Ren}, \citenamefont
  {Kamba}, \citenamefont {Cao},\ and\ \citenamefont
  {Nelson}}]{sivarajahthz2019}%
  \BibitemOpen
  \bibfield  {author} {\bibinfo {author} {\bibfnamefont {P.}~\bibnamefont
  {Sivarajah}}, \bibinfo {author} {\bibfnamefont {A.}~\bibnamefont
  {Steinbacher}}, \bibinfo {author} {\bibfnamefont {B.}~\bibnamefont
  {Dastrup}}, \bibinfo {author} {\bibfnamefont {J.}~\bibnamefont {Lu}},
  \bibinfo {author} {\bibfnamefont {M.}~\bibnamefont {Xiang}}, \bibinfo
  {author} {\bibfnamefont {W.}~\bibnamefont {Ren}}, \bibinfo {author}
  {\bibfnamefont {S.}~\bibnamefont {Kamba}}, \bibinfo {author} {\bibfnamefont
  {S.}~\bibnamefont {Cao}}, \ and\ \bibinfo {author} {\bibfnamefont {K.~A.}\
  \bibnamefont {Nelson}},\ }\href {\doibase 10.1063/1.5083849} {\bibfield
  {journal} {\bibinfo  {journal} {Journal of Applied Physics}\ }\textbf
  {\bibinfo {volume} {125}},\ \bibinfo {pages} {213103} (\bibinfo {year}
  {2019})}\BibitemShut {NoStop}%
\bibitem [{\citenamefont {Bia\l{}ek}\ \emph
  {et~al.}(2020{\natexlab{a}})\citenamefont {Bia\l{}ek}, \citenamefont
  {Magrez},\ and\ \citenamefont {Ansermet}}]{bialekspinwave2020}%
  \BibitemOpen
  \bibfield  {author} {\bibinfo {author} {\bibfnamefont {M.}~\bibnamefont
  {Bia\l{}ek}}, \bibinfo {author} {\bibfnamefont {A.}~\bibnamefont {Magrez}}, \
  and\ \bibinfo {author} {\bibfnamefont {J.-P.}\ \bibnamefont {Ansermet}},\
  }\href {\doibase 10.1103/PhysRevB.101.024405} {\bibfield  {journal} {\bibinfo
   {journal} {Phys. Rev. B}\ }\textbf {\bibinfo {volume} {101}},\ \bibinfo
  {pages} {024405} (\bibinfo {year} {2020}{\natexlab{a}})}\BibitemShut
  {NoStop}%
\bibitem [{\citenamefont {Keffer}\ and\ \citenamefont
  {Kittel}(1952)}]{keffer_theory_1952}%
  \BibitemOpen
  \bibfield  {author} {\bibinfo {author} {\bibfnamefont {F.}~\bibnamefont
  {Keffer}}\ and\ \bibinfo {author} {\bibfnamefont {C.}~\bibnamefont
  {Kittel}},\ }\href {\doibase 10.1103/PhysRev.85.329} {\bibfield  {journal}
  {\bibinfo  {journal} {Physical Review}\ }\textbf {\bibinfo {volume} {85}},\
  \bibinfo {pages} {329} (\bibinfo {year} {1952})}\BibitemShut {NoStop}%
\bibitem [{\citenamefont {Gomonay}\ \emph {et~al.}(2018)\citenamefont
  {Gomonay}, \citenamefont {Baltz}, \citenamefont {Brataas},\ and\
  \citenamefont {Tserkovnyak}}]{gomonay_antiferromagnetic_2018}%
  \BibitemOpen
  \bibfield  {author} {\bibinfo {author} {\bibfnamefont {O.}~\bibnamefont
  {Gomonay}}, \bibinfo {author} {\bibfnamefont {V.}~\bibnamefont {Baltz}},
  \bibinfo {author} {\bibfnamefont {A.}~\bibnamefont {Brataas}}, \ and\
  \bibinfo {author} {\bibfnamefont {Y.}~\bibnamefont {Tserkovnyak}},\ }\href
  {\doibase 10.1038/s41567-018-0049-4} {\bibfield  {journal} {\bibinfo
  {journal} {Nature Physics}\ }\textbf {\bibinfo {volume} {14}} (\bibinfo
  {year} {2018}),\ 10.1038/s41567-018-0049-4}\BibitemShut {NoStop}%
\bibitem [{\citenamefont {Jungwirth}\ \emph {et~al.}(2018)\citenamefont
  {Jungwirth}, \citenamefont {Sinova}, \citenamefont {Manchon}, \citenamefont
  {Marti}, \citenamefont {Wunderlich},\ and\ \citenamefont
  {Felser}}]{jungwirth_the_2018}%
  \BibitemOpen
  \bibfield  {author} {\bibinfo {author} {\bibfnamefont {T.}~\bibnamefont
  {Jungwirth}}, \bibinfo {author} {\bibfnamefont {J.}~\bibnamefont {Sinova}},
  \bibinfo {author} {\bibfnamefont {A.}~\bibnamefont {Manchon}}, \bibinfo
  {author} {\bibfnamefont {X.}~\bibnamefont {Marti}}, \bibinfo {author}
  {\bibfnamefont {J.}~\bibnamefont {Wunderlich}}, \ and\ \bibinfo {author}
  {\bibfnamefont {C.}~\bibnamefont {Felser}},\ }\href {\doibase
  10.1038/s41567-018-0063-6} {\bibfield  {journal} {\bibinfo  {journal} {Nature
  Physics}\ }\textbf {\bibinfo {volume} {14}} (\bibinfo {year} {2018}),\
  10.1038/s41567-018-0063-6}\BibitemShut {NoStop}%
\bibitem [{\citenamefont {Jungfleisch}\ \emph {et~al.}(2018)\citenamefont
  {Jungfleisch}, \citenamefont {Zhang},\ and\ \citenamefont
  {Hoffmann}}]{jungfleisch_perspectives_2018}%
  \BibitemOpen
  \bibfield  {author} {\bibinfo {author} {\bibfnamefont {M.~B.}\ \bibnamefont
  {Jungfleisch}}, \bibinfo {author} {\bibfnamefont {W.}~\bibnamefont {Zhang}},
  \ and\ \bibinfo {author} {\bibfnamefont {A.}~\bibnamefont {Hoffmann}},\
  }\href {\doibase https://doi.org/10.1016/j.physleta.2018.01.008} {\bibfield
  {journal} {\bibinfo  {journal} {Physics Letters A}\ }\textbf {\bibinfo
  {volume} {382}},\ \bibinfo {pages} {865 } (\bibinfo {year}
  {2018})}\BibitemShut {NoStop}%
\bibitem [{\citenamefont {Baltz}\ \emph {et~al.}(2018)\citenamefont {Baltz},
  \citenamefont {Manchon}, \citenamefont {Tsoi}, \citenamefont {Moriyama},
  \citenamefont {Ono},\ and\ \citenamefont
  {Tserkovnyak}}]{baltz_antiferromagnetic_2018}%
  \BibitemOpen
  \bibfield  {author} {\bibinfo {author} {\bibfnamefont {V.}~\bibnamefont
  {Baltz}}, \bibinfo {author} {\bibfnamefont {A.}~\bibnamefont {Manchon}},
  \bibinfo {author} {\bibfnamefont {M.}~\bibnamefont {Tsoi}}, \bibinfo {author}
  {\bibfnamefont {T.}~\bibnamefont {Moriyama}}, \bibinfo {author}
  {\bibfnamefont {T.}~\bibnamefont {Ono}}, \ and\ \bibinfo {author}
  {\bibfnamefont {Y.}~\bibnamefont {Tserkovnyak}},\ }\href {\doibase
  10.1103/RevModPhys.90.015005} {\bibfield  {journal} {\bibinfo  {journal}
  {Rev. Mod. Phys.}\ }\textbf {\bibinfo {volume} {90}},\ \bibinfo {pages}
  {015005} (\bibinfo {year} {2018})}\BibitemShut {NoStop}%
\bibitem [{\citenamefont {Mergenthaler}\ \emph {et~al.}(2017)\citenamefont
  {Mergenthaler}, \citenamefont {Liu}, \citenamefont {Le~Roy}, \citenamefont
  {Ares}, \citenamefont {Thompson}, \citenamefont {Bogani}, \citenamefont
  {Luis}, \citenamefont {Blundell}, \citenamefont {Lancaster}, \citenamefont
  {Ardavan}, \citenamefont {Briggs}, \citenamefont {Leek},\ and\ \citenamefont
  {Laird}}]{mergenthaler_strong_2017}%
  \BibitemOpen
  \bibfield  {author} {\bibinfo {author} {\bibfnamefont {M.}~\bibnamefont
  {Mergenthaler}}, \bibinfo {author} {\bibfnamefont {J.}~\bibnamefont {Liu}},
  \bibinfo {author} {\bibfnamefont {J.~J.}\ \bibnamefont {Le~Roy}}, \bibinfo
  {author} {\bibfnamefont {N.}~\bibnamefont {Ares}}, \bibinfo {author}
  {\bibfnamefont {A.~L.}\ \bibnamefont {Thompson}}, \bibinfo {author}
  {\bibfnamefont {L.}~\bibnamefont {Bogani}}, \bibinfo {author} {\bibfnamefont
  {F.}~\bibnamefont {Luis}}, \bibinfo {author} {\bibfnamefont {S.~J.}\
  \bibnamefont {Blundell}}, \bibinfo {author} {\bibfnamefont {T.}~\bibnamefont
  {Lancaster}}, \bibinfo {author} {\bibfnamefont {A.}~\bibnamefont {Ardavan}},
  \bibinfo {author} {\bibfnamefont {G.~A.~D.}\ \bibnamefont {Briggs}}, \bibinfo
  {author} {\bibfnamefont {P.~J.}\ \bibnamefont {Leek}}, \ and\ \bibinfo
  {author} {\bibfnamefont {E.~A.}\ \bibnamefont {Laird}},\ }\href {\doibase
  10.1103/PhysRevLett.119.147701} {\bibfield  {journal} {\bibinfo  {journal}
  {Physical Review Letters}\ }\textbf {\bibinfo {volume} {119}},\ \bibinfo
  {pages} {147701} (\bibinfo {year} {2017})}\BibitemShut {NoStop}%
\bibitem [{\citenamefont {Xiao}\ \emph {et~al.}(2019)\citenamefont {Xiao},
  \citenamefont {Yan}, \citenamefont {Zhang}, \citenamefont {Grigoryan},
  \citenamefont {Hu}, \citenamefont {Guo},\ and\ \citenamefont
  {Xia}}]{xiao_magnon_2019}%
  \BibitemOpen
  \bibfield  {author} {\bibinfo {author} {\bibfnamefont {Y.}~\bibnamefont
  {Xiao}}, \bibinfo {author} {\bibfnamefont {X.~H.}\ \bibnamefont {Yan}},
  \bibinfo {author} {\bibfnamefont {Y.}~\bibnamefont {Zhang}}, \bibinfo
  {author} {\bibfnamefont {V.~L.}\ \bibnamefont {Grigoryan}}, \bibinfo {author}
  {\bibfnamefont {C.~M.}\ \bibnamefont {Hu}}, \bibinfo {author} {\bibfnamefont
  {H.}~\bibnamefont {Guo}}, \ and\ \bibinfo {author} {\bibfnamefont
  {K.}~\bibnamefont {Xia}},\ }\href {\doibase 10.1103/PhysRevB.99.094407}
  {\bibfield  {journal} {\bibinfo  {journal} {Physical Review B}\ }\textbf
  {\bibinfo {volume} {99}},\ \bibinfo {pages} {094407} (\bibinfo {year}
  {2019})}\BibitemShut {NoStop}%
\bibitem [{\citenamefont {Johansen}\ and\ \citenamefont
  {Brataas}(2018)}]{johansen_nonlocal_2018}%
  \BibitemOpen
  \bibfield  {author} {\bibinfo {author} {\bibfnamefont {{\O}.}~\bibnamefont
  {Johansen}}\ and\ \bibinfo {author} {\bibfnamefont {A.}~\bibnamefont
  {Brataas}},\ }\href {\doibase 10.1103/PhysRevLett.121.087204} {\bibfield
  {journal} {\bibinfo  {journal} {Phys. Rev. Lett.}\ }\textbf {\bibinfo
  {volume} {121}},\ \bibinfo {pages} {087204} (\bibinfo {year}
  {2018})}\BibitemShut {NoStop}%
\bibitem [{\citenamefont {Higuchi}\ and\ \citenamefont
  {Kuwata-Gonokami}(2016)}]{higuchicontrolofantiferromagnetic2016}%
  \BibitemOpen
  \bibfield  {author} {\bibinfo {author} {\bibfnamefont {T.}~\bibnamefont
  {Higuchi}}\ and\ \bibinfo {author} {\bibfnamefont {M.}~\bibnamefont
  {Kuwata-Gonokami}},\ }\href {\doibase 10.1038/ncomms10720} {\bibfield
  {journal} {\bibinfo  {journal} {Nature Communications}\ }\textbf {\bibinfo
  {volume} {7}} (\bibinfo {year} {2016}),\ 10.1038/ncomms10720}\BibitemShut
  {NoStop}%
\bibitem [{\citenamefont {N{\v e}mec}\ \emph {et~al.}(2018)\citenamefont {N{\v
  e}mec}, \citenamefont {Fiebig}, \citenamefont {Kampfrath},\ and\
  \citenamefont {Kimel}}]{nemec_antiferromagnetic_2018}%
  \BibitemOpen
  \bibfield  {author} {\bibinfo {author} {\bibfnamefont {P.}~\bibnamefont {N{\v
  e}mec}}, \bibinfo {author} {\bibfnamefont {M.}~\bibnamefont {Fiebig}},
  \bibinfo {author} {\bibfnamefont {T.}~\bibnamefont {Kampfrath}}, \ and\
  \bibinfo {author} {\bibfnamefont {A.~V.}\ \bibnamefont {Kimel}},\ }\href
  {\doibase 10.1038/s41567-018-0051-x} {\bibfield  {journal} {\bibinfo
  {journal} {Nature Physics}\ }\textbf {\bibinfo {volume} {14}},\ \bibinfo
  {pages} {229} (\bibinfo {year} {2018})}\BibitemShut {NoStop}%
\bibitem [{\citenamefont {Haigh}\ \emph {et~al.}(2015)\citenamefont {Haigh},
  \citenamefont {Langenfeld}, \citenamefont {Lambert}, \citenamefont
  {Baumberg}, \citenamefont {Ramsay}, \citenamefont {Nunnenkamp},\ and\
  \citenamefont {Ferguson}}]{haigh_magneto-optical_2015}%
  \BibitemOpen
  \bibfield  {author} {\bibinfo {author} {\bibfnamefont {J.~A.}\ \bibnamefont
  {Haigh}}, \bibinfo {author} {\bibfnamefont {S.}~\bibnamefont {Langenfeld}},
  \bibinfo {author} {\bibfnamefont {N.~J.}\ \bibnamefont {Lambert}}, \bibinfo
  {author} {\bibfnamefont {J.~J.}\ \bibnamefont {Baumberg}}, \bibinfo {author}
  {\bibfnamefont {A.~J.}\ \bibnamefont {Ramsay}}, \bibinfo {author}
  {\bibfnamefont {A.}~\bibnamefont {Nunnenkamp}}, \ and\ \bibinfo {author}
  {\bibfnamefont {A.~J.}\ \bibnamefont {Ferguson}},\ }\href {\doibase
  10.1103/PhysRevA.92.063845} {\bibfield  {journal} {\bibinfo  {journal}
  {Physical Review A}\ }\textbf {\bibinfo {volume} {92}},\ \bibinfo {pages}
  {063845} (\bibinfo {year} {2015})}\BibitemShut {NoStop}%
\bibitem [{\citenamefont {Osada}\ \emph
  {et~al.}(2018{\natexlab{a}})\citenamefont {Osada}, \citenamefont {Gloppe},
  \citenamefont {Hisatomi}, \citenamefont {Noguchi}, \citenamefont {Yamazaki},
  \citenamefont {Nomura}, \citenamefont {Nakamura},\ and\ \citenamefont
  {Usami}}]{osada_brillouin_2018}%
  \BibitemOpen
  \bibfield  {author} {\bibinfo {author} {\bibfnamefont {A.}~\bibnamefont
  {Osada}}, \bibinfo {author} {\bibfnamefont {A.}~\bibnamefont {Gloppe}},
  \bibinfo {author} {\bibfnamefont {R.}~\bibnamefont {Hisatomi}}, \bibinfo
  {author} {\bibfnamefont {A.}~\bibnamefont {Noguchi}}, \bibinfo {author}
  {\bibfnamefont {R.}~\bibnamefont {Yamazaki}}, \bibinfo {author}
  {\bibfnamefont {M.}~\bibnamefont {Nomura}}, \bibinfo {author} {\bibfnamefont
  {Y.}~\bibnamefont {Nakamura}}, \ and\ \bibinfo {author} {\bibfnamefont
  {K.}~\bibnamefont {Usami}},\ }\href {\doibase 10.1103/PhysRevLett.120.133602}
  {\bibfield  {journal} {\bibinfo  {journal} {Physical Review Letters}\
  }\textbf {\bibinfo {volume} {120}},\ \bibinfo {pages} {133602} (\bibinfo
  {year} {2018}{\natexlab{a}})}\BibitemShut {NoStop}%
\bibitem [{\citenamefont {Osada}\ \emph
  {et~al.}(2018{\natexlab{b}})\citenamefont {Osada}, \citenamefont {Gloppe},
  \citenamefont {Nakamura},\ and\ \citenamefont {Usami}}]{osada_orbital_2018}%
  \BibitemOpen
  \bibfield  {author} {\bibinfo {author} {\bibfnamefont {A.}~\bibnamefont
  {Osada}}, \bibinfo {author} {\bibfnamefont {A.}~\bibnamefont {Gloppe}},
  \bibinfo {author} {\bibfnamefont {Y.}~\bibnamefont {Nakamura}}, \ and\
  \bibinfo {author} {\bibfnamefont {K.}~\bibnamefont {Usami}},\ }\href
  {\doibase 10.1088/1367-2630/aae4b1} {\bibfield  {journal} {\bibinfo
  {journal} {New Journal of Physics}\ }\textbf {\bibinfo {volume} {20}},\
  \bibinfo {pages} {103018} (\bibinfo {year} {2018}{\natexlab{b}})}\BibitemShut
  {NoStop}%
\bibitem [{\citenamefont {Liu}\ \emph {et~al.}(2016)\citenamefont {Liu},
  \citenamefont {Zhang}, \citenamefont {Tang},\ and\ \citenamefont
  {Flatt\'e}}]{liu_optomagnonics_2016}%
  \BibitemOpen
  \bibfield  {author} {\bibinfo {author} {\bibfnamefont {T.}~\bibnamefont
  {Liu}}, \bibinfo {author} {\bibfnamefont {X.}~\bibnamefont {Zhang}}, \bibinfo
  {author} {\bibfnamefont {H.~X.}\ \bibnamefont {Tang}}, \ and\ \bibinfo
  {author} {\bibfnamefont {M.~E.}\ \bibnamefont {Flatt\'e}},\ }\href {\doibase
  10.1103/PhysRevB.94.060405} {\bibfield  {journal} {\bibinfo  {journal}
  {Physical Review B}\ }\textbf {\bibinfo {volume} {94}},\ \bibinfo {pages}
  {060405(R)} (\bibinfo {year} {2016})}\BibitemShut {NoStop}%
\bibitem [{\citenamefont {Almpanis}(2018)}]{almpanisdielectric2018}%
  \BibitemOpen
  \bibfield  {author} {\bibinfo {author} {\bibfnamefont {E.}~\bibnamefont
  {Almpanis}},\ }\href {\doibase 10.1103/PhysRevB.97.184406} {\bibfield
  {journal} {\bibinfo  {journal} {Phys. Rev. B}\ }\textbf {\bibinfo {volume}
  {97}},\ \bibinfo {pages} {184406} (\bibinfo {year} {2018})}\BibitemShut
  {NoStop}%
\bibitem [{\citenamefont {Pantazopoulos}\ \emph {et~al.}(2017)\citenamefont
  {Pantazopoulos}, \citenamefont {Stefanou}, \citenamefont {Almpanis},\ and\
  \citenamefont {Papanikolaou}}]{pantazopoulosphotomagnonic2017}%
  \BibitemOpen
  \bibfield  {author} {\bibinfo {author} {\bibfnamefont {P.~A.}\ \bibnamefont
  {Pantazopoulos}}, \bibinfo {author} {\bibfnamefont {N.}~\bibnamefont
  {Stefanou}}, \bibinfo {author} {\bibfnamefont {E.}~\bibnamefont {Almpanis}},
  \ and\ \bibinfo {author} {\bibfnamefont {N.}~\bibnamefont {Papanikolaou}},\
  }\href {\doibase 10.1103/PhysRevB.96.104425} {\bibfield  {journal} {\bibinfo
  {journal} {Phys. Rev. B}\ }\textbf {\bibinfo {volume} {96}},\ \bibinfo
  {pages} {104425} (\bibinfo {year} {2017})}\BibitemShut {NoStop}%
\bibitem [{\citenamefont {Pantazopoulos}\ \emph {et~al.}(2019)\citenamefont
  {Pantazopoulos}, \citenamefont {Tsakmakidis}, \citenamefont {Almpanis},
  \citenamefont {Zouros},\ and\ \citenamefont
  {Stefanou}}]{pantazopouloshighefficiency2019}%
  \BibitemOpen
  \bibfield  {author} {\bibinfo {author} {\bibfnamefont {P.~A.}\ \bibnamefont
  {Pantazopoulos}}, \bibinfo {author} {\bibfnamefont {K.~L.}\ \bibnamefont
  {Tsakmakidis}}, \bibinfo {author} {\bibfnamefont {E.}~\bibnamefont
  {Almpanis}}, \bibinfo {author} {\bibfnamefont {G.~P.}\ \bibnamefont
  {Zouros}}, \ and\ \bibinfo {author} {\bibfnamefont {N.}~\bibnamefont
  {Stefanou}},\ }\href {\doibase 10.1088/1367-2630/ab3ad9} {\bibfield
  {journal} {\bibinfo  {journal} {New Journal of Physics}\ }\textbf {\bibinfo
  {volume} {21}},\ \bibinfo {pages} {095001} (\bibinfo {year}
  {2019})}\BibitemShut {NoStop}%
\bibitem [{\citenamefont {Viola~Kusminskiy}\ \emph {et~al.}(2016)\citenamefont
  {Viola~Kusminskiy}, \citenamefont {Tang},\ and\ \citenamefont
  {Marquardt}}]{viola_kusminskiy_coupled_2016}%
  \BibitemOpen
  \bibfield  {author} {\bibinfo {author} {\bibfnamefont {S.}~\bibnamefont
  {Viola~Kusminskiy}}, \bibinfo {author} {\bibfnamefont {H.~X.}\ \bibnamefont
  {Tang}}, \ and\ \bibinfo {author} {\bibfnamefont {F.}~\bibnamefont
  {Marquardt}},\ }\href {\doibase 10.1103/PhysRevA.94.033821} {\bibfield
  {journal} {\bibinfo  {journal} {Physical Review A}\ }\textbf {\bibinfo
  {volume} {94}},\ \bibinfo {pages} {033821} (\bibinfo {year}
  {2016})}\BibitemShut {NoStop}%
\bibitem [{\citenamefont {Graf}\ \emph {et~al.}(2018)\citenamefont {Graf},
  \citenamefont {Pfeifer}, \citenamefont {Marquardt},\ and\ \citenamefont
  {Viola~Kusminskiy}}]{graf_cavity_2018}%
  \BibitemOpen
  \bibfield  {author} {\bibinfo {author} {\bibfnamefont {J.}~\bibnamefont
  {Graf}}, \bibinfo {author} {\bibfnamefont {H.}~\bibnamefont {Pfeifer}},
  \bibinfo {author} {\bibfnamefont {F.}~\bibnamefont {Marquardt}}, \ and\
  \bibinfo {author} {\bibfnamefont {S.}~\bibnamefont {Viola~Kusminskiy}},\
  }\href {\doibase 10.1103/PhysRevB.98.241406} {\bibfield  {journal} {\bibinfo
  {journal} {Physical Review B}\ }\textbf {\bibinfo {volume} {98}},\ \bibinfo
  {pages} {241406(R)} (\bibinfo {year} {2018})}\BibitemShut {NoStop}%
\bibitem [{\citenamefont {Sharma}\ \emph {et~al.}(2018)\citenamefont {Sharma},
  \citenamefont {Blanter},\ and\ \citenamefont {Bauer}}]{sharma_optical_2018}%
  \BibitemOpen
  \bibfield  {author} {\bibinfo {author} {\bibfnamefont {S.}~\bibnamefont
  {Sharma}}, \bibinfo {author} {\bibfnamefont {Y.~M.}\ \bibnamefont {Blanter}},
  \ and\ \bibinfo {author} {\bibfnamefont {G.~E.~W.}\ \bibnamefont {Bauer}},\
  }\href {\doibase 10.1103/PhysRevLett.121.087205} {\bibfield  {journal}
  {\bibinfo  {journal} {Physical Review Letters}\ }\textbf {\bibinfo {volume}
  {121}},\ \bibinfo {pages} {087205} (\bibinfo {year} {2018})}\BibitemShut
  {NoStop}%
\bibitem [{\citenamefont {Bittencourt}\ \emph {et~al.}(2019)\citenamefont
  {Bittencourt}, \citenamefont {Feulner},\ and\ \citenamefont
  {Kusminskiy}}]{bittencourt_magnon_2019}%
  \BibitemOpen
  \bibfield  {author} {\bibinfo {author} {\bibfnamefont {V.~A. S.~V.}\
  \bibnamefont {Bittencourt}}, \bibinfo {author} {\bibfnamefont
  {V.}~\bibnamefont {Feulner}}, \ and\ \bibinfo {author} {\bibfnamefont
  {S.~V.}\ \bibnamefont {Kusminskiy}},\ }\href {\doibase
  10.1103/PhysRevA.100.013810} {\bibfield  {journal} {\bibinfo  {journal}
  {Physical Review A}\ }\textbf {\bibinfo {volume} {100}},\ \bibinfo {pages}
  {013810} (\bibinfo {year} {2019})}\BibitemShut {NoStop}%
\bibitem [{\citenamefont {Aspelmeyer}\ \emph {et~al.}(2014)\citenamefont
  {Aspelmeyer}, \citenamefont {Kippenberg},\ and\ \citenamefont
  {Marquardt}}]{aspelmeyer_cavity_2014}%
  \BibitemOpen
  \bibfield  {author} {\bibinfo {author} {\bibfnamefont {M.}~\bibnamefont
  {Aspelmeyer}}, \bibinfo {author} {\bibfnamefont {T.~J.}\ \bibnamefont
  {Kippenberg}}, \ and\ \bibinfo {author} {\bibfnamefont {F.}~\bibnamefont
  {Marquardt}},\ }\href {\doibase 10.1103/RevModPhys.86.1391} {\bibfield
  {journal} {\bibinfo  {journal} {Reviews of Modern Physics}\ }\textbf
  {\bibinfo {volume} {86}},\ \bibinfo {pages} {1391} (\bibinfo {year}
  {2014})}\BibitemShut {NoStop}%
\bibitem [{\citenamefont {Powell}\ and\ \citenamefont
  {Spicer}(1970)}]{powell_optical_1970}%
  \BibitemOpen
  \bibfield  {author} {\bibinfo {author} {\bibfnamefont {R.~J.}\ \bibnamefont
  {Powell}}\ and\ \bibinfo {author} {\bibfnamefont {W.~E.}\ \bibnamefont
  {Spicer}},\ }\href {\doibase 10.1103/PhysRevB.2.2182} {\bibfield  {journal}
  {\bibinfo  {journal} {Physical Review B}\ }\textbf {\bibinfo {volume} {2}},\
  \bibinfo {pages} {2182} (\bibinfo {year} {1970})}\BibitemShut {NoStop}%
\bibitem [{\citenamefont {Dodge}(1984)}]{dodge_refractive_1984}%
  \BibitemOpen
  \bibfield  {author} {\bibinfo {author} {\bibfnamefont {M.~J.}\ \bibnamefont
  {Dodge}},\ }\href {\doibase 10.1364/AO.23.001980} {\bibfield  {journal}
  {\bibinfo  {journal} {Applied Optics}\ }\textbf {\bibinfo {volume} {23}},\
  \bibinfo {pages} {1980} (\bibinfo {year} {1984})}\BibitemShut {NoStop}%
\bibitem [{\citenamefont {Jahn}(1973)}]{jahn_linear_1973}%
  \BibitemOpen
  \bibfield  {author} {\bibinfo {author} {\bibfnamefont {I.~R.}\ \bibnamefont
  {Jahn}},\ }\href {\doibase 10.1002/pssb.2220570225} {\bibfield  {journal}
  {\bibinfo  {journal} {Physica Status Solidi (b)}\ }\textbf {\bibinfo {volume}
  {57}},\ \bibinfo {pages} {681} (\bibinfo {year} {1973})}\BibitemShut
  {NoStop}%
\bibitem [{\citenamefont {Kittel}(1963)}]{kittelQuantumTheorySolids1963}%
  \BibitemOpen
  \bibfield  {author} {\bibinfo {author} {\bibfnamefont {C.}~\bibnamefont
  {Kittel}},\ }\href@noop {} {\emph {\bibinfo {title} {Quantum {{Theory}} of
  {{Solids}}}}},\ \bibinfo {edition} {2nd}\ ed.\ (\bibinfo  {publisher} {{John
  Wiley \& Sons}},\ \bibinfo {address} {{New York}},\ \bibinfo {year}
  {1963})\BibitemShut {NoStop}%
\bibitem [{\citenamefont {Holstein}\ and\ \citenamefont
  {Primakoff}(1940)}]{holstein_field_1940}%
  \BibitemOpen
  \bibfield  {author} {\bibinfo {author} {\bibfnamefont {T.}~\bibnamefont
  {Holstein}}\ and\ \bibinfo {author} {\bibfnamefont {H.}~\bibnamefont
  {Primakoff}},\ }\href {\doibase 10.1103/PhysRev.58.1098} {\bibfield
  {journal} {\bibinfo  {journal} {Physical Review}\ }\textbf {\bibinfo {volume}
  {58}},\ \bibinfo {pages} {1098} (\bibinfo {year} {1940})}\BibitemShut
  {NoStop}%
\bibitem [{\citenamefont {Parvini}\ \emph {et~al.}()\citenamefont {Parvini},
  \citenamefont {Bittencourt},\ and\ \citenamefont
  {Viola~Kusminskiy}}]{parviniSupplementalMaterial}%
  \BibitemOpen
  \bibfield  {author} {\bibinfo {author} {\bibfnamefont {T.}~\bibnamefont
  {Parvini}}, \bibinfo {author} {\bibfnamefont {V.~A. S.~V.}\ \bibnamefont
  {Bittencourt}}, \ and\ \bibinfo {author} {\bibfnamefont {S.}~\bibnamefont
  {Viola~Kusminskiy}},\ }\href@noop {} {}\bibinfo {note} {{See Supplemental
  Material for: (I) the details on the diagonalization of the AFM Hamiltonian
  in terms of the Bogoliubov modes following
  Refs.~\cite{kamra_noninteger-spin_2017, johansen_nonlocal_2018}; (II)
  Derivation of the optomagnonic Hamiltonian for antiferromagnetic systems
  starting with the Hamiltonian from Refs.~\cite{cottam_on_1975,
  cottamLightScatteringMagnetic1986, landauElectrodynamicsContinuousMedia1984};
  (III) symmetry considerations for deriving the properties of the optomagnonic
  coupling; (IV) Detailed derivation of the cavity spectra from the
  Heisenberg-Langevin equations \cite{gardinerQuantumNoiseHandbook2000,
  spillaneIdealityFiberTaperCoupledMicroresonator2003,
  caiObservationCriticalCoupling2000, purdyStrongOptomechanicalSqueezing2013,
  safavi-naeiniElectromagneticallyInducedTransparency2011a,
  weis2010optomechanically} and role of cavity-induced magnon-magnon
  interactions on dynamical quantities.}}\BibitemShut {Stop}%
\bibitem [{\citenamefont {Kamra}\ \emph {et~al.}(2017)\citenamefont {Kamra},
  \citenamefont {Agrawal},\ and\ \citenamefont
  {Belzig}}]{kamra_noninteger-spin_2017}%
  \BibitemOpen
  \bibfield  {author} {\bibinfo {author} {\bibfnamefont {A.}~\bibnamefont
  {Kamra}}, \bibinfo {author} {\bibfnamefont {U.}~\bibnamefont {Agrawal}}, \
  and\ \bibinfo {author} {\bibfnamefont {W.}~\bibnamefont {Belzig}},\ }\href
  {\doibase 10.1103/PhysRevB.96.020411} {\bibfield  {journal} {\bibinfo
  {journal} {Physical Review B}\ }\textbf {\bibinfo {volume} {96}},\ \bibinfo
  {pages} {020411(R)} (\bibinfo {year} {2017})}\BibitemShut {NoStop}%
\bibitem [{\citenamefont {Kamra}\ and\ \citenamefont
  {Belzig}(2017)}]{kamra_spin_2017}%
  \BibitemOpen
  \bibfield  {author} {\bibinfo {author} {\bibfnamefont {A.}~\bibnamefont
  {Kamra}}\ and\ \bibinfo {author} {\bibfnamefont {W.}~\bibnamefont {Belzig}},\
  }\href {\doibase 10.1103/PhysRevLett.119.197201} {\bibfield  {journal}
  {\bibinfo  {journal} {Physical Review Letters}\ }\textbf {\bibinfo {volume}
  {119}},\ \bibinfo {pages} {197201} (\bibinfo {year} {2017})}\BibitemShut
  {NoStop}%
\bibitem [{\citenamefont {Machado}\ \emph {et~al.}(2017)\citenamefont
  {Machado}, \citenamefont {Ribeiro}, \citenamefont {Holanda}, \citenamefont
  {Rodr\'{\i}guez-Su\'arez}, \citenamefont {Azevedo},\ and\ \citenamefont
  {Rezende}}]{machado_spin-flop_2017}%
  \BibitemOpen
  \bibfield  {author} {\bibinfo {author} {\bibfnamefont {F.~L.~A.}\
  \bibnamefont {Machado}}, \bibinfo {author} {\bibfnamefont {P.~R.~T.}\
  \bibnamefont {Ribeiro}}, \bibinfo {author} {\bibfnamefont {J.}~\bibnamefont
  {Holanda}}, \bibinfo {author} {\bibfnamefont {R.~L.}\ \bibnamefont
  {Rodr\'{\i}guez-Su\'arez}}, \bibinfo {author} {\bibfnamefont
  {A.}~\bibnamefont {Azevedo}}, \ and\ \bibinfo {author} {\bibfnamefont
  {S.~M.}\ \bibnamefont {Rezende}},\ }\href {\doibase
  10.1103/PhysRevB.95.104418} {\bibfield  {journal} {\bibinfo  {journal} {Phys.
  Rev. B}\ }\textbf {\bibinfo {volume} {95}},\ \bibinfo {pages} {104418}
  (\bibinfo {year} {2017})}\BibitemShut {NoStop}%
\bibitem [{\citenamefont {Cottam}(1975)}]{cottam_on_1975}%
  \BibitemOpen
  \bibfield  {author} {\bibinfo {author} {\bibfnamefont {M.~G.}\ \bibnamefont
  {Cottam}},\ }\href {\doibase 10.1088/0022-3719/8/12/019} {\bibfield
  {journal} {\bibinfo  {journal} {Journal of Physics C: Solid State Physics}\
  }\textbf {\bibinfo {volume} {8}},\ \bibinfo {pages} {1933} (\bibinfo {year}
  {1975})}\BibitemShut {NoStop}%
\bibitem [{\citenamefont {Cottam}\ and\ \citenamefont
  {Lockwood}(1986)}]{cottamLightScatteringMagnetic1986}%
  \BibitemOpen
  \bibfield  {author} {\bibinfo {author} {\bibfnamefont {M.~G.}\ \bibnamefont
  {Cottam}}\ and\ \bibinfo {author} {\bibfnamefont {D.~J.}\ \bibnamefont
  {Lockwood}},\ }\href@noop {} {\emph {\bibinfo {title} {Light {{Scattering}}
  in {{Magnetic Solids}}}}}\ (\bibinfo  {publisher} {{John Wiley \& Sons}},\
  \bibinfo {address} {{New York}},\ \bibinfo {year} {1986})\BibitemShut
  {NoStop}%
\bibitem [{\citenamefont {Lockwood}\ and\ \citenamefont
  {Cottam}(2012)}]{lockwood_magnetooptic_2012}%
  \BibitemOpen
  \bibfield  {author} {\bibinfo {author} {\bibfnamefont {D.~J.}\ \bibnamefont
  {Lockwood}}\ and\ \bibinfo {author} {\bibfnamefont {M.~G.}\ \bibnamefont
  {Cottam}},\ }\href {\doibase 10.1063/1.4733682} {\bibfield  {journal}
  {\bibinfo  {journal} {Low Temperature Physics}\ }\textbf {\bibinfo {volume}
  {38}},\ \bibinfo {pages} {549} (\bibinfo {year} {2012})}\BibitemShut
  {NoStop}%
\bibitem [{\citenamefont {Ariai}\ \emph {et~al.}(1982)\citenamefont {Ariai},
  \citenamefont {Bates}, \citenamefont {Cottam},\ and\ \citenamefont
  {Smith}}]{ariai_effects_1982}%
  \BibitemOpen
  \bibfield  {author} {\bibinfo {author} {\bibfnamefont {J.}~\bibnamefont
  {Ariai}}, \bibinfo {author} {\bibfnamefont {P.~A.}\ \bibnamefont {Bates}},
  \bibinfo {author} {\bibfnamefont {M.~G.}\ \bibnamefont {Cottam}}, \ and\
  \bibinfo {author} {\bibfnamefont {S.~R.~P.}\ \bibnamefont {Smith}},\ }\href
  {\doibase 10.1088/0022-3719/15/12/023} {\bibfield  {journal} {\bibinfo
  {journal} {Journal of Physics C: Solid State Physics}\ }\textbf {\bibinfo
  {volume} {15}},\ \bibinfo {pages} {2767} (\bibinfo {year}
  {1982})}\BibitemShut {NoStop}%
\bibitem [{\citenamefont {Bi}\ \emph {et~al.}(2008)\citenamefont {Bi},
  \citenamefont {Taussig}, \citenamefont {Kim}, \citenamefont {Wang},
  \citenamefont {Dionne}, \citenamefont {Bono}, \citenamefont {Persson},
  \citenamefont {Ceder},\ and\ \citenamefont {Ross}}]{bi_structural_2008}%
  \BibitemOpen
  \bibfield  {author} {\bibinfo {author} {\bibfnamefont {L.}~\bibnamefont
  {Bi}}, \bibinfo {author} {\bibfnamefont {A.~R.}\ \bibnamefont {Taussig}},
  \bibinfo {author} {\bibfnamefont {H.-S.}\ \bibnamefont {Kim}}, \bibinfo
  {author} {\bibfnamefont {L.}~\bibnamefont {Wang}}, \bibinfo {author}
  {\bibfnamefont {G.~F.}\ \bibnamefont {Dionne}}, \bibinfo {author}
  {\bibfnamefont {D.}~\bibnamefont {Bono}}, \bibinfo {author} {\bibfnamefont
  {K.}~\bibnamefont {Persson}}, \bibinfo {author} {\bibfnamefont
  {G.}~\bibnamefont {Ceder}}, \ and\ \bibinfo {author} {\bibfnamefont {C.~A.}\
  \bibnamefont {Ross}},\ }\href {\doibase 10.1103/PhysRevB.78.104106}
  {\bibfield  {journal} {\bibinfo  {journal} {Phys. Rev. B}\ }\textbf {\bibinfo
  {volume} {78}},\ \bibinfo {pages} {104106} (\bibinfo {year}
  {2008})}\BibitemShut {NoStop}%
\bibitem [{\citenamefont {Sharma}\ \emph {et~al.}(2019)\citenamefont {Sharma},
  \citenamefont {Rameshti}, \citenamefont {Blanter},\ and\ \citenamefont
  {Bauer}}]{sharma_2019}%
  \BibitemOpen
  \bibfield  {author} {\bibinfo {author} {\bibfnamefont {S.}~\bibnamefont
  {Sharma}}, \bibinfo {author} {\bibfnamefont {B.~Z.}\ \bibnamefont
  {Rameshti}}, \bibinfo {author} {\bibfnamefont {Y.~M.}\ \bibnamefont
  {Blanter}}, \ and\ \bibinfo {author} {\bibfnamefont {G.~E.~W.}\ \bibnamefont
  {Bauer}},\ }\href {\doibase 10.1103/PhysRevB.99.214423} {\bibfield  {journal}
  {\bibinfo  {journal} {Physical Review B}\ }\textbf {\bibinfo {volume} {99}},\
  \bibinfo {pages} {214423} (\bibinfo {year} {2019})}\BibitemShut {NoStop}%
\bibitem [{\citenamefont {Brahms}\ and\ \citenamefont
  {Stamper-Kurn}(2010)}]{brahms_spin_2010}%
  \BibitemOpen
  \bibfield  {author} {\bibinfo {author} {\bibfnamefont {N.}~\bibnamefont
  {Brahms}}\ and\ \bibinfo {author} {\bibfnamefont {D.~M.}\ \bibnamefont
  {Stamper-Kurn}},\ }\href {\doibase 10.1103/PhysRevA.82.041804} {\bibfield
  {journal} {\bibinfo  {journal} {Physical Review A}\ }\textbf {\bibinfo
  {volume} {82}},\ \bibinfo {pages} {041804(R)} (\bibinfo {year}
  {2010})}\BibitemShut {NoStop}%
\bibitem [{\citenamefont {Kohler}\ \emph {et~al.}(2017)\citenamefont {Kohler},
  \citenamefont {Spethmann}, \citenamefont {Schreppler},\ and\ \citenamefont
  {Stamper-Kurn}}]{kohler_cavity-assisted_2017}%
  \BibitemOpen
  \bibfield  {author} {\bibinfo {author} {\bibfnamefont {J.}~\bibnamefont
  {Kohler}}, \bibinfo {author} {\bibfnamefont {N.}~\bibnamefont {Spethmann}},
  \bibinfo {author} {\bibfnamefont {S.}~\bibnamefont {Schreppler}}, \ and\
  \bibinfo {author} {\bibfnamefont {D.~M.}\ \bibnamefont {Stamper-Kurn}},\
  }\href {\doibase 10.1103/PhysRevLett.118.063604} {\bibfield  {journal}
  {\bibinfo  {journal} {Physical Review Letters}\ }\textbf {\bibinfo {volume}
  {118}},\ \bibinfo {pages} {063604} (\bibinfo {year} {2017})}\BibitemShut
  {NoStop}%
\bibitem [{\citenamefont {Landini}\ \emph {et~al.}(2018)\citenamefont
  {Landini}, \citenamefont {Dogra}, \citenamefont {Kroeger}, \citenamefont
  {Hruby}, \citenamefont {Donner},\ and\ \citenamefont
  {Esslinger}}]{landini_formation_2018}%
  \BibitemOpen
  \bibfield  {author} {\bibinfo {author} {\bibfnamefont {M.}~\bibnamefont
  {Landini}}, \bibinfo {author} {\bibfnamefont {N.}~\bibnamefont {Dogra}},
  \bibinfo {author} {\bibfnamefont {K.}~\bibnamefont {Kroeger}}, \bibinfo
  {author} {\bibfnamefont {L.}~\bibnamefont {Hruby}}, \bibinfo {author}
  {\bibfnamefont {T.}~\bibnamefont {Donner}}, \ and\ \bibinfo {author}
  {\bibfnamefont {T.}~\bibnamefont {Esslinger}},\ }\href {\doibase
  10.1103/PhysRevLett.120.223602} {\bibfield  {journal} {\bibinfo  {journal}
  {Physical Review Letters}\ }\textbf {\bibinfo {volume} {120}},\ \bibinfo
  {pages} {223602} (\bibinfo {year} {2018})}\BibitemShut {NoStop}%
\bibitem [{\citenamefont {Kroeze}\ \emph {et~al.}(2018)\citenamefont {Kroeze},
  \citenamefont {Guo}, \citenamefont {Vaidya}, \citenamefont {Keeling},\ and\
  \citenamefont {Lev}}]{kroeze_spinor_2018}%
  \BibitemOpen
  \bibfield  {author} {\bibinfo {author} {\bibfnamefont {R.~M.}\ \bibnamefont
  {Kroeze}}, \bibinfo {author} {\bibfnamefont {Y.}~\bibnamefont {Guo}},
  \bibinfo {author} {\bibfnamefont {V.~D.}\ \bibnamefont {Vaidya}}, \bibinfo
  {author} {\bibfnamefont {J.}~\bibnamefont {Keeling}}, \ and\ \bibinfo
  {author} {\bibfnamefont {B.~L.}\ \bibnamefont {Lev}},\ }\href {\doibase
  10.1103/PhysRevLett.121.163601} {\bibfield  {journal} {\bibinfo  {journal}
  {Physical Review Letters}\ }\textbf {\bibinfo {volume} {121}},\ \bibinfo
  {pages} {163601} (\bibinfo {year} {2018})}\BibitemShut {NoStop}%
\bibitem [{\citenamefont {Mivehvar}\ \emph {et~al.}(2019)\citenamefont
  {Mivehvar}, \citenamefont {Ritsch},\ and\ \citenamefont
  {Piazza}}]{mivehvar_cavity-quantum-electrodynamical_2019}%
  \BibitemOpen
  \bibfield  {author} {\bibinfo {author} {\bibfnamefont {F.}~\bibnamefont
  {Mivehvar}}, \bibinfo {author} {\bibfnamefont {H.}~\bibnamefont {Ritsch}}, \
  and\ \bibinfo {author} {\bibfnamefont {F.}~\bibnamefont {Piazza}},\ }\href
  {\doibase 10.1103/PhysRevLett.122.113603} {\bibfield  {journal} {\bibinfo
  {journal} {Physical Review Letters}\ }\textbf {\bibinfo {volume} {122}},\
  \bibinfo {pages} {113603} (\bibinfo {year} {2019})}\BibitemShut {NoStop}%
\bibitem [{\citenamefont {Barak}\ \emph {et~al.}(1978)\citenamefont {Barak},
  \citenamefont {Jaccarino},\ and\ \citenamefont
  {Rezende}}]{barak_magnetic_nodate}%
  \BibitemOpen
  \bibfield  {author} {\bibinfo {author} {\bibfnamefont {J.}~\bibnamefont
  {Barak}}, \bibinfo {author} {\bibfnamefont {V.}~\bibnamefont {Jaccarino}}, \
  and\ \bibinfo {author} {\bibfnamefont {S.}~\bibnamefont {Rezende}},\ }\href
  {\doibase 10.1016/0304-8853(78)90087-2} {\bibfield  {journal} {\bibinfo
  {journal} {Journal of Magnetism and Magnetic Materials}\ }\textbf {\bibinfo
  {volume} {9}},\ \bibinfo {pages} {323 } (\bibinfo {year} {1978})}\BibitemShut
  {NoStop}%
\bibitem [{\citenamefont {Ohlmann}\ and\ \citenamefont
  {Tinkham}(1961)}]{ohlmann_antiferromagnetic_1961}%
  \BibitemOpen
  \bibfield  {author} {\bibinfo {author} {\bibfnamefont {R.~C.}\ \bibnamefont
  {Ohlmann}}\ and\ \bibinfo {author} {\bibfnamefont {M.}~\bibnamefont
  {Tinkham}},\ }\href {\doibase 10.1103/PhysRev.123.425} {\bibfield  {journal}
  {\bibinfo  {journal} {Physical Review}\ }\textbf {\bibinfo {volume} {123}},\
  \bibinfo {pages} {425} (\bibinfo {year} {1961})}\BibitemShut {NoStop}%
\bibitem [{\citenamefont {Satoh}\ \emph {et~al.}(2010)\citenamefont {Satoh},
  \citenamefont {Cho}, \citenamefont {Iida}, \citenamefont {Shimura},
  \citenamefont {Kuroda}, \citenamefont {Ueda}, \citenamefont {Ueda},
  \citenamefont {Ivanov}, \citenamefont {Nori},\ and\ \citenamefont
  {Fiebig}}]{satoh_spin_2010}%
  \BibitemOpen
  \bibfield  {author} {\bibinfo {author} {\bibfnamefont {T.}~\bibnamefont
  {Satoh}}, \bibinfo {author} {\bibfnamefont {S.-J.}\ \bibnamefont {Cho}},
  \bibinfo {author} {\bibfnamefont {R.}~\bibnamefont {Iida}}, \bibinfo {author}
  {\bibfnamefont {T.}~\bibnamefont {Shimura}}, \bibinfo {author} {\bibfnamefont
  {K.}~\bibnamefont {Kuroda}}, \bibinfo {author} {\bibfnamefont
  {H.}~\bibnamefont {Ueda}}, \bibinfo {author} {\bibfnamefont {Y.}~\bibnamefont
  {Ueda}}, \bibinfo {author} {\bibfnamefont {B.~A.}\ \bibnamefont {Ivanov}},
  \bibinfo {author} {\bibfnamefont {F.}~\bibnamefont {Nori}}, \ and\ \bibinfo
  {author} {\bibfnamefont {M.}~\bibnamefont {Fiebig}},\ }\href {\doibase
  10.1103/PhysRevLett.105.077402} {\bibfield  {journal} {\bibinfo  {journal}
  {Physical Review Letters}\ }\textbf {\bibinfo {volume} {105}},\ \bibinfo
  {pages} {077402} (\bibinfo {year} {2010})}\BibitemShut {NoStop}%
\bibitem [{\citenamefont {Kampfrath}\ \emph {et~al.}(2011)\citenamefont
  {Kampfrath}, \citenamefont {Sell}, \citenamefont {Klatt}, \citenamefont
  {Pashkin}, \citenamefont {M{\"a}hrlein}, \citenamefont {Dekorsy},
  \citenamefont {Wolf}, \citenamefont {Fiebig}, \citenamefont {Leitenstorfer},\
  and\ \citenamefont {Huber}}]{kampfrath_coherent_2011}%
  \BibitemOpen
  \bibfield  {author} {\bibinfo {author} {\bibfnamefont {T.}~\bibnamefont
  {Kampfrath}}, \bibinfo {author} {\bibfnamefont {A.}~\bibnamefont {Sell}},
  \bibinfo {author} {\bibfnamefont {G.}~\bibnamefont {Klatt}}, \bibinfo
  {author} {\bibfnamefont {A.}~\bibnamefont {Pashkin}}, \bibinfo {author}
  {\bibfnamefont {S.}~\bibnamefont {M{\"a}hrlein}}, \bibinfo {author}
  {\bibfnamefont {T.}~\bibnamefont {Dekorsy}}, \bibinfo {author} {\bibfnamefont
  {M.}~\bibnamefont {Wolf}}, \bibinfo {author} {\bibfnamefont {M.}~\bibnamefont
  {Fiebig}}, \bibinfo {author} {\bibfnamefont {A.}~\bibnamefont
  {Leitenstorfer}}, \ and\ \bibinfo {author} {\bibfnamefont {R.}~\bibnamefont
  {Huber}},\ }\href {\doibase 10.1038/nphoton.2010.259} {\bibfield  {journal}
  {\bibinfo  {journal} {Nature Photonics}\ }\textbf {\bibinfo {volume} {5}},\
  \bibinfo {pages} {31} (\bibinfo {year} {2011})}\BibitemShut {NoStop}%
\bibitem [{\citenamefont {Zhou}\ \emph {et~al.}(2018)\citenamefont {Zhou},
  \citenamefont {Gao},\ and\ \citenamefont {Fu}}]{zhou_giant_2018}%
  \BibitemOpen
  \bibfield  {author} {\bibinfo {author} {\bibfnamefont {S.}~\bibnamefont
  {Zhou}}, \bibinfo {author} {\bibfnamefont {Y.}~\bibnamefont {Gao}}, \ and\
  \bibinfo {author} {\bibfnamefont {S.}~\bibnamefont {Fu}},\ }\href {\doibase
  10.1140/epjb/e2017-80263-8} {\bibfield  {journal} {\bibinfo  {journal} {The
  European Physical Journal B}\ }\textbf {\bibinfo {volume} {91}},\ \bibinfo
  {pages} {41} (\bibinfo {year} {2018})}\BibitemShut {NoStop}%
\bibitem [{\citenamefont {Weis}\ \emph {et~al.}(2010)\citenamefont {Weis},
  \citenamefont {Rivi{\`e}re}, \citenamefont {Del{\'e}glise}, \citenamefont
  {Gavartin}, \citenamefont {Arcizet}, \citenamefont {Schliesser},\ and\
  \citenamefont {Kippenberg}}]{weis2010optomechanically}%
  \BibitemOpen
  \bibfield  {author} {\bibinfo {author} {\bibfnamefont {S.}~\bibnamefont
  {Weis}}, \bibinfo {author} {\bibfnamefont {R.}~\bibnamefont {Rivi{\`e}re}},
  \bibinfo {author} {\bibfnamefont {S.}~\bibnamefont {Del{\'e}glise}}, \bibinfo
  {author} {\bibfnamefont {E.}~\bibnamefont {Gavartin}}, \bibinfo {author}
  {\bibfnamefont {O.}~\bibnamefont {Arcizet}}, \bibinfo {author} {\bibfnamefont
  {A.}~\bibnamefont {Schliesser}}, \ and\ \bibinfo {author} {\bibfnamefont
  {T.~J.}\ \bibnamefont {Kippenberg}},\ }\href {\doibase
  10.1126/science.1195596} {\bibfield  {journal} {\bibinfo  {journal}
  {Science}\ }\textbf {\bibinfo {volume} {330}},\ \bibinfo {pages} {1520}
  (\bibinfo {year} {2010})}\BibitemShut {NoStop}%
\bibitem [{\citenamefont {Xiong}\ and\ \citenamefont
  {Wu}(2018)}]{xiong_fundamentals_2018}%
  \BibitemOpen
  \bibfield  {author} {\bibinfo {author} {\bibfnamefont {H.}~\bibnamefont
  {Xiong}}\ and\ \bibinfo {author} {\bibfnamefont {Y.}~\bibnamefont {Wu}},\
  }\href {\doibase 10.1063/1.5027122} {\bibfield  {journal} {\bibinfo
  {journal} {Applied Physics Reviews}\ }\textbf {\bibinfo {volume} {5}},\
  \bibinfo {pages} {031305} (\bibinfo {year} {2018})}\BibitemShut {NoStop}%
\bibitem [{\citenamefont {Genes}\ \emph {et~al.}(2008)\citenamefont {Genes},
  \citenamefont {Vitali},\ and\ \citenamefont
  {Tombesi}}]{genessimultaneous2008}%
  \BibitemOpen
  \bibfield  {author} {\bibinfo {author} {\bibfnamefont {C.}~\bibnamefont
  {Genes}}, \bibinfo {author} {\bibfnamefont {D.}~\bibnamefont {Vitali}}, \
  and\ \bibinfo {author} {\bibfnamefont {P.}~\bibnamefont {Tombesi}},\ }\href
  {\doibase 10.1088/1367-2630/10/9/095009} {\bibfield  {journal} {\bibinfo
  {journal} {New Journal of Physics}\ }\textbf {\bibinfo {volume} {10}},\
  \bibinfo {pages} {095009} (\bibinfo {year} {2008})}\BibitemShut {NoStop}%
\bibitem [{\citenamefont {Ockeloen-Korppi}\ \emph {et~al.}(2019)\citenamefont
  {Ockeloen-Korppi}, \citenamefont {Gely}, \citenamefont {Damsk\"agg},
  \citenamefont {Jenkins}, \citenamefont {Steele},\ and\ \citenamefont
  {Sillanp\"a\"a}}]{ockeloensidebandcooling2019}%
  \BibitemOpen
  \bibfield  {author} {\bibinfo {author} {\bibfnamefont {C.~F.}\ \bibnamefont
  {Ockeloen-Korppi}}, \bibinfo {author} {\bibfnamefont {M.~F.}\ \bibnamefont
  {Gely}}, \bibinfo {author} {\bibfnamefont {E.}~\bibnamefont {Damsk\"agg}},
  \bibinfo {author} {\bibfnamefont {M.}~\bibnamefont {Jenkins}}, \bibinfo
  {author} {\bibfnamefont {G.~A.}\ \bibnamefont {Steele}}, \ and\ \bibinfo
  {author} {\bibfnamefont {M.~A.}\ \bibnamefont {Sillanp\"a\"a}},\ }\href
  {\doibase 10.1103/PhysRevA.99.023826} {\bibfield  {journal} {\bibinfo
  {journal} {Phys. Rev. A}\ }\textbf {\bibinfo {volume} {99}},\ \bibinfo
  {pages} {023826} (\bibinfo {year} {2019})}\BibitemShut {NoStop}%
\bibitem [{\citenamefont {Sommer}\ and\ \citenamefont
  {Genes}(2019)}]{sommerpartialoptomechanical2019}%
  \BibitemOpen
  \bibfield  {author} {\bibinfo {author} {\bibfnamefont {C.}~\bibnamefont
  {Sommer}}\ and\ \bibinfo {author} {\bibfnamefont {C.}~\bibnamefont {Genes}},\
  }\href {\doibase 10.1103/PhysRevLett.123.203605} {\bibfield  {journal}
  {\bibinfo  {journal} {Phys. Rev. Lett.}\ }\textbf {\bibinfo {volume} {123}},\
  \bibinfo {pages} {203605} (\bibinfo {year} {2019})}\BibitemShut {NoStop}%
\bibitem [{\citenamefont {McGee}\ \emph {et~al.}(2013)\citenamefont {McGee},
  \citenamefont {Meiser}, \citenamefont {Regal}, \citenamefont {Lehnert},\ and\
  \citenamefont {Holland}}]{mcgeemechanicalresonators2013}%
  \BibitemOpen
  \bibfield  {author} {\bibinfo {author} {\bibfnamefont {S.~A.}\ \bibnamefont
  {McGee}}, \bibinfo {author} {\bibfnamefont {D.}~\bibnamefont {Meiser}},
  \bibinfo {author} {\bibfnamefont {C.~A.}\ \bibnamefont {Regal}}, \bibinfo
  {author} {\bibfnamefont {K.~W.}\ \bibnamefont {Lehnert}}, \ and\ \bibinfo
  {author} {\bibfnamefont {M.~J.}\ \bibnamefont {Holland}},\ }\href {\doibase
  10.1103/PhysRevA.87.053818} {\bibfield  {journal} {\bibinfo  {journal} {Phys.
  Rev. A}\ }\textbf {\bibinfo {volume} {87}},\ \bibinfo {pages} {053818}
  (\bibinfo {year} {2013})}\BibitemShut {NoStop}%
\bibitem [{\citenamefont {Lukin}\ \emph {et~al.}(2000)\citenamefont {Lukin},
  \citenamefont {Yelin},\ and\ \citenamefont
  {Fleischhauer}}]{lukinentanglementof2000}%
  \BibitemOpen
  \bibfield  {author} {\bibinfo {author} {\bibfnamefont {M.~D.}\ \bibnamefont
  {Lukin}}, \bibinfo {author} {\bibfnamefont {S.~F.}\ \bibnamefont {Yelin}}, \
  and\ \bibinfo {author} {\bibfnamefont {M.}~\bibnamefont {Fleischhauer}},\
  }\href {\doibase 10.1103/PhysRevLett.84.4232} {\bibfield  {journal} {\bibinfo
   {journal} {Phys. Rev. Lett.}\ }\textbf {\bibinfo {volume} {84}},\ \bibinfo
  {pages} {4232} (\bibinfo {year} {2000})}\BibitemShut {NoStop}%
\bibitem [{\citenamefont {Fleischhauer}\ and\ \citenamefont
  {Lukin}(2000)}]{fleischhauerdarkstate2000}%
  \BibitemOpen
  \bibfield  {author} {\bibinfo {author} {\bibfnamefont {M.}~\bibnamefont
  {Fleischhauer}}\ and\ \bibinfo {author} {\bibfnamefont {M.~D.}\ \bibnamefont
  {Lukin}},\ }\href {\doibase 10.1103/PhysRevLett.84.5094} {\bibfield
  {journal} {\bibinfo  {journal} {Phys. Rev. Lett.}\ }\textbf {\bibinfo
  {volume} {84}},\ \bibinfo {pages} {5094} (\bibinfo {year}
  {2000})}\BibitemShut {NoStop}%
\bibitem [{\citenamefont {Fleischhauer}\ and\ \citenamefont
  {Lukin}(2002)}]{fleischhauerquantummemory2002}%
  \BibitemOpen
  \bibfield  {author} {\bibinfo {author} {\bibfnamefont {M.}~\bibnamefont
  {Fleischhauer}}\ and\ \bibinfo {author} {\bibfnamefont {M.~D.}\ \bibnamefont
  {Lukin}},\ }\href {\doibase 10.1103/PhysRevA.65.022314} {\bibfield  {journal}
  {\bibinfo  {journal} {Phys. Rev. A}\ }\textbf {\bibinfo {volume} {65}},\
  \bibinfo {pages} {022314} (\bibinfo {year} {2002})}\BibitemShut {NoStop}%
\bibitem [{\citenamefont {Gorshkov}\ \emph {et~al.}(2007)\citenamefont
  {Gorshkov}, \citenamefont {Andr\'e}, \citenamefont {Lukin},\ and\
  \citenamefont {S\o{}rensen}}]{gorshkovphotonstorage2007}%
  \BibitemOpen
  \bibfield  {author} {\bibinfo {author} {\bibfnamefont {A.~V.}\ \bibnamefont
  {Gorshkov}}, \bibinfo {author} {\bibfnamefont {A.}~\bibnamefont {Andr\'e}},
  \bibinfo {author} {\bibfnamefont {M.~D.}\ \bibnamefont {Lukin}}, \ and\
  \bibinfo {author} {\bibfnamefont {A.~S.}\ \bibnamefont {S\o{}rensen}},\
  }\href {\doibase 10.1103/PhysRevA.76.033804} {\bibfield  {journal} {\bibinfo
  {journal} {Phys. Rev. A}\ }\textbf {\bibinfo {volume} {76}},\ \bibinfo
  {pages} {033804} (\bibinfo {year} {2007})}\BibitemShut {NoStop}%
\bibitem [{\citenamefont {Ma}\ \emph {et~al.}(2017)\citenamefont {Ma},
  \citenamefont {Slattery},\ and\ \citenamefont {Tang}}]{maopticalquantum2017}%
  \BibitemOpen
  \bibfield  {author} {\bibinfo {author} {\bibfnamefont {L.}~\bibnamefont
  {Ma}}, \bibinfo {author} {\bibfnamefont {O.}~\bibnamefont {Slattery}}, \ and\
  \bibinfo {author} {\bibfnamefont {X.}~\bibnamefont {Tang}},\ }\href {\doibase
  10.1088/2040-8986/19/4/043001} {\bibfield  {journal} {\bibinfo  {journal}
  {Journal of Optics}\ }\textbf {\bibinfo {volume} {19}},\ \bibinfo {pages}
  {043001} (\bibinfo {year} {2017})}\BibitemShut {NoStop}%
\bibitem [{\citenamefont {Kondoh}\ and\ \citenamefont
  {Takeda}(1964)}]{kondoh1964observation}%
  \BibitemOpen
  \bibfield  {author} {\bibinfo {author} {\bibfnamefont {H.}~\bibnamefont
  {Kondoh}}\ and\ \bibinfo {author} {\bibfnamefont {T.}~\bibnamefont
  {Takeda}},\ }\href {\doibase 10.1143/JPSJ.19.2041} {\bibfield  {journal}
  {\bibinfo  {journal} {Journal of the Physical Society of Japan}\ }\textbf
  {\bibinfo {volume} {19}},\ \bibinfo {pages} {2041} (\bibinfo {year}
  {1964})}\BibitemShut {NoStop}%
\bibitem [{\citenamefont {Baruchel}\ \emph {et~al.}(1981)\citenamefont
  {Baruchel}, \citenamefont {Schlenker}, \citenamefont {Kurosawa},\ and\
  \citenamefont {Saito}}]{baruchel_antiferromagnetic_1981}%
  \BibitemOpen
  \bibfield  {author} {\bibinfo {author} {\bibfnamefont {J.}~\bibnamefont
  {Baruchel}}, \bibinfo {author} {\bibfnamefont {M.}~\bibnamefont {Schlenker}},
  \bibinfo {author} {\bibfnamefont {K.}~\bibnamefont {Kurosawa}}, \ and\
  \bibinfo {author} {\bibfnamefont {S.}~\bibnamefont {Saito}},\ }\href
  {\doibase 10.1080/01418638108222351} {\bibfield  {journal} {\bibinfo
  {journal} {Philosophical Magazine B}\ }\textbf {\bibinfo {volume} {43}},\
  \bibinfo {pages} {853} (\bibinfo {year} {1981})}\BibitemShut {NoStop}%
\bibitem [{\citenamefont {Bia\l{}ek}\ \emph
  {et~al.}(2020{\natexlab{b}})\citenamefont {Bia\l{}ek}, \citenamefont
  {Magrez},\ and\ \citenamefont {Ansermet}}]{bialekspinwavecoupling2020}%
  \BibitemOpen
  \bibfield  {author} {\bibinfo {author} {\bibfnamefont {M.}~\bibnamefont
  {Bia\l{}ek}}, \bibinfo {author} {\bibfnamefont {A.}~\bibnamefont {Magrez}}, \
  and\ \bibinfo {author} {\bibfnamefont {J.-P.}\ \bibnamefont {Ansermet}},\
  }\href {\doibase 10.1103/PhysRevB.101.024405} {\bibfield  {journal} {\bibinfo
   {journal} {Phys. Rev. B}\ }\textbf {\bibinfo {volume} {101}},\ \bibinfo
  {pages} {024405} (\bibinfo {year} {2020}{\natexlab{b}})}\BibitemShut
  {NoStop}%
\bibitem [{\citenamefont {Landau}\ and\ \citenamefont
  {Lifshitz}(1984)}]{landauElectrodynamicsContinuousMedia1984}%
  \BibitemOpen
  \bibfield  {author} {\bibinfo {author} {\bibfnamefont {L.~D.}\ \bibnamefont
  {Landau}}\ and\ \bibinfo {author} {\bibfnamefont {E.~M.}\ \bibnamefont
  {Lifshitz}},\ }\href@noop {} {\emph {\bibinfo {title} {Electrodynamics of
  {{Continuous Media}}}}},\ \bibinfo {edition} {2nd}\ ed.\ (\bibinfo
  {publisher} {{Pergamon}},\ \bibinfo {address} {Oxford, UK},\ \bibinfo {year}
  {1984})\BibitemShut {NoStop}%
\bibitem [{\citenamefont {Cai}\ \emph {et~al.}(2000)\citenamefont {Cai},
  \citenamefont {Painter},\ and\ \citenamefont
  {Vahala}}]{caiObservationCriticalCoupling2000}%
  \BibitemOpen
  \bibfield  {author} {\bibinfo {author} {\bibfnamefont {M.}~\bibnamefont
  {Cai}}, \bibinfo {author} {\bibfnamefont {O.}~\bibnamefont {Painter}}, \ and\
  \bibinfo {author} {\bibfnamefont {K.~J.}\ \bibnamefont {Vahala}},\ }\href
  {\doibase 10.1103/PhysRevLett.85.74} {\bibfield  {journal} {\bibinfo
  {journal} {Phys. Rev. Lett.}\ }\textbf {\bibinfo {volume} {85}},\ \bibinfo
  {pages} {74} (\bibinfo {year} {2000})}\BibitemShut {NoStop}%
\bibitem [{\citenamefont {Spillane}\ \emph {et~al.}(2003)\citenamefont
  {Spillane}, \citenamefont {Kippenberg}, \citenamefont {Painter},\ and\
  \citenamefont
  {Vahala}}]{spillaneIdealityFiberTaperCoupledMicroresonator2003}%
  \BibitemOpen
  \bibfield  {author} {\bibinfo {author} {\bibfnamefont {S.~M.}\ \bibnamefont
  {Spillane}}, \bibinfo {author} {\bibfnamefont {T.~J.}\ \bibnamefont
  {Kippenberg}}, \bibinfo {author} {\bibfnamefont {O.~J.}\ \bibnamefont
  {Painter}}, \ and\ \bibinfo {author} {\bibfnamefont {K.~J.}\ \bibnamefont
  {Vahala}},\ }\href {\doibase 10.1103/PhysRevLett.91.043902} {\bibfield
  {journal} {\bibinfo  {journal} {Phys. Rev. Lett.}\ }\textbf {\bibinfo
  {volume} {91}},\ \bibinfo {pages} {043902} (\bibinfo {year}
  {2003})}\BibitemShut {NoStop}%
\bibitem [{\citenamefont {{Safavi-Naeini}}\ \emph {et~al.}(2011)\citenamefont
  {{Safavi-Naeini}}, \citenamefont {Alegre}, \citenamefont {Chan},
  \citenamefont {Eichenfield}, \citenamefont {Winger}, \citenamefont {Lin},
  \citenamefont {Hill}, \citenamefont {Chang},\ and\ \citenamefont
  {Painter}}]{safavi-naeiniElectromagneticallyInducedTransparency2011a}%
  \BibitemOpen
  \bibfield  {author} {\bibinfo {author} {\bibfnamefont {A.~H.}\ \bibnamefont
  {{Safavi-Naeini}}}, \bibinfo {author} {\bibfnamefont {T.~P.~M.}\ \bibnamefont
  {Alegre}}, \bibinfo {author} {\bibfnamefont {J.}~\bibnamefont {Chan}},
  \bibinfo {author} {\bibfnamefont {M.}~\bibnamefont {Eichenfield}}, \bibinfo
  {author} {\bibfnamefont {M.}~\bibnamefont {Winger}}, \bibinfo {author}
  {\bibfnamefont {Q.}~\bibnamefont {Lin}}, \bibinfo {author} {\bibfnamefont
  {J.~T.}\ \bibnamefont {Hill}}, \bibinfo {author} {\bibfnamefont {D.~E.}\
  \bibnamefont {Chang}}, \ and\ \bibinfo {author} {\bibfnamefont
  {O.}~\bibnamefont {Painter}},\ }\href {\doibase 10.1038/nature09933}
  {\bibfield  {journal} {\bibinfo  {journal} {Nature}\ }\textbf {\bibinfo
  {volume} {472}},\ \bibinfo {pages} {69} (\bibinfo {year} {2011})}\BibitemShut
  {NoStop}%
\bibitem [{\citenamefont {Gardiner}\ and\ \citenamefont
  {Zoller}(2000)}]{gardinerQuantumNoiseHandbook2000}%
  \BibitemOpen
  \bibfield  {author} {\bibinfo {author} {\bibfnamefont {C.~W.}\ \bibnamefont
  {Gardiner}}\ and\ \bibinfo {author} {\bibfnamefont {P.}~\bibnamefont
  {Zoller}},\ }\href@noop {} {\emph {\bibinfo {title} {Quantum {{Noise}}: {{A
  Handbook}} of {{Markovian}} and {{Non}}-{{Markovian Quantum Stochastic
  Methods}} with {{Applications}} to {{Quantum Optics}}}}},\ \bibinfo {edition}
  {2nd}\ ed.\ (\bibinfo  {publisher} {{Springer}},\ \bibinfo {address}
  {{Berlin}},\ \bibinfo {year} {2000})\BibitemShut {NoStop}%
\bibitem [{\citenamefont {Purdy}\ \emph {et~al.}(2013)\citenamefont {Purdy},
  \citenamefont {Yu}, \citenamefont {Peterson}, \citenamefont {Kampel},\ and\
  \citenamefont {Regal}}]{purdyStrongOptomechanicalSqueezing2013}%
  \BibitemOpen
  \bibfield  {author} {\bibinfo {author} {\bibfnamefont {T.~P.}\ \bibnamefont
  {Purdy}}, \bibinfo {author} {\bibfnamefont {P.-L.}\ \bibnamefont {Yu}},
  \bibinfo {author} {\bibfnamefont {R.~W.}\ \bibnamefont {Peterson}}, \bibinfo
  {author} {\bibfnamefont {N.~S.}\ \bibnamefont {Kampel}}, \ and\ \bibinfo
  {author} {\bibfnamefont {C.~A.}\ \bibnamefont {Regal}},\ }\href {\doibase
  10.1103/PhysRevX.3.031012} {\bibfield  {journal} {\bibinfo  {journal} {Phys.
  Rev. X}\ }\textbf {\bibinfo {volume} {3}},\ \bibinfo {pages} {031012}
  (\bibinfo {year} {2013})}\BibitemShut {NoStop}%
\end{thebibliography}%

\end{document}